\documentclass[journal]{IEEEtran}

\usepackage{algorithm}
\usepackage{algpseudocode}
\algrenewcommand\algorithmicrequire{\textbf{Input:}}
\algrenewcommand\algorithmicensure{\textbf{Output:}}

\usepackage{cite}
\usepackage{amsmath}
\usepackage{amssymb}
\usepackage{dsfont}
\usepackage[mathcal]{euscript}
\usepackage{empheq}
\usepackage{graphicx}
\usepackage{graphics}
\usepackage{bm}
\usepackage{psfrag}
\usepackage{mathrsfs}
\usepackage{bbm}
\usepackage{color}
\usepackage{cancel}
\usepackage{amsthm}

\input{epsf}
\newtheorem{theorem}{Theorem}
\newtheorem{lemma}{Lemma}

\newtheorem{remark}{Remark}

\newtheorem{property}{Property}

\newcommand*{\QEDB}{\hfill\ensuremath{\square}}%

\renewcommand{\P}{\mathbb{P}}
\newcommand{\E}{\mathbb{E}}

\newcommand{\beq}{\begin{equation}}
\newcommand{\eeq}{\end{equation}}
\newcommand{\beqa}{\begin{eqnarray}}
\newcommand{\eeqa}{\end{eqnarray}}
\newcommand{\dfz}{\triangleq}

\title{Local Tomography of Large Networks\\ under the Low-Observability Regime}

\author{Augusto Santos$^{\ast}$, Vincenzo Matta$^\dagger$, and Ali H. Sayed$^{\star}$

\thanks{Manuscript received May 23, 2018; revised September 9, 2019; accepted September 19, 2019. A. H. Sayed was supported in part by the NSF under Grant CCF-1524250 and Grant ECCS-1407712.}

\thanks{$\ast$ A. Santos was with the Adaptive Systems Laboratory, EPFL, CH-1015 Lausanne, Switzerland (email: augusto.pt@gmail.com, augusto.santos@ist.utl.pt).}
\thanks{$\dagger$ V. Matta is with DIEM, University of Salerno, 84084 Fisciano, Italy (email: vmatta@unisa.it).}
\thanks{$\star$ A.~H.~Sayed is with the \'Ecole Polytechnique F\'ed\'erale de Lausanne (EPFL), CH-1015 Lausanne, Switzerland (email: ali.sayed@epfl.ch).}
\thanks{Communicated by I. Kontoyiannis, Associate Editor At Large.}
\thanks{Color versions of one or more of the figures in this article are available online at http://ieeexplore.ieee.org.}
\thanks{Digital Object Identifier 10.1109/TIT.2019.2945033}

}

\begin{document}


%
%
\maketitle


\begin{abstract}
This article studies the problem of reconstructing the topology of a network of interacting agents via observations of the state-evolution of the agents.
We focus on the large-scale network setting with the additional constraint of {\em partial} observations, where only a small fraction of the agents can be feasibly observed.
The goal is to infer the underlying subnetwork of interactions and we refer to this problem as {\em local tomography}.
In order to study the large-scale setting, we adopt a proper stochastic formulation where the unobserved part of the network is modeled as an Erd\H{o}s-R\'enyi random graph, while the observable subnetwork is left arbitrary.
The main result of this work is to establish that, under this setting, local tomography is actually possible with high probability, provided that certain conditions on the network model are met (such as stability and symmetry of the network combination matrix).
Remarkably, such conclusion is established under the {\em low-observability regime}, where the cardinality of the observable subnetwork is fixed, while the size of the overall network scales to infinity.
\end{abstract}


\begin{IEEEkeywords}
\textnormal{Topology inference, network tomography, graph learning, low-observability, local tomography, large-scale networks, Erd\H{o}s-R\'enyi model, random graphs, diffusion networks.}
\end{IEEEkeywords}

\section{Introduction}


In networked dynamical systems~\cite{Barrat,liggett,Queues,porterdynamical} the state of the agents comprising the network evolves over time and is affected by peer-to-peer interactions.
In general, information about the profile of interactions is unavailable. It is the goal of {\em network tomography} to infer network connectivity from observing the evolution of the graph nodes. Problems of this type arise in many domains where knowledge of the underlying topology linking the agents is critical for better inference and control mechanisms. For example, in distributed processing over networks, the underlying topology is critical for the performance delivered by the distributed strategies such as consensus~\cite{XiaoBoydSCL2004,TsitsiklisBertsekasAthansTAC1986,BoydGhoshPrabhakarShahTIT2006,DimakisKarMouraRabbatScaglioneProcIEEE2010,MouraetalTSP2011,MouraetalTSP2012,KarMouraJSTSP2011,BracaMaranoMattaTSP2008,BracaMaranoMattaWillettTSP2010,SayedProcIEEE2014,ChenSayedTIT2015part1,ChenSayedTIT2015part2} and diffusion~\cite{LopesSayedTSP2008,CattivelliSayedTSP2010,CattivelliSayedTSP2011,SayedTuChenZhaoTowficSPmag2013,MattaBracaMaranoSayedTIT2016,MattaBracaMaranoSayedTSIPN2016}; in the context of {\em epidemics}, it is well-known that the network topology may foster or hinder the outbreak of diseases or opinions~\cite{topoepidemics}; in the context of {\em brain functionality}, it is also known that the connectivity among brain regions impacts the efficiency and robustness of the brain dynamics~\cite{Morone} and can help explain brain functional disorders~\cite{alzheimer,parkinson}; in {\em cyber-security} applications it is important to determine and understand the underlying network structure to devise effective counter-measures~\cite{botnetopology}; and tomography is also a relevant problem in economics~\cite{Moneta} (in the context of causal inference) and physics applications~\cite{PhysRevE.lai}. Depending on the particular application, an appropriate model for the underlying networked dynamical system must be selected. In this work, we focus on a linear stochastic dynamical system that will be detailed in Sec.~\ref{sec:mainrespreview} --- see also Sec.~\ref{subsec:motivation} for an example.


\begin{figure} [t]
\begin{center}
\includegraphics[scale= 0.55]{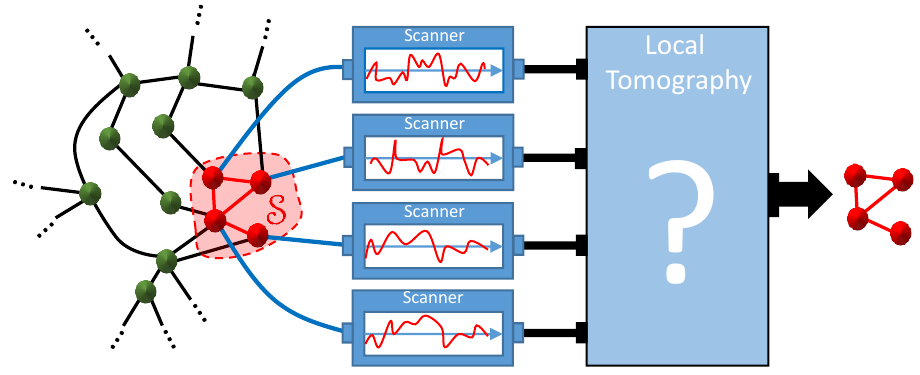}
\caption{Illustration of the {\em local tomography problem}. The goal is to design a mechanism to recover the underlying subnetwork topology by appropriately processing the observables, i.e., the state-evolution of the observable nodes.}
\label{fig:tomography}
\end{center}
\end{figure}


This article focuses on the large-scale network setting, where one can typically observe and/or process limited portions of the network. More formally, we address a {\em local} tomography problem: a subset of the agents is observed and their subnetwork of interactions is inferred from these observations. Figure~\ref{fig:tomography} depicts the local tomography paradigm. There are three main reasons that cause this observability limitation in the large-scale network setting:

$\bullet$ {\em Accessibility-limit.} Some portions of the network are not accessible and, hence, unobservable. Moreover, in many large-scale settings the existence of some sources of interactions (i.e., unobserved network links) might be unknown.

$\bullet$ {\em Probing-limit.} The acquisition of data and storage capacities can be smaller than the scale of the network.

$\bullet$ {\em Processing-limit.} The complexity of the data-mining further constrains the size of the data that can be processed.

For instance, one may probe the activity of a subset of nodes -- as it is unfeasible to track the activity of all the nodes in a large-scale network -- in order to reconstruct its underlying profile of interactions. This requires that we partially observe the system and extract information about its underlying subnetwork of interactions.


Under the aforementioned local tomography setting, the problem of inferring the subnetwork topology across the observed agents becomes exceedingly challenging or even ill-posed. It is therefore important to devise nontrivial conditions (if any) under which the problem is still well-posed, i.e., the information about the topology can be effectively inferred from the observable samples.
In this article, we show that under an appropriate setting, the problem of local tomography becomes well-posed with high probability in the thermodynamic limit: when the number of interacting agents $N$ grows, the (fixed) subnetwork topology associated with the observed agents can be perfectly recovered.
We refer to such framework as ``low-observability'' to emphasize that we are interested in studying the local tomography problem in the thermodynamic limit of large networks while the observed part is fixed and finite. Besides ascertaining conditions under which the problem is well-posed with high probability, we further derive a procedure on the space of observables to recover the subnetwork topology.
Finally, as an application of these results, we devise a strategy that shows how to learn the topology {\em sequentially}, by partitioning the observable network into small patches, and launching successive instances of the local tomography algorithm on these patches.


\subsection{Preview of the Main Result}
\label{sec:mainrespreview}
Network tomography is associated with retrieving the underlying network structure of a distributed dynamical system via observation of the output measurements of the constituent elements.
The typical formulation of the network tomography problem involves two main objects: $i)$ the statistical model that governs the laws of evolution of the (stochastic) dynamical system of interest; $ii)$ and a set of observables.
In this article, we consider a stochastic dynamical system described by a first-order Vector Auto-Regressive (VAR) or diffusion model, which is a popular theoretical model arising across many disciplines. Under this model, $N$ entities corresponding to the network agents interact over time $n$ according to the following law:
\beq
\boxed{
\bm{y}_n = A\,\bm{y}_{n-1}+\, \beta \bm{x}_n
}
\label{eq:VARmodel}
\eeq
Here, $A$ is a stable $N\times N$ matrix with nonnegative entries, and
\beqa
\bm{y}_n&=&[\bm{y}_1(n),\bm{y}_2(n),\ldots,\bm{y}_N(n)]^{\top},\\
\bm{x}_n&=&[\bm{x}_1(n),\bm{x}_2(n),\ldots,\bm{x}_N(n)]^{\top},
\eeqa
with the vector $\bm{y}_n$ collecting the state (or output measurements) at time $n$ of the $N$ agents comprising the network; and~$\bm{x}_n$ representing a random input (e.g., a source of noise or streaming data) at time $n$. The ensemble $\{\bm{x}_i(n)\}$ are independent and identically distributed (i.i.d.) both spatially (i.e., w.r.t. to index $i$) and temporally (i.e., w.r.t. to index $n$). Without loss of generality, we assume that the random variables $\bm{x}_i(n)$ have zero mean and unit variance.

\begin{figure} [t]
\begin{center}
\includegraphics[scale= 0.4]{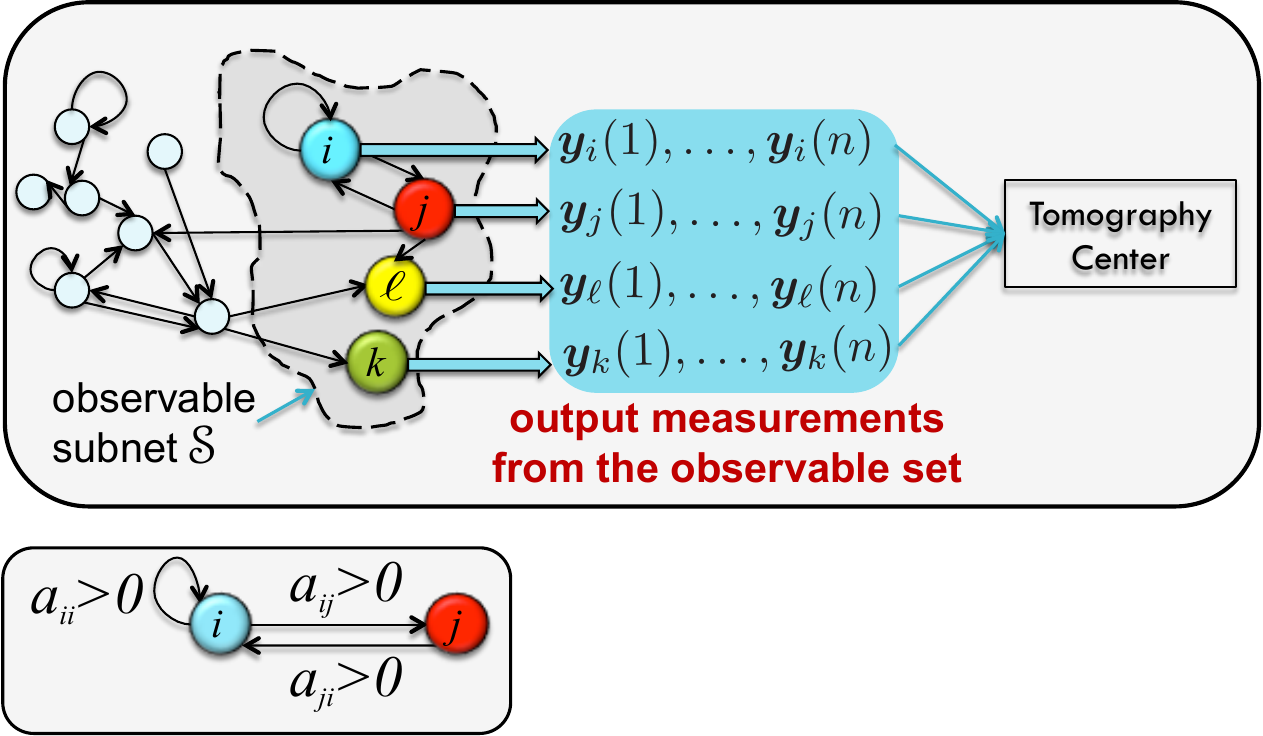}
\caption{{\bf Uppermost panel}. Illustration of the {\em local tomography problem} in connection with~(\ref{eq:1storder}).
The observable set is $\mathcal{S}=\{i,j,\ell,k\}$. The measurements delivered by these observable nodes are collected by a tomography center. {\bf Lowermost panel}. How the combination weights determine the profile of interaction (i.e., the topology).}
\label{fig:combination}
\end{center}
\end{figure}

The support-graph of $A$ reflects the underlying connections among the agents. Indeed, we have from~(\ref{eq:VARmodel}) that:
\beq\label{eq:1storder}
\bm{y}_{i}(n)=\sum\limits_{\ell=1}^N a_{i \ell} \, \bm{y}_{\ell}(n - 1) + \, \beta \bm{x}_i(n),
\eeq
which shows that, in order to update its output at time $n$, agent $i$ {\em combines} the outputs of other agents from time $n-1$. In particular, agent $i$ scales the output of agent $\ell$ by using a combination weight $a_{i\ell}$. Note that the output of agent $\ell$ is employed by agent $i$  only if $a_{i\ell}\neq 0$.
The uppermost panel in Figure~\ref{fig:combination} offers a pictorial view of the local tomography problem, whereas the lowermost panel illustrates the role of the combination weights in determining the mutual influences between nodes.

Model~\eqref{eq:1storder} further shows how the observation $\bm{y}_{i}(n)$ is affected by the source value $\bm{x}_{i}(n)$, which is available locally at agent $i$ at time $n$. 

The problem of support-graph recovery addressed in this work is generally referred to as network {\em tomography} in the literature because only {\em indirect} observations are available.
In our framework, only output measurements from a {\em subset} of the nodes are accessible, and no information is available about the unobserved nodes including their number or connectivity. We refer to this paradigm as \emph{local tomography}.
Under this challenging framework, the goal is identifiability of the topology linking the observable agents. That is, we consider the problem of inferring the topology associated with a subset $\mathcal{S}$ of {\em observable} interacting agents, by measuring only the outputs produced by such agents.

Let us ignore for a while the restriction of partial observability. It is tempting (and actually not that uncommon in the literature) to estimate the connections between the network agents by measuring the correlation between their output measurements.
There is, however, one critical issue related to the use of the correlation measure for topology inference, arising from the {\em streaming} nature of the data.
In general, when an external observer starts collecting output measurements, the network would have been in operation since some time already.
Therefore, after a transient phase, over a connected network abiding by a stable linear stochastic dynamical system as described in~\eqref{eq:VARmodel}, {\em all} agent pairs will become correlated.
In order to illustrate this point in greater detail, let us introduce the correlation matrix at time $n$, namely, $R_0(n)\dfz\E[\bm{y}_n \bm{y}_{n}^T]$.
When $A$ is symmetric (which is the case considered in this work), using~(\ref{eq:VARmodel}), and neglecting {\em transient} terms associated with the initial state $\bm{y}_0$, we have that:
\beq
\boxed{
R_0(n)= \beta^2\sum_{i=0}^{n-1} A^{2 i}
\stackrel{n\rightarrow\infty}{\longrightarrow} R_0= \beta^2(I - A^2)^{-1}
}
\label{eq:Lyapunov}
\eeq
where the latter series is guaranteed to converge whenever $A$ is a stable matrix -- all its eigenvalues lie inside the unit disc.
Assuming that the system is observed at steady-state (i.e., that the system is in operation since some time), we must focus on the limiting correlation matrix, $R_0$.
However, we immediately see from (5) that, {\em even if the correlation matrix were perfectly known}, direct retrieval of the support graph of $A$ from $R_0$ is obfuscated by the fact that the correlation matrix depends on (a superposition of) powers of $A$, and {\em not only} on $A$.
Moreover, {\em  even with full observation}, when inversion of the matrix $R_0$ can be performed, in view of equation~\eqref{eq:Lyapunov}, one would retrieve $A^2$, and {\em not} $A$.
Then, since the mapping from $A^2$ to (the support graph of) $A$ is not bijective in general, one would be faced with the inverse problem of retrieving the support graph of $A$ from $A^2$ -- such inverse problem can still be explored by properly reinforcing some sparsity constraints, refer, e.g., to~\cite{pasdeloup}.



Tomography relies primarily in developing a scheme to properly process the observables -- e.g., the state evolution of the interconnected agents -- so as to infer the underlying network structure. The above {\em na\"{i}ve} scheme, based purely on the correlation $R_0$, can be improved by introducing the one-lag correlation matrix, which, in view of~(\ref{eq:VARmodel}), takes the form:
\beq
R_1(n)\dfz\E[\bm{y}_n \bm{y}_{n-1}^T]=AR_0(n-1) \stackrel{n\rightarrow\infty}{\longrightarrow} R_1=A R_0.
\label{eq:R1nR0n}
\eeq
Therefore, we obtain the following relationship:
\beq
\boxed{
A=R_1R_0^{-1}
}
\label{eq:fullinvert}
\eeq
In principle, since there exist many ways to estimate $R_0$ and $R_1$ consistently as $n\rightarrow\infty$, expression~(\ref{eq:fullinvert}) reveals one possible strategy to estimate $A$ (and hence its support-graph) from the observations.

Topology estimation based on relation~\eqref{eq:fullinvert} is viable whenever {\em full-observation} of the system is permitted. Under a {\em partial} observability restriction, however, when only a subset $\mathcal{S}$ of the network is accessible, only the covariance submatrices associated with the observable agents, denoted by $\left[R_0\right]_{\mathcal{S}}$ and $\left[R_1\right]_{\mathcal{S}}$, are available.
One is certainly free to introduce a truncated version of~(\ref{eq:fullinvert}), say, as:
\begin{equation}
\boxed{
\widehat{A}_{\mathcal{S}}=\left[R_1\right]_{\mathcal{S}}\left(\left[R_0\right]_{\mathcal{S}}\right)^{-1}
}
\label{eq:truncinvert}
\end{equation}
It is clear from basic linear algebra that~(\ref{eq:truncinvert}) is in general distinct from the {\em ground truth} matrix $A_{\mathcal{S}}=\left[R_1 R^{-1}_0\right]_{\mathcal{S}}$, namely, from the combination matrix corresponding to the subnetwork connecting the observable agents $\mathcal{S}$ whose support must be inferred.

Despite this difference, it has been shown in the recent work~\cite{tomo,tomo_icassp} that the support of the observable network can still be recovered ({\em consistent tomography}) through the truncated estimator in~(\ref{eq:truncinvert}), under certain conditions that can be summarized as follows: $i)$ the overall network graph is drawn from a connected Erd\H{o}s-R\'enyi random graph with vanishing connection probability; $ii)$ the cardinality of the observed subnetwork grows linearly with the size of the overall network; $iii)$ the matrix $A$ is a symmetric combination matrix belonging to a certain class.



The work~\cite{tomo} leads to several insightful conclusions about network tomography for Erd\H{o}s-R\'enyi models. In this work, we pursue the same network tomography problem albeit for a different more demanding network setting, explained below, which will require new arguments and lead to new results that extend the results from~\cite{tomo}. In particular, the proof techniques used here will rely on graph theoretical techniques and on special graph constructs to arrive at the important conclusion that the na\"{i}ve truncated estimator~\eqref{eq:truncinvert} is still able to deliver consistent tomography under partial network observations and more relaxed requirements. 

It is worth noting that this work establishes a strong consistency result (Theorem 1) that does not require an independence condition. The main features of the framework proposed in the current manuscript in comparison with \cite{tomo} can be summarized as follows:

\noindent
{\em --- Topology of the accessible network portion.}  We assume that the subnetwork connecting the observable agents has an {\em arbitrary} topology, which is modeled through a {\em deterministic} graph. This subgraph is the object of inference. The remaining unobserved part is assumed to be drawn from an Erd\H{o}s-R\'enyi {\em random} graph. The overall network construction is therefore referred to as a {\em partial} Erd\H{o}s-R\'enyi model. It is useful to interpret and motivate this model in the classical ``signal plus noise'' paradigm in the following sense.
For what concerns the object of the inference (i.e., the support graph of the observable nodes), it is modeled as an arbitrary deterministic signal.
For what concerns the undesired component (i.e., the unobserved subnet), it is modeled as a {\em noisy} component.
To get insightful results, we must choose some model for this random component. In the absence of any prior information, it is meaningful to opt for a uniform model, namely, the Erd\H{o}s-R\'enyi {\em random} graph, where the presence/absence of each edge is determined through a sequence of i.i.d. Bernoulli experiments.
Accordingly, few connections (i.e., high sparsity in the unobserved portion) take on the meaning of a controlled noise level. In contrast, in~\cite{tomo} it was assumed that the {\em overall} network (observed portion $+$ unobserved portion) is drawn from an Erd\H{o}s-R\'enyi {\em random} graph. Such a construction poses limitations on the subgraph that we wish to identify, which cannot be selected in an arbitrary fashion any longer.
Moreover, the network construction used in~\cite{tomo} assumes a {\em vanishing fraction of connected nodes} within the observable set; a condition that is removed in the current analysis.

\noindent
{\em --- Cardinality of the accessible network portion.}
In~\cite{tomo}, it was assumed that the cardinality $S$ of the observed subset~$\mathcal{S}$ scales linearly with $N$, so that the ratio $S/N$ converges to some positive fraction~$\xi\in \left(\left.0,1\right]\right.$ as $N\rightarrow\infty$.
In contrast, we assume here that the subnetwork of observable nodes~$\mathcal{S}$ is fixed. This means that, in our framework, we focus on retrieving the support of a subnetwork $\mathcal{S}$ that is embedded in a network that becomes infinitely larger as $N\rightarrow\infty$, i.e., the size of the unobserved component becomes asymptotically dominant. The resulting regime is accordingly referred to as a {\em low-observability} regime.
Such a model is particularly relevant, for example, in circumstances where we have a large network, and we are constrained to perform probing actions at few accessible locations.
We remark that the case where the ratio $S/N$ goes to zero is not addressed (nor can be obtained from the results) in~\cite{tomo}.

\noindent
{\em --- Consistent tomography.}
The main result of the present work (Theorem~$1$ further ahead) is to establish that consistent tomography is achievable under the aforementioned setting.
We shall prove that, if the unobserved network is drawn from an Erd\H{o}s-R\'enyi random graph with connection probability
\beq
\boxed{
p_N=\frac{\log N + c_N}{N}
}
\label{eq:pconncond}
\eeq
where $c_N$ is a sequence that diverges as $N\rightarrow\infty$, satisfying the condition:
\beq
\boxed{
\frac{[\log(\log N + c_N)]^2}{\log N}\rightarrow 0
}
\label{eq:cnconst}
\eeq
then the (arbitrary) support graph of $A_{\mathcal{S}}$ can be recovered through the truncated estimator $\widehat{A}_{\mathcal{S}}$, with probability tending to one as $N\rightarrow\infty$.
More specifically, in this work we are able to establish consistency in that each entry of the support graph is recovered perfectly in the limit as $N\rightarrow \infty$.
In~\cite{tomo}, where the object of estimation has cardinality growing with $N$, consistency is not formulated in terms of an entry-by-entry recovery. Instead, consistency there is formulated in terms of two macroscopic indicators, namely, the fraction of correctly classified interacting pairs, and the fraction of correctly classified non-interacting pairs. Both fractions are proved to converge to one as $N$ grows to infinity.

Another difference with respect to~\cite{tomo} relates to the connection probability of the Erd\H{o}s-R\'enyi graph.
Having a sufficiently small $p_N$ translates into a sufficient degree of sparsity.
In other words, if we interpret the unobserved network as a noisy component, the noise in the system cannot exceed a certain threshold to grant perfect reconstruction.
For the setting considered in~\cite{tomo}, a connection probability vanishing as in~(\ref{eq:pconncond}) was sufficient to achieve consistent tomography, without additional constraints on $c_N$.
On the other hand, for the results of this work to hold, we need the additional constraint in~(\ref{eq:cnconst}), which corresponds to invoking slightly more sparsity.

We remark that the results of this work allow drawing some useful conclusions also in relation to the setting addressed in~\cite{tomo}, namely, in relation to the case of a full Erd\H{o}s-R\'enyi construction with $S$ growing linearly with $N$.
We find it convenient to postpone the comments on this particular issue to Sec.~\ref{sec:comparison}, because some technical details are necessary for a proper explanation.

In summary, in this work we address a network tomography problem in the {\em low-observability} regime, where the cardinality of the accessible network portion is fixed and the number of unobserved nodes scales to infinity. This challenging regime cannot be dealt with by using the analysis provided in~\cite{tomo}. In comparison with~\cite{tomo}, where the accessible network was Erd\H{o}s-R\'enyi with vanishing connection probability, in this work we consider an {\em arbitrary} topology for the accessible network. Having the flexibility of an arbitrary topology is critical because, in the low-observability regime, an Erd\H{o}s-R\'enyi accessible subgraph with vanishing connection probability would become a trivial subgraph that is totally-disconnected as $N\rightarrow \infty$. Moreover, the result in~\cite{tomo} holds in a weak sense: it shows that the fraction of correctly identified edges converges to one in the limit and an independence approximation is required to control the error rate. Remarkably, in this work consistency holds in a strong sense, meaning that the whole subgraph is exactly recovered as $N\rightarrow \infty$, and no independence approximation is required.

\subsection{Related work}

The existing approaches to network reconstruction can be categorized based on two major features:
\\
\noindent
--- $\mathscr{F}_1$: Class of networked dynamical systems governing the state-evolution of the agents, e.g., the diffusion model in~\eqref{eq:VARmodel}, and related observables, e.g., the process $\bm{y}_n$ in~\eqref{eq:VARmodel}.
\\
\noindent
--- $\mathscr{F}_2$: Topology-retrieval methods that should exploit the relation between the observables and the underlying support-topology.
Such methods are sensitive to the dynamics and the observables arising from the model in $\mathscr{F}_1$.

Regarding $\mathscr{F}_1$, most works focus on linear systems. Nonlinear dynamics are often dealt with by linearizing via considering variational characterizations of the dynamics (under small-noise regimes)~\cite{Ching2017ReconstructingLI, Napoletani, Noise_Jien_Ren} or by appropriately increasing the dimension of the observable space~\cite{ScienceRobustNetInference, Koopman_Gon}. In the context of linear (or linearized) systems, particular attention is paid to autoregressive diffusion models~\cite{Moneta, Geigeretal15, pasdeloup, SantiagoTopo, MeiMoura}.

For what concerns $\mathscr{F}_2$, the majority of the literature considers methods aimed at identifying commonalities between correlation constructs and graph topologies. We now make a brief summary of the available results as regards the existing topology-retrieval methods that are more closely related to our setting. To get some ideal benchmark, it is useful to start with the full observation case, and then focus on the case of interest of {\em partial} observation.

\noindent
--- {\em Tomography under full observations}.
In~\cite{Kiyavash1}, the authors introduce {\em directed information graphs}, which are used to reveal the dependencies in networks of interacting processes linked by causal dynamics.
Such a setting is enlarged in~\cite{Kiyavash2}, where a metric is proposed to learn causal relationships by means of {\em functional dependencies}, over a (possibly nonlinear) dynamical network.
Causal graph processes are exploited in~\cite{MeiMoura}, where an algorithm (with a detailed convergence analysis) is proposed for topology recovery.
Recently, the inverse problem of recovering the topology via correlation structures has been addressed through optimization-based methods, by reinforcing some (application-dependent) structural constraints such as sparsity, stability, symmetry.
For instance, in~\cite{pasdeloup, SantiagoTopo}, since the combination matrix and the correlation matrix share the same eigenvectors, the set of candidate topologies is reduced by computing these eigenvectors, and the inverse problem is then tackled with optimization methods under sparsity constraints.

An account of topology inference from node measurements (still under the {\em full} observations regime) is offered in~\cite{MaterassiSalapakaTAC2012}, where a {\em general} linear model is considered and an approach based on Wiener filtering is proposed to infer the topology.

However, as already noted in~\cite{MaterassiSalapakaTAC2012} a Wiener filtering approach is redundant, since exact topology recovery can be obtained (with full observations) through the estimator in~(\ref{eq:fullinvert}).
As it is well known, this solution admits the following useful interpretation: the combination weights $\{a_{i j}\}_{j=1}^N$ obtained through~(\ref{eq:fullinvert}) are the coefficients of the {\em best one-step linear predictor} (a.k.a., in the context of causal analysis, as {\em Granger estimator}), i.e., they yield the minimum expected squared error in estimating $\bm{y}_{i}(n)$ from the past samples $\{\bm{y}_{j}(n-1)\}_{j=1}^N$ -- see, e.g.,~\cite{Sayed2008adaptive}.
We remark that the case where~(\ref{eq:fullinvert}) is applied with correlation matrices {\em estimated empirically from the measurements} provides the best one-step linear predictor in a least-squares sense (i.e., when the {\em expected} squared error is replaced by the {\em empirical} squared error evaluated on the measurements collected over time).
However, all the aforementioned results pertain to the case where node measurements from the whole network are available.
It is instead necessary to consider the case when only partial observation of the network is permitted.

\noindent
--- {\em Tomography under partial observations, identifiability}.
The case of partial observations is addressed in~\cite{MaterassiSalapakaCDC2012, KiyavashPolytrees}, for cases when the network graph is a polytree. 

The case of more general topologies is instead addressed in~\cite{Geigeretal15, MaterassiSalapakaCDC2015}, where technical conditions for exact or partial topology identification  are provided.
It is useful to contrast such identifiability conditions with the approach pursued in the present work.
Basically, the identifiability conditions offered in~\cite{Geigeretal15, MaterassiSalapakaCDC2015} act at a ``microscopic'' level, namely, they need a detailed knowledge of the topology and/or the statistical model (e.g., type of noise, joint distribution of the observable data).
For these reasons, the approach is not practical for large-scale network settings (which are the main focus of this work).

In contrast, in this work we pursue a statistical asymptotic approach that is genuinely tailored to the large-scale setting: the conditions on the network topology are described at a {\em macroscopic} level through average descriptive indicators, such as the connection probability between any two nodes. Under these conditions, we focus on establishing an achievability result that holds (in a {\em statistical} sense) as the size of the network scales to infinity.

\noindent
--- {\em Tomography under partial observations, methods}.
As already noted, the classic, exact solution to the topology problem under full observation is provided by~(\ref{eq:fullinvert}), and arises from the solution of a one-step linear prediction problem~\cite{MaterassiSalapakaTAC2012, Geigeretal15}.
Under partial observations, we propose to keep the same approach, except that the best one-step linear prediction is enforced {\em on the observable nodes only}.
As a matter of fact, the combination weights estimated through~(\ref{eq:truncinvert}) provide the best one-step linear prediction of the {\em observable} measurement $\bm{y}_{i}(n)$ (for $i\in\mathcal{S}$) from the past {\em observable} measurements $\{\bm{y}_{j}(n-1)\}_{j\in\mathcal{S}} $.
We remark that this solution, which can still be interpreted as a Granger estimator, is widely adopted in causal inference from time series, when one ignores and/or neglects the existence of latent components.
However, there is in principle no guarantee that such an estimator can provide reliable tomography.
Our main goal is to establish that it actually can, under the demanding setting illustrated in Sec.~\ref{sec:mainrespreview}.

{\em --- Connections with graphical models}.
In a nutshell, a graphical model can be described as a collection of random variables indexed by nodes in a network, whose pairwise relationships (which determine the topology, i.e., the undirected graph) are encoded in a Markov random field.
One of the fundamental problems in graphical models is to retrieve the network topology by collecting measurements from the network nodes.
It is useful to comment on some fundamental differences, as well as useful commonalities, between the graphical model setting and our problem.

In the standard graphical model formulation (and, hence, in the vast majority of the available related results) the network evolution over time (e.g., the dynamical system in~(\ref{eq:VARmodel})) is not taken into account. Rather, the samples $\left\{{\bm y}_n\right\}_{n\in\mathbb{N}}$ are assumed independent across the index $n$. This difference has at least two relevant implications.

The first difference pertains to the type of estimators used for topology retrieval.
For example, in a Gaussian graphical model, the inverse of the correlation matrix~$R_0^{-1}$, a.k.a. {\em concentration matrix}, contains full information of the graph topology: the $(i,j)$-th entry of the concentration matrix is nonzero if, and only if, nodes $i$ and $j$ are connected.
In contrast, we see from the Granger estimator in~(\ref{eq:fullinvert}) that in our case an additional operation is needed (namely, multiplication with the one-lag correlation matrix $R_1$) to obtain the matrix that contains the topology information (in our case, the combination matrix, $A$). This difference is an inherent consequence of the system dynamics described by the first-order VAR model in~(\ref{eq:VARmodel}).
Second, the dynamical system ruling the network evolution usually enforces some degree of dependence between subsequent measurements. For this reason, while in our case the observations collected over time are correlated, in the standard graphical model formulation the samples upon which the topology inference is based are usually assumed statistically independent.
Keeping in mind these fundamental distinctions, we now list some recent works about topology recovery on graphical models.


The idea of studying the large-network behavior through an Erd\H{o}s-R\'enyi model has been applied in~\cite{AnandkumarWalkSummabilityJMLR}, where the emergence of ``large'' paths over the random graph (a property that we will use in our treatment) has been exploited for topology inference. However, reference~\cite{AnandkumarWalkSummabilityJMLR} addresses the case of full observations.
Instead, for the case of partial observations, in~\cite{AnandkumarValluvanAOS} an efficient method is proposed, which is suited for the case of large-girth graphs, such as, e.g., the bipartite Ramanujan graphs and the random Cayley graphs.
In~\cite{ChandrasekaranParriloWillsky}, still for the case of partial observations, an inference method is proposed under the assumption that the connection matrix is sparse, whereas the error matrix associated to the latent-variables component exhibits a low-rank property.

In summary, contrasted with recent results about topology recovery on graphical models, the results obtained in the present work constitute an advance because: $i)$ we deal with a dynamical system, see~(\ref{eq:VARmodel}); $ii)$ the partial observations setting considered in the present work relies on assumptions different from those used in~\cite{ChandrasekaranParriloWillsky,AnandkumarValluvanAOS}: in our case the unobserved component is Erd\H{o}s-R\'enyi, but the subnetwork of observable nodes is deterministic and arbitrary, and the combination matrix obeys transparent conditions borrowed from the adaptive networks literature.
We believe that the possibility of working with dynamical models, the arbitrariness of the monitored subnetwork, as well as the direct physical meaning of the conditions on the combination matrix, provide useful novel insights on the problem of topology inference under partial observations.

To sum up, our major contribution lies in establishing technical guarantees for graph structural consistency of the Granger estimator applied to the subset of observable nodes. This is formally stated in the main result of this paper, Theorem~$1$.

Short versions of this work were reported in~\cite{tomo_isit,tomo_dsw}.

\subsection{Motivating Example: Adaptive Diffusion Networks}\label{subsec:motivation}
A network of $N$ agents observes a spatially and temporally i.i.d. sequence of zero-mean and unit-variance streaming data $\{\bm{x}_i(n)\}$, for $n=1,2,\ldots$, and $i=1,2,\ldots,N$.
Here, the letter $n$ refers to the time index while the letter $i$ refers to the node index. In order to track drifts in the phenomenon they are monitoring, the network agents implement an adaptive diffusion strategy~\cite{SayedProcIEEE2014, Sayed, MattaBracaMaranoSayedTIT2016}, where each individual agent relies on local cooperation with its neighbors.
One useful form is the combine-then-adapt (CTA) rule, which has been studied in some detail in these references. It involves two steps: a combination step followed by an adaptation step.

During the first step, agent $i$ {\em combines} the data of its neighbors through a sequence of convex (i.e., nonnegative and adding up to one) combination weights $w_{i\ell}$, for $\ell=1,2,\dots,N$. The combination step produces the intermediate variable:
\beq
\bm{v}_i(n-1)=\sum_{\ell=1}^N w_{i\ell}\,\bm{y}_{\ell}(n-1).
\label{eq:combine}
\eeq
Next, during the {\em adaptation} step, agent $i$ updates its output variable by comparing against the incoming streaming data $\bm{x}_i(n)$, and updating its state by using a small step-size $\mu\in(0,1)$:
\beq
\bm{y}_i(n)=\bm{v}_i(n-1) + \mu[\bm{x}_i(n) - \bm{v}_i(n-1)].
\label{eq:adapt}
\eeq
Merging~(\ref{eq:combine}) and~(\ref{eq:adapt}) into a single step yields:
\beq
\bm{y}_i(n)=(1-\mu)\sum_{\ell=1}^N w_{i\ell}\, \bm{y}_{\ell}(n-1) + \mu\,\bm{x}_i(n).
\label{eq:CTA1}
\eeq
It is convenient for our purposes to introduce a combination matrix, which we denote by $A$, whose entries are obtained by scaling the weights $w_{ij}$ as follows:
\beq
a_{ij}\dfz(1-\mu)w_{ij}.
\label{eq:aweightsmatdef}
\eeq
With this definition, we see immediately that~(\ref{eq:CTA1}) corresponds to~(\ref{eq:VARmodel}) or~\eqref{eq:1storder} with $\beta=\mu$. Note that under this diffusion framework, the matrix $A$ is naturally nonnegative and if we assume symmetry, its normalized counterpart $A/(1-\mu)$ is doubly stochastic.


\section{Notation and Definitions}
\noindent
We list our notation and some definitions used in later sections for ease of reference.

\subsection{Symbols}\label{sec:symbols}
We represent sets and events by upper-case calligraphic letters, and the corresponding normal font letter will be used to denote the set cardinality. For instance, the cardinality of set ${\cal S}$ is $S$. The complement of a set $\mathcal{S}$ is denoted by $\mathcal{S}'$.

Standard canonical sets follow a different convention, for instance, the set of natural numbers is denoted by~$\mathbb{N}=\{1,2,3,\ldots\}$, and the set of $N\times N$ symmetric matrices 
with nonnegative entries by $\mathbb{S}^{N\times N}_{+}$.

We use boldface letters to denote random variables, and normal font letters for their realizations. Capital letters refer to matrices, small letters to both vectors and scalars. Sometimes we violate the latter convention, for instance, we denote the total number of network agents by $N$.

Given an $N\times N$ matrix $Z$, the submatrix that lies in the rows of $Z$ indexed by the set $\mathcal{S}\subseteq\{1,2,\ldots, N\}$ and in the columns indexed by the set $\mathcal{T}\subseteq\{1,2,\ldots,N\}$, is denoted by $Z_{\mathcal{S} \mathcal{T}}$, or alternatively by $[Z]_{\mathcal{S} \mathcal{T}}$. When $\mathcal{S}=\mathcal{T}$, the submatrix $Z_{\mathcal{S} \mathcal{T}}$ will be abbreviated as $Z_{\mathcal{S}}$ or $\left[Z\right]_{\mathcal{S}}$. In the indexing of the submatrix we will retain the index set of the original matrix. For example, if $\mathcal{S}=\{2,3\}$ and $\mathcal{T}=\{2,4,5\}$, we have that the submatrix $M=Z_{\mathcal{S} \mathcal{T}}$ is a $2\times 3$ matrix, indexed as follows:
\beq
M=
\left(
\begin{array}{llll}
z_{22}&z_{24}&z_{25}\\
z_{32}&z_{34}&z_{35}
\end{array}
\right)
=\left(
\begin{array}{llll}
m_{22}&m_{24}&m_{25}\\
m_{32}&m_{34}&m_{35}
\end{array}
\right).
\eeq
This notation is crucial in our treatment, since it will allow us to identify nodes without cumbersome and redundant double-index notation.

Finally, $\mathds{1}$ denotes a column vector with all its entries equal to one; $0_{N\times N}$ denotes an $N\times N$ matrix with all its entries equal to zero; $\mathbb{I}_{\mathcal{E}}$ denotes the indicator function, which is equal to one if condition $\mathcal{E}$ is true, and is equal to zero, otherwise; the $N\times N$ identity matrix is denoted by $I$; and $\log(\cdot)$ denotes the natural logarithm.

\subsection{Graph notation}
\label{sec:graphnotation}

The set of all undirected graphs that can be defined on a set of nodes (vertex set) $\mathcal{V}$ is denoted by $\mathcal{G}(\mathcal{V})$.
When $N$ is the number of nodes, the notation $\mathcal{G}(N)$ implies that the vertex set is $\mathcal{V}=\{1,2,\ldots,N\}$.

When dealing with a graph $G\in\mathcal{G}(N)$, its connection structure (i.e., the edges of the graph) can be described through its $N\times N$ adjacency matrix. The $(i,j)$-th entry of the adjacency matrix of the graph $G$ will be denoted by the lower-case symbol $g_{ij}$, with $g_{ij}=1$ if the nodes $i$ and $j$ are connected, and $g_{ij}=0$ otherwise. Henceforth, we assume that~$g_{ii}=1$, i.e., all nodes exhibit self-loops. This reflects the fact that usually each agent uses information from its own output measurement to update its state.

Given $G\in\mathcal{G}(N)$, and a subset $\mathcal{S}\subseteq\{1,2,\ldots,N\}$, the subgraph corresponding to $\mathcal{S}$ is denoted by $G_{\mathcal{S}}\in\mathcal{G}(\mathcal{S})$. The support graph of a matrix $A$ is denoted by $G(A)$. The $(i,j)$-th entry of its adjacency matrix is $\mathbb{I}_{\{a_{ij}>0\}}$, namely, nodes $i$ and $j$ are connected on $G(A)$ if, and only, if $a_{ij}$ is strictly positive.

A path from $i$ to $j$ is a sequence of edges where the first edge originates from $i$ and the last edge terminates at $j$. The existence of a path of length $r$ can be expressed as:
\beq
g_{i n_1} g_{n_1 n_2} \ldots g_{n_{r-1} j} =1,
\eeq
for a certain sequence of vertices $n_1, n_2,\ldots, n_{r-1}$ belonging to $\mathcal{V}$.
According to this definition, a path can also pass multiple times through the same node, or can linger for one or more steps at the same node when it has a self-loop.

The set of neighbors of the node $i$ (including $i$ itself) in the undirected graph $G$ will be denoted by $\mathcal{N}_i(G)$.
The degree of the node $i$ is the cardinality of $\mathcal{N}_i(G)$, whereas $d_{\max}(G)$ is the maximum degree in $G$.
Likewise, the $r$-th order neighborhood of the node $i$ (including $i$ itself) is denoted by $\mathcal{N}^{(r)}_i(G)$, and is formally given by:
\beq
\mathcal{N}^{(r)}_i(G)=\{j\in \mathcal{V}\,:\, \delta_{i,j}(G)\leq r\},
\eeq
where $\delta_{i,j}(G)$ is the distance between the nodes $i$ and $j$ on the graph $G$, i.e., the length of the {\em shortest} path linking $i$ and $j$.

A {\em random} graph $\bm{G}$ obeys the Erd\H{o}s-R\'enyi model if each edge of $\bm{G}$ is drawn, independently from the other edges, with identical probability $p_N$. Equivalently stated, the adjacency random variables $\bm{g}_{ij}$, for $i=1,2,\ldots,N$ and $i< j$, are independent and identically distributed (i.i.d.) Bernoulli variables. The notation $\bm{G}\sim\mathscr{G}^{\star}(N,p_N)$ signifies that the graph $\bm{G}$ belongs to the Erd\H{o}s-R\'enyi class with connection probability that vanishes as $N\rightarrow \infty$, and that obeys the following scaling law:
\beq
p_N=\frac{\log N + c_N}{N},
\label{eq:pnscale}
\eeq
where $c_N\rightarrow\infty$ as $N\rightarrow\infty$ (in an arbitrary way, provided that $p_N\rightarrow 0$).
It is a well-known result that random graphs belonging to the family $\mathscr{G}^{\star}(N,p_N)$ are connected with high probability~\cite{erd}.

\begin{remark}\label{re:probspace}
As a note of clarity, in the forthcoming treatment, we assume that all random variables find domain in a common probability space $\left(\Omega, \mathcal{F},\mathbb{P}\right)$, where $\Omega$ is the set of realizations, $\mathcal{F}$ is the sigma-algebra of measurable sets and $\mathbb{P}$ is the probability measure. For instance, the event
\begin{equation}\label{eq:distancesm}
\left\{\omega\in\Omega \,:\, \delta_{i,j}\left({\bm G}(\omega)\right)\leq r\right\}\in \mathcal{F},
\end{equation}
represents the set of realizations $\omega\in \Omega$ yielding a graph ${\bm G}(\omega)$ whose distance between the (fixed) nodes $i$ and $j$ does not exceed $r$. To render a more compact notation, we henceforth omit the realization~$\omega$ in the characterization of the events. For instance, in the case of the event~\eqref{eq:distancesm} we represent it rather as
\begin{equation}\label{eq:event}
\left\{\delta_{i,j}\left({\bm G}\right)\leq r\right\},
\end{equation}
where the random quantities are emphasized by the boldface letter -- in the event in~\eqref{eq:event}, the only random object is the graph ${\bm G}$ as it is the only boldface variable.
\end{remark}

\begin{figure} [t]
\begin{center}
\includegraphics[scale= 0.6]{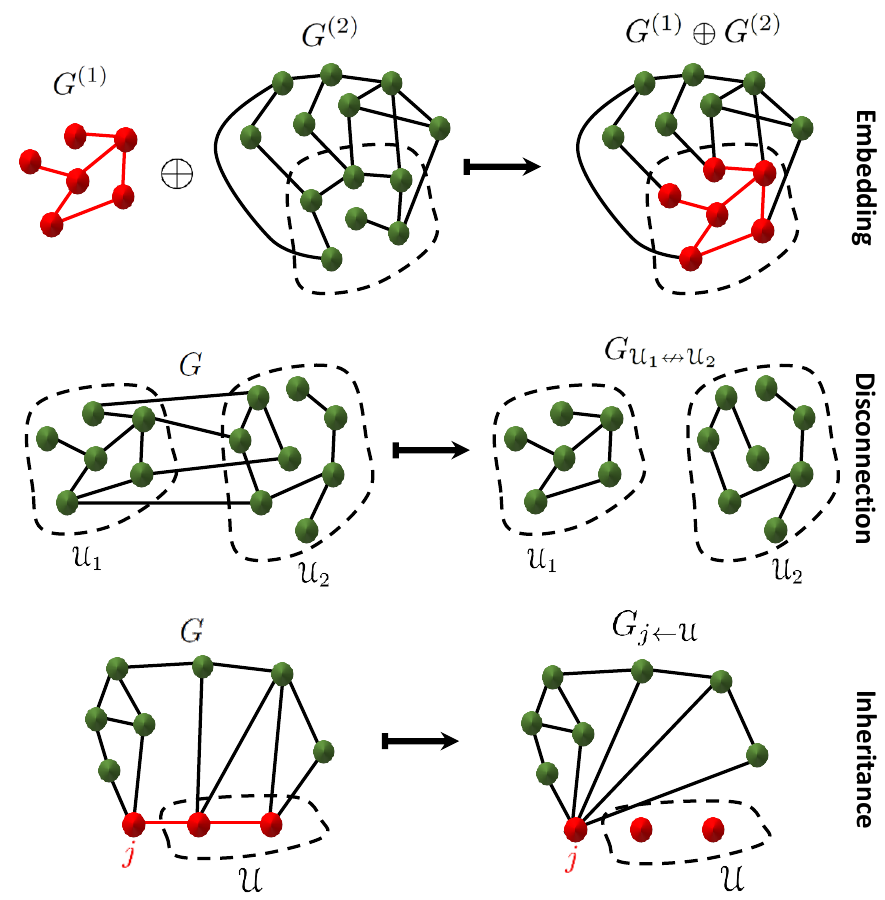}
\caption{Summary of the graph operations defined in Sec.~\ref{sec:graphop} related to embedding, disconnection, and inheritance.}\label{fig:summary}
\end{center}
\end{figure}


\subsection{Useful Graph Operations}
\label{sec:graphop}
In our exposition, we will be performing certain operations over graphs, as well as evaluate certain functions such as comparing distances between nodes over distinct graphs. Therefore, it is useful to introduce the following graph operations for later use (which are illustrated in Figure~\ref{fig:summary}):
\begin{enumerate}

\item
{\em Graph embedding}.
Given a vertex set $\mathcal{V}$, and a subset thereof, $\mathcal{S}\subset \mathcal{V}$, the embedding of a graph $G^{(1)}\in\mathcal{G}(\mathcal{S})$ into the {\em larger} graph $G^{(2)}\in\mathcal{G}(\mathcal{V})$ will be denoted by:\footnote{In order to avoid confusion, we remark that the symbol $\oplus$ is also used, in the graph literature, to denote a different kind of operation called ``ring sum''. However, we prefer to denote our embedding operation by the same summation symbol  to emphasize the ``signal+noise'' structure that is relevant in our application.}
\beq
G=G^{(1)}\oplus\,G^{(2)},\qquad G\in\mathcal{G}(\mathcal{V}),
\label{eq:gembed}
\eeq 
and results in a graph with the following properties: $i)$ the connections between nodes in $\mathcal{S}$ that are present in $G^{(2)}$ are cancelled; $ii)$ the nodes in the vertex set $\mathcal{S}$ of graph $G^{(1)}$ are mapped into the corresponding nodes of graph $G^{(2)}$, and so are the pertinent connections.
We stress that the connections from $\mathcal{S}'$ to $\mathcal{S}$ are determined by the graph $G^{(2)}$.
We notice that the operation in~(\ref{eq:gembed}) is not commutative (because the first graph is embedded into the second graph, and not vice versa), and that the output graph $G$ does not depend on the connections existing in $G^{(2)}$ among nodes belonging to the set $\mathcal{S}$.



\item
{\em Local disconnection}.
Given a graph $G\in\mathcal{G}(\mathcal{V})$, the notation:
\beq
G_{\mathcal{U}_1\nleftrightarrow \mathcal{U}_2}\,\in\mathcal{G}(\mathcal{V}),
\eeq
describes the graph that is obtained from $G$ by removing all the edges that connect nodes in $\mathcal{U}_1$ to nodes in $\mathcal{U}_2$, namely, all the connections between $\mathcal{U}_1$ and $\mathcal{U}_2$.

\item
{\em Connections inheritance}.
The notation:
\beq
G_{j\leftarrow \mathcal{U}}\,\in\mathcal{G}(V),
\eeq
describes the graph that is obtained from $G$ through the following chain of operations: $i)$ all edges within $\mathcal{U}$ are removed; $ii)$ all edges connecting nodes in $\mathcal{U}$ to the rest of the network are removed; $iii)$ all connections from $\mathcal{U}$ to the rest of the network are {\em inherited by node $j$}.
\end{enumerate}

All the above graph operations preserve self-loops unless otherwise stated.

\section{Problem Formulation}\label{sec:probfor}

Consider a graph $G(\mathcal{V})$ and assume we are able to observe data from a subset $\mathcal{S}\subset \mathcal{V}$ of the nodes. From these observations, we would like to devise a procedure that allows us to discover the connections among the nodes in $\mathcal{S}$, under the assumption that the structure of the graph in the complement set, $\mathcal{S}'$, and as well as the connections between $\mathcal{S}$ and $\mathcal{S}'$, will be random, following i.i.d. drawing of the pertinent edges.
The desired construction can be formally described as follows.

Let $G^{\textnormal{(obs)}}\in\mathcal{G}(\mathcal{S})$ be a deterministic graph on the observable set $\mathcal{S}$, with some unknown topology (which is not restricted in any way), and let $\bm{G}^{\textnormal{(unobs)}}\sim\mathscr{G}^{\star}(N,p_N)$ be an Erd\H{o}s-R\'enyi random graph on $N$ nodes.
We assume that the overall network graph, $\bm{G}$, is of the form:
\beq
\boxed{
\bm{G}=G^{\textnormal{(obs)}}\oplus \bm{G}^{\textnormal{(unobs)}}
}
\label{eq:partialERconstruct}
\eeq
Specifically, the connections within the observable set $\mathcal{S}$ are described through the graph $G^{\textnormal{(obs)}}$, while the connections within $\mathcal{S}'$, as well as the connections between $\mathcal{S}'$ and $\mathcal{S}$, are described through the graph $\bm{G}^{\textnormal{(unobs)}}$. Note that $\bm{G}^{\textnormal{(unobs)}}$ is an Erd\H{o}s-R\'enyi random graph on $N$ nodes, but its subgraph $\bm{G}^{\textnormal{(unobs)}}_{\mathcal{S}}$ is replaced by $G^{\textnormal{(obs)}}$ in characterizing $\bm{G}$, in view of~\eqref{eq:partialERconstruct} (refer also to Figure~\ref{fig:summary}). Therefore, the structure of $\bm{G}^{\textnormal{(unobs)}}$ within the observable subnet becomes immaterial. Equation~(\ref{eq:partialERconstruct}) highlights the ``signal+noise'' construction, with the boldface notation emphasizing the random (i.e., noisy) component that corresponds to the unobserved network portion, and with the normal font emphasizing the deterministic component that corresponds to the arbitrary topology of the observed network portion.

\begin{figure} [t]
\begin{center}
\includegraphics[scale= 0.6]{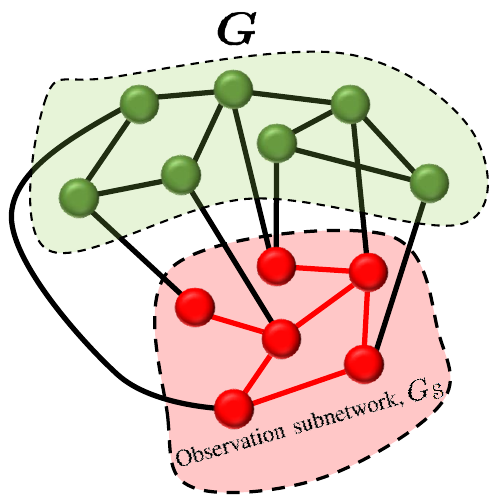}
\caption{Observations are collected from the red subgraph $G_{\mathcal{S}}$ (comprised by the red nodes along with the red edges). The black edges, i.e., the edges connecting green-green and green-red nodes are assumed to be drawn randomly with probability $p_N$.}\label{fig:partial}
\end{center}
\end{figure}

The aforementioned construction will be referred to as a {\em partial} Erd\H{o}s-R\'enyi graph.
The class of partial Erd\H{o}s-R\'enyi random graphs with a deterministic graph component $G^{\textnormal{(obs)}}$ placed on the set $\mathcal{S}$, will be formally represented by the notation $\mathscr{G}^{\star}(N,p_N,G^{\textnormal{(obs)}})$.
We shall often refer to the observable graph over $\mathcal{S}$ by the simpler notation $G_{\mathcal{S}}$. As such, we can also write,
\beq
\boxed{
\bm{G}\sim\mathscr{G}^{\star}(N,p_N,G_{\mathcal{S}})
}
\eeq
to denote partial Erd\H{o}s-R\'enyi random graphs with deterministic component $G_{\mathcal{S}}$. We assume that ${\bm G}$ (and hence $G_{\mathcal{S}}$) is unknown. In this context, the goal of local tomography is to estimate $G_{\mathcal{S}}$ via observing the state evolution of the observable agents in $\mathcal{S}$.
Figure~\ref{fig:partial} illustrates the partial Erd\H{o}s-R\'enyi construction just described.

Before formulating the tomography problem, we observe that, under condition~(\ref{eq:pconncond}), the partial Erd\H{o}s-R\'enyi graph is asymptotically connected with high probability for any choice of the subgraph $G_{\mathcal{S}}$, as stated in the following lemma.
\begin{lemma}[Connectivity of partial Erd\H{o}s-R\'enyi graphs]\label{lem:connected}
Given any graph~$G_{\mathcal{S}}\in \mathcal{G}(\mathcal{S})$, the partial Erd\H{o}s-R\'enyi graph
\beq
\bm{G}\sim\mathscr{G}^{\star}(N,p_N,G_{\mathcal{S}})
\eeq
is connected with high probability, i.e.,
\beq
\lim_{N\rightarrow \infty}\mathbb{P}\left[\bm{G}\mbox{ is connected}\right]=1.
\eeq
\end{lemma}

\begin{IEEEproof}
See Appendix~\ref{app:lemmas}.
\end{IEEEproof}


\subsection{Combination assignment.}

A combination assignment is a rule that builds the combination matrix $A$ as a function of the underlying graph $G$.
We assign (positive) weights to the edges of $G$ and denote the resulting matrix of weights by $A$.
Some useful and popular choices are the Laplacian and the Metropolis rules, which arise naturally in the context of adaptive diffusion networks~\cite{Sayed}, and are defined as follows\footnote{Strictly speaking, in the network literature the Laplacian and Metropolis rules are defined with weights that add up to one, which would correspond to~(\ref{eq:LapMat}) and~(\ref{eq:MetroMat}) without the multiplying factor $\rho$. The multiplying factor $\rho$, which provides the matrix stability, is usually left separate and not absorbed into the combination matrix.
For instance, in the case of diffusion networks~\eqref{eq:aweightsmatdef} we have $\rho=1-\mu$, where $\mu$ is the step-size. In our treatment, it is more convenient to include this scaling factor into the combination matrix, as done in~(\ref{eq:LapMat}) and~(\ref{eq:MetroMat}).}.
Let~$\rho\in\left(0,1\right)$ and $\lambda\in \left.\left(0,1\right.\right]$:\\

\noindent
\textbf{Laplacian rule.}
\begin{equation}\label{eq:LapMat}
a_{ij}=\rho\times
\left\{\begin{array}{ll} \displaystyle{\frac{\lambda g_{ij}}{d_{\max}}}, & \mbox{ for }i\neq j
\\
\\
1 - \displaystyle{\frac{\lambda}{d_{\max}} \sum_{\ell\neq i} g_{i\ell}}, & \mbox{ for }i=j
\end{array}\right..
\end{equation}

\noindent
\textbf{Metropolis rule.}
\begin{equation}\label{eq:MetroMat}
a_{ij}=\rho\times
\left\{\begin{array}{ll} \displaystyle{\frac{g_{ij}}{\max(d_i,d_j)}}, & \mbox{ for }i\neq j
\\
\\
1 - \displaystyle{\sum_{\ell\neq i} \frac{g_{i\ell}}{\max(d_i,d_\ell)}}, & \mbox{ for }i=j\end{array}\right.,
\end{equation}
where~$d_i$ is the degree of agent $i$ and $d_{\max}$ is the maximum degree in the network. It is useful to remark that different combination rules may have different impact on the performance of the topology estimators, as we will see in the examples presented in Sec.~\ref{sec:algorithmkey}.


In this paper, we shall focus on the family of nonnegative symmetric combination policies introduced in~\cite{tomo}, and whose characterizing properties we recall next.


%

\begin{property}[Bounded-norm]\label{pr:stability}
The maximum row-sum norm,
\begin{equation}
\left|\left|A\right|\right|_{\infty}\overset{\Delta}=\max_{i}\sum_{\ell=1}^N |a_{i\ell}|,
\end{equation}
is upper bounded by some $\rho<1$.\QEDB
\end{property}
For nonnegative symmetric matrices, Property~$1$ becomes:
\beq
\boxed{
\left|\left|A\right|\right|_{\infty}=\max_{i=1,2,\dots N}\sum_{\ell=1}^N a_{i\ell}=\max_{i=1,2,\dots N}\sum_{\ell=1}^N a_{\ell i}\leq \rho
}
\label{eq:prop1stab}
\eeq
From Property~$1$ we see that (most of) the combination weights $a_{i\ell}$ typically vanish as $N$ gets large, since a finite mass of value at most $\rho$ must be allocated across an ever-increasing number of neighbors -- on an Erd\H{o}s-R\'enyi graph, the average number of neighbors scales as $N p_N$, and in the regime considered in this paper we have~$Np_N\rightarrow \infty$ in view of~\eqref{eq:pnscale}.

The next property identifies a useful class of combination policies, for which degeneracy to zero of the combination weights is prevented by proper scaling. As highlighted below, such property is broad enough to encompass typical combination rules, such as the Laplacian~\eqref{eq:LapMat} and the Metropolis~\eqref{eq:MetroMat} rules.
\begin{property}[Non-degeneracy under $(N p_N)$-scaling]\label{pr:nondegeneracy}
Consider a combination policy applied to a partial Erd\H{o}s-R\'enyi graph $\bm{G}\sim\mathscr{G}^{\star}(N,p_N,G_{\mathcal{S}})$.
The combination policy belongs to class $\mathscr{C}_{\tau}$ for some $\tau > 0$, if for all $i, j = 1, 2, ..., N$ with $i \neq j$:
\beq
\boxed{
\P[Np_N \bm{a}_{ij}> \tau | \bm{g}_{ij}=1] \geq 1 - \epsilon_N
}
\label{eq:bound}
\eeq
where $\epsilon_N$ goes to zero as $N\rightarrow\infty$. In other words, if two nodes $i$ and $j$ are connected (corresponding to $\bm{g}_{ij}=1$), then the scaled combination coefficient $N p_N \bm{a}_{ij}$ lies above a certain threshold value denoted by $\tau$, with high probability, for large $N$.\QEDB
\end{property}

It is useful to remark that, since condition~(\ref{eq:bound}) is applied to a {\em partial} Erd\H{o}s-R\'enyi construction, the nodes belonging to the observable set $\mathcal{S}$ are connected in a deterministic fashion. This means that, for $i,j\in \mathcal{S}$, the random variable $\bm{g}_{ij}=g_{ij}$ is in fact deterministic. In this case, the condition in~(\ref{eq:bound}) should be rewritten, for any connected pair $(i,j)$ in $G_{\mathcal{S}}$, as:
\beq
\P[Np_N \bm{a}_{ij}> \tau | g_{ij}=1]=\P[Np_N \bm{a}_{ij}> \tau] \geq 1 - \epsilon_N,
\label{eq:boundet}
\eeq
because conditioning on a deterministic event becomes immaterial.

We now introduce a sufficient condition under which a combination rule fulfills Property~$2$. The relevance of this condition is that it can be readily verified for certain typical combination assignments such as Laplacian and Metropolis rules, and it automatically provides one value of $\tau$ to identify the class $\mathscr{C}_{\tau}$.
\begin{lemma}[Useful policies belonging to $\mathscr{C}_{\tau}$]
Any policy obeying the inequality
\beq
\bm{a}_{ij} \geq \frac{\gamma}{d_{\max}(\bm{G})} \;\bm{g}_{ij}
\label{eq:dmaxa}
\eeq
for all $i\neq j$ and for some $\gamma >0$, satisfies Property~$2$ with the choice $\tau=\gamma/e$.
\label{lem:policies}
\end{lemma}
\begin{IEEEproof}
See Appendix~\ref{app:lemmas}.
\end{IEEEproof}
We denote by $\mathscr{C}_{\rho,\tau}$ the class of weight-assignment policies for which Properties $1$ and $2$ hold simultaneously.

Using the definition of the Laplacian rule in~(\ref{eq:LapMat}), it is readily verified that this rule possesses Property $1$ and fulfills~(\ref{eq:dmaxa}) with the choice $\gamma=\rho \lambda$.
Likewise, using~(\ref{eq:MetroMat}), it is readily verified that the Metropolis rule possesses Property $1$ and fulfills~(\ref{eq:dmaxa}) with the choice $\gamma=\rho$.
As a result, both policies belong to the class $\mathscr{C}_{\rho,\tau}$, with the following choices of $\tau$ (the meaning of the subscripts should be obvious):
\beq
\tau_L=\frac{\rho\lambda}{e},
\qquad
\tau_M=\frac{\rho}{e}.
\label{eq:taus}
\eeq

Before proving the main result of this work, it is useful to illustrate the physical meaning of Property~$2$ in connection with the network tomography problem.
We introduce the $S\times S$ error matrix that quantifies how much the truncated estimator in~(\ref{eq:truncinvert}) differs from the true sub-matrix ${\bm A}_{\mathcal{S}}$, namely,
\beq
\boxed{
\bm{E}_{\mathcal{S}}\dfz  \widehat{\bm{A}}_{\mathcal{S}} - \bm{A}_{\mathcal{S}}
}
\label{eq:errmatfirstdef0}
\eeq
The magnified $(i,j)$-th entry of the truncated estimator in~(\ref{eq:truncinvert}), $N p_N [\widehat{\bm{A}}_{\mathcal{S}}]_{ij}$, can be written as:
\beq
\left\{
\begin{array}{lll}
\underbrace{N p_N \bm{a}_{ij}}_{\textnormal{not vanishing}}
+
N p_N \bm{e}_{ij}, &\textnormal{if $i$ and $j$ are connected},\\
\\
N p_N \bm{e}_{ij}, &\textnormal{otherwise},
\end{array}
\right.
\label{eq:aobsarray2}
\eeq
where $e_{ij}$ is an error quantity, and the qualification of being ``not vanishing'' is a consequence of Property~$2$.
According to~(\ref{eq:aobsarray2}), if we want the nonzero entries $N p_N \widehat{\bm{a}}_{ij}$ to stand out from the error floor, when $i$ and $j$ are interacting, or to be bounded above (by $\tau$), when $i$ and $j$ are non-interacting, as $N$ grows large, we must be able to control the impact of the error term $N p_N \bm{e}_{ij}$.

We are now ready to summarize the main problem treated in this article.

\noindent
\textbf{Local Tomography.} Let $\bm{A}$ be an $N\times N$ matrix obtained from any combination assignment belonging to the class $\mathscr{C}_{\rho,\tau}$, over a graph $\bm{G}\sim\mathscr{G}^{\star}(N,p_N,G_{\mathcal{S}})$ on $N$ nodes with a given (arbitrary) subgraph~$G_{\mathcal{S}}$. Let~$\left\{\left[\bm{y}_n\right]_{\mathcal{S}}\right\}_{n\in\mathbb{N}}$ be the state-evolution associated with the observable subset of agents~$\mathcal{S}$ and obeying the stochastic dynamical law
\begin{equation}
\bm{y}_n= \bm{A}\bm{y}_{n-1}+\beta \bm{x}_{n}.
\end{equation}

\noindent
\textbf{Problem:} given~$\left\{\left[\bm{y}_n\right]_{\mathcal{S}}\right\}_{n\in\mathbb{N}}$, can we determine~$G_{\mathcal{S}}$?

\section{Main result}\label{sec:sketch}
The main result of this work is to establish that the truncated estimator $\widehat{A}_{\mathcal{S}}=\left[R_1\right]_{\mathcal{S}} \left(\left[R_0\right]_{\mathcal{S}}\right)^{-1}$ introduced in~(\ref{eq:truncinvert}), contains enough information to recover the {\em true} support graph, $G_{\mathcal{S}}$, of the combination matrix $A_{\mathcal{S}}$ that corresponds to the observable subset, $\mathcal{S}$. More specifically, we establish that a positive threshold $\tau$ exists, such that the graph obtained by comparing the entries of $\widehat{A}_{\mathcal{S}}$ against this threshold matches, with high probability, the {\em true} support graph $G_{\mathcal{S}}$.
This implies that the topology of the subnetwork $G_{\mathcal{S}}$ can be fully recovered, with high probability, via the output measurements~$\left\{\left[\bm{y}_n\right]_{\mathcal{S}}\right\}_{n\in\mathbb{N}}$ used to construct $\left[R_{0}\right]_{\mathcal{S}}$ and $\left[R_{1}\right]_{\mathcal{S}}$, namely, via the observable nodes only.
Even if broader observation is permitted, the result enables the possibility of surmounting the curse of dimensionality by processing smaller $S\times S$ matrices $\left[R_{1}\right]_{\mathcal{S}}$ and $\left[R_{0}\right]_{\mathcal{S}}$, instead of large-scale $N\times N$ matrices~$R_{0}$ and $R_{1}$ -- wherein one of the operations involves the often expensive inversion of a large matrix $R_{0}$ -- and yet attaining exact recovery with high probability.

Before stating the main theorem, let us introduce a useful thresholding operator. We consider a matrix~$M\in \mathbb{S}^{S \times S}_{+}$, whose $(i,j)$-th entry is $m_{ij}$, with $i\in \mathcal{S}$ and $j\in \mathcal{S}$.
The thresholding operator compares the off-diagonal entries against some threshold~$\tau>0$, and produces as output a graph, $\Gamma_{\tau}(M)\in\mathcal{G}(\mathcal{S})$, whose adjacency matrix has $(i,j)$-th entry equal to
\begin{equation}
\mathbb{I}_{\left\{m_{ij} > \tau\right\}}, \qquad \forall i\neq j.
\label{eq:threshop}
\end{equation}
In other words, the thresholding operator returns a graph whereby two nodes $i$ and $j$ are connected only if the $(i,j)$-th entry of the matrix $M$ exceeds the threshold. We assume all entries of the main diagonal of $\bm{G}$ equal to one.
We are now ready to state the main theorem.

\begin{theorem}[Exact recovery of $G_{\mathcal{S}}$]
\label{co:main}
Let $G_{\mathcal{S}}$ be a given deterministic graph (with arbitrary topology) on $S$ nodes, and let $\bm{G}\sim\mathscr{G}^{\star}\left(N,p_N,G_{\mathcal{S}}\right)$ be a partial Erd\H{o}s-R\'enyi random graph where the sequence $c_N$ that determines the connection probability, $p_N=(1/N)(\log N + c_N)$, obeys the condition:
\beq
\frac{[\log(\log N + c_N)]^2}{\log N}\rightarrow 0.
\label{eq:cnconststatement}
\eeq
Let also $\bm{A}$ be a combination matrix with support graph $\bm{G}$. If there exists $\tau>0$ and $0 < \rho < 1$ such that $\bm{A}$ belongs to $\mathscr{C}_{\rho,\tau}$, then the following results hold.
\begin{itemize}
\item[$i)$]
If $i,j\in \mathcal{S}$ are interacting, then the $(i,j)$-th magnified entry of the truncated estimator, $N p_N [\widehat{\bm{A}}_{\mathcal{S}}]_{ij}$, exceeds the threshold $\tau$ with high probability as $N\rightarrow\infty$.
\item[$ii)$]
If $i,j\in \mathcal{S}$ are non-interacting, then the $(i,j)$-th magnified entry of the error matrix, $N p_N \bm{e}_{ij}$, converges to zero in probability.
\item[$iii)$]
The graph obtained by applying the thresholding operator in~(\ref{eq:threshop}) to the magnified truncated estimator, $Np_N \widehat{\bm{A}}_{\mathcal{S}}$, matches the true support graph, $G_{\mathcal{S}}$, with high probability as $N\rightarrow\infty$, namely,
\begin{equation}\label{eq:thmthreshold}
\boxed{
\lim_{N\rightarrow \infty}\mathbb{P}[\Gamma_{\tau}(Np_N \widehat{\bm{A}}_{\mathcal{S}}) = G_{\mathcal{S}} ]= 1.
}
\end{equation}
\end{itemize}
\end{theorem}

\begin{IEEEproof} See Appendix~\ref{app:Theor1}.
\end{IEEEproof}

\subsection{Outline of the main proof}
We offer here an outline of the proof of Theorem~\ref{co:main}.
The detailed proof is reported in Appendix~\ref{app:Theor1} and related appendices~\ref{sec:proof},~\ref{sec:HC}, and~~\ref{app:smalldist}.

First, we use the fact (proved in~\cite{tomo}) that the entries of the error matrix in~(\ref{eq:errmatfirstdef0}) are nonnegative, implying, for $i,j\in \mathcal{S}$:
\beq
Np_N [\hat{\bm{A}}_{\mathcal{S}}]_{ij}=Np_N \bm{a}_{ij} + N p_N \bm{e}_{ij}\geq Np_N \bm{a}_{ij}.
\label{eq:errnonneg}
\eeq
In view of Property~$2$, Eq.~(\ref{eq:errnonneg}) implies that, when $i$ and $j$ are interacting nodes, the quantity $Np_N [\hat{\bm{A}}_{\mathcal{S}}]_{ij}$ exceeds a positive threshold $\tau$ with high probability and, hence, part $i)$ of Theorem~\ref{co:main} is proved.
If, in addition, we show that the magnified error $Np_N \bm{e}_{ij}$ converges to zero in probability over the non-interacting pairs, i.e., if we prove part $ii)$, then $Np_N \left[\widehat{\bm{A}}_{\mathcal{S}}\right]_{ij}\rightarrow 0$ (and, hence, $Np_N \left[\widehat{\bm{A}}_{\mathcal{S}}\right]_{ij}$ stays below $\tau$ as $N$ goes to infinity) over the non-interacting pairs. Now, in order to prove part $iii)$, we need to show that it is possible to classify correctly, as $N\rightarrow\infty$, {\em each} pair of nodes by comparing the truncated estimator $\widehat{\bm{A}}_{\mathcal{S}}$ against the threshold $\tau$ (or any smaller value): if $Np_N [\widehat{\bm{A}}_{\mathcal{S}}]_{ij}>\tau$, then classify $(i,j)$ as an interacting pair, otherwise classify it as non-interacting. Since the cardinality of the observable set is finite, parts $i)$ and $ii)$ imply part $iii)$ by direct application of the union bound. Accordingly, it remains to examine part $ii)$.

\begin{figure} [t]
\begin{center}
\includegraphics[scale= 0.45]{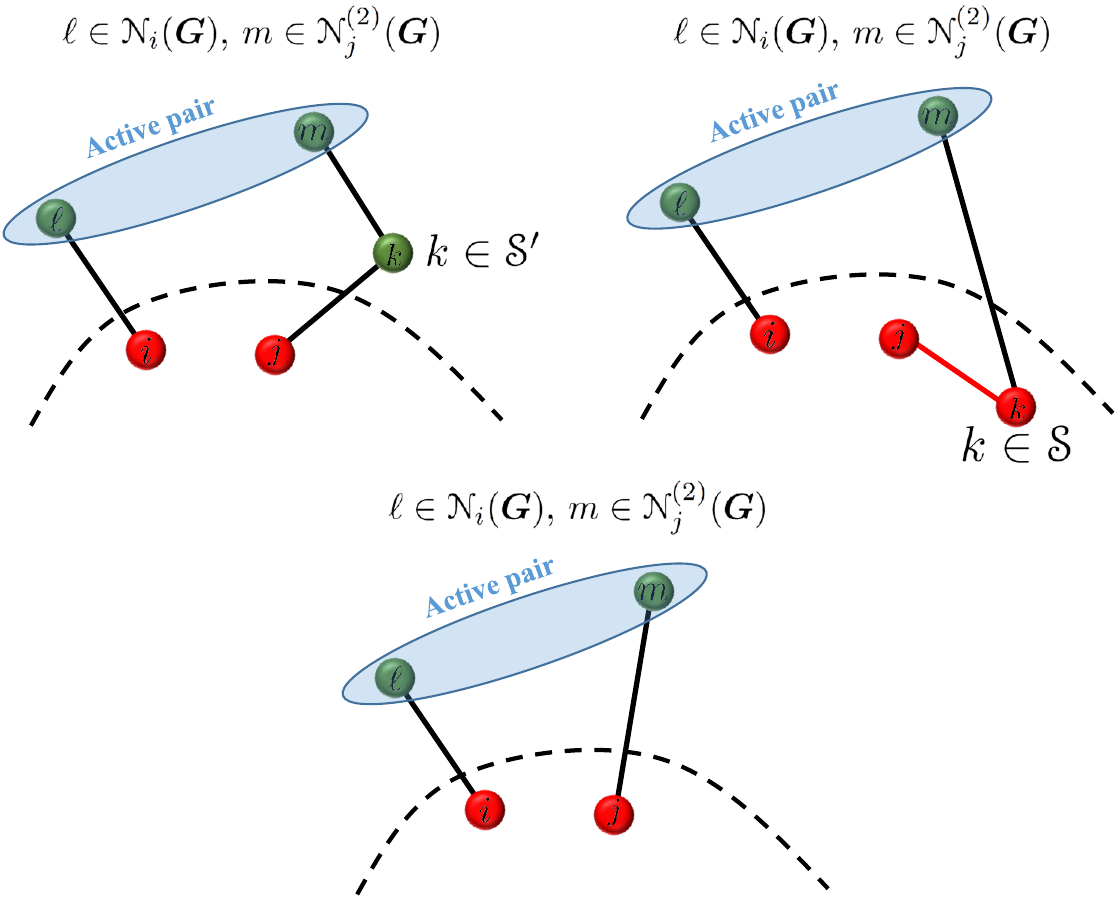}
\caption{Graphical illustration of the active pairs. Note that the neighborhood constraints $\ell\in \mathcal{N}_i(\bm{G})$ and $m\in\mathcal{N}^{(2)}_j(\bm{G})$ refer to the partial Erd\H{o}s-R\'enyi graph~$\bm{G}$ and hence the intermediate node $k$ may belong to $\mathcal{S}$.}
\label{fig:active}
\end{center}
\end{figure}

Using one result in~\cite{tomo}, we rewrite the entries of the error matrix in~(\ref{eq:errmatfirstdef0}) as:
\beq
e_{ij} = \sum_{\ell,m\in \mathcal{S}'} a_{i \ell} h_{\ell m} b_{m j},\qquad i,j\in \mathcal{S},
\label{eq:relerr0}
\eeq
where $B\dfz A^2$ and $H\dfz\left(I_{\mathcal{S}'}-B_{\mathcal{S}'}\right)^{-1}$.
The error in~(\ref{eq:relerr0}) is determined by three main factors, namely: $i)$ $a_{i\ell}$, which is nonzero only if node $i$ (from subset $\mathcal{S}$) and agent $\ell$ (from subset $\mathcal{S}'$) are neighbors; $ii)$ $b_{m j}$, which is nonzero only if node $m$ (from subnet $\mathcal{S}'$) and agent $j$ (from subset $\mathcal{S}$) are second-order neighbors (i.e., connected in one or two steps); $iii)$ the term $h_{\ell m}$, which is the $(\ell,m)$-th entry of the matrix $H$.
Clearly, in~(\ref{eq:relerr0}), the relevant entries $h_{\ell m}$ are those that are ``activated'' by nonzero values of $a_{i\ell}$ and $b_{m j}$. The $(\ell,m)$ pairs for which $a_{i\ell}$ and $b_{mj}$ are nonzero will be accordingly referred to as ``active pairs''. Refer to Figure~\ref{fig:active} for a graphical illustration of the active pairs.

\begin{figure} [t]
\begin{center}
\includegraphics[scale= 0.5]{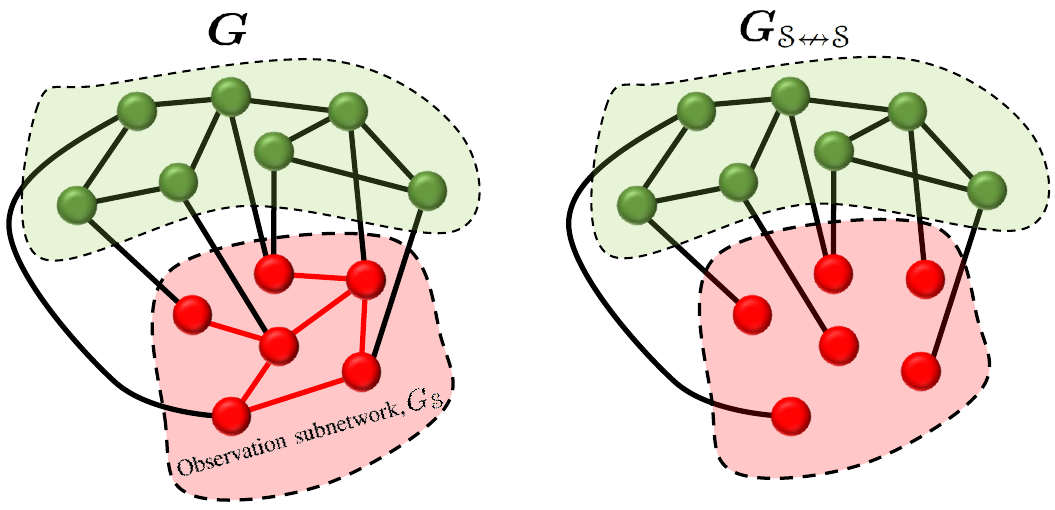}
\caption{The matrix $H$ in~(\ref{eq:BHmats}) does not depend on the combination submatrix $A_{\mathcal{S}}$.
If two nodes $\ell,m\in \mathcal{S}'$ are distant in the graph $G_{\mathcal{S}\nleftrightarrow \mathcal{S}}$ displayed on the right, i.e., if $\delta_{\ell,m}(G_{\mathcal{S}\nleftrightarrow \mathcal{S}}) \gg 1$, then $h_{\ell m}$ is small. This does not imply, in general, that $\delta_{\ell,m}(G)\gg 1$, namely, that the same nodes are distant over the (original) graph $G$ displayed on the left.}\label{fig:partial2}
\end{center}
\end{figure}

It is also clear from~(\ref{eq:relerr0}) that, in order to get a small error, small values of $h_{\ell m}$ (over the active pairs) are desirable.
In Theorem~\ref{lem:distancia} -- see Appendix~\ref{sec:proof} -- we are able to show that, for large $N$, vanishing values of $h_{\ell m}$ are obtained when the distance between nodes $\ell$ and $m$ gets large.
In particular, Theorem~\ref{lem:distancia} states that the distance between $\ell$ and $m$ that plays an important role in the magnitude of $H$ is the one {\em evaluated on the transformed graph} $G_{\mathcal{S}\nleftrightarrow \mathcal{S}}$ (see Figure~\ref{fig:partial2}), where the observable graph $G_{\mathcal{S}}$ is replaced by an empty graph.
As observed in the proof of Theorem~\ref{lem:distancia} the magnitude of $h_{\ell m}$ is not contingent on the particular topology $G_{\mathcal{S}}$. As a result, removing the dependency on $G_{\mathcal{S}}$ is crucial to get a {\em universal} result, namely, a result that holds {\em for any arbitrary} $G_{\mathcal{S}}$.

Since, loosely speaking, Theorem~\ref{lem:distancia} implies that the error is small if the distance on the transformed graph $G_{\mathcal{S}\nleftrightarrow \mathcal{S}}$ is large, the remaining part of the proof consists of showing that the distance over the active $(\ell,m)$ pairs is large with high probability, namely, that small distances are rare as $N$ goes to infinity.
Now, such a conclusion can be proved for a {\em pure} Erd\H{o}s-R\'enyi graph, $\mathscr{G}^{\star}(N,p_N)$, as shown in Lemma~\ref{lemma:dist2new} -- see Appendix~\ref{app:smalldist}.
However, proving the same result for a {\em partial} Erd\H{o}s-R\'enyi graph, $\mathscr{G}^{\star}(N,p_N,G_{\mathcal{S}})$, presents a nontrivial difficulty related to the fact that the partial Erd\H{o}s-R\'enyi graph is not homogeneous\footnote{Actually, we will see in Appendix~\ref{app:Theor1} that a further source of inhomogeneity arises, which is related to the terms $a_{i\ell}$ and $b_{m j}$. The homogenization procedure that we are going to introduce is able to solve this further inhomogeneity.} because the observable subgraph $G_{\mathcal{S}}$ can be {\em arbitrary}.
In order to overcome this difficulty, we will carefully implement a procedure of homogenization and coupling -- see Appendix~\ref{sec:HC}.
The homogenization procedure amounts to carefully replacing the partial Erd\H{o}s-R\'enyi random graph $\bm{G}$ by an Erd\H{o}s-R\'enyi graph $\widetilde{\bm{G}}$ that is coupled with $\bm{G}$ in the following sense: if small distances are rare over the classic (hence homogeneous) Erd\H{o}s-R\'enyi graph $\widetilde{\bm{G}}$, then small distances are also rare over the {\em coupled} partial (hence inhomogeneous) Erd\H{o}s-R\'enyi graph $\bm{G}$.
In summary, the homogenization-and-coupling is a useful formal tool that is used to reduce the {\em inhomogeneous} case to a (simpler) {\em homogeneous} graph.
%
%
%


\section{Applying Theorem~\ref{co:main}}\label{sec:algorithmkey}

Theorem~\ref{co:main} asserts the possibility of performing local tomography over large-scale diffusion networks as it asserts the existence of a threshold $\tau$ such that the entries of the na\"{i}ve estimator $\widehat{{\bm A}}_{\mathcal{S}}$ provide correct reconstruction of the observable network with high-probability. In particular, introducing a detection threshold:
\beq
\eta\dfz\tau/(N p_N),
\label{eq:dethresh}
\eeq
the topology of the observable network can be recovered for sufficiently large $N$ as follows: If $[\widehat{{\bm A}}_{\mathcal{S}}]_{ij} > \eta$, then classify nodes $i$ and $j$ as interacting, otherwise classify them as non-interacting.
The inverse dependence of $\eta$ on $Np_N$ can be explained as follows. Since we compare the {\em unnormalized} matrix entries, $[\widehat{{\bm A}}_{\mathcal{S}}]_{ij}$, against the threshold $\eta$, and since in view of Theorem~$1$ these entries (over the connected pairs) scale essentially as $1/(Np_N)$, a proper classification threshold should exhibit the same type of scaling.

From a practical perspective, it is necessary to select an appropriate value for $\eta$, in order to correctly classify the interacting and non-interacting pairs.
In this connection, prior information on the dynamical system in~\eqref{eq:VARmodel} can be useful to set the detection threshold.
Indeed, we see from~(\ref{eq:dethresh}) that knowledge is needed about: $i)$ the average degree $Np_N$; and $ii)$ the parameter $\tau$ that characterizes the class $\mathscr{C}_{\rho,\tau}$ where the combination matrix stems from.
Let us assume that such a knowledge is available.
Now, using the values of $\tau$ reported in~(\ref{eq:taus}), for the Laplacian and Metropolis rules we have, respectively:
\begin{equation}
\eta_L=\frac{\rho\lambda}{e N p_N},\qquad  \eta_M=\frac{\rho}{e N p_N}.
\label{eq:etas}
\end{equation}
We observe that Theorem~\ref{co:main} is an asymptotic (in the size $N$) result. For a numerical practical application of this result, it is useful to make three remarks.

First, given a detection threshold $\eta$ that guarantees exact asymptotic classification, any threshold smaller than $\eta$ still guarantees exact asymptotic classification. In order to explain why, let us consider two thresholds $\eta_1$ and $\eta_2$, with $\eta_1<\eta_2$, and assume that $\eta_2$ is known to provide exact asymptotic classification.
Then we have that: $i)$ for interacting pairs, if $[\widehat{{\bm A}}_{\mathcal{S}}]_{ij}$ is higher than $\eta_2$, then it is obviously higher than $\eta_1$, implying correct classification also with threshold $\eta_1$; $ii)$ for non-interacting pairs, Theorem~$1$ ensures that, asymptotically, $[\widehat{{\bm A}}_{\mathcal{S}}]_{ij}$ will be smaller than {\em any} small value $\epsilon$, implying correct classification also with threshold $\eta_1$. In other words, $\eta_1$ also provides exact classification.

Second, we observe that a combination rule can fulfill Property~$2$ for different values of $\tau$.
For example, assume that we proved that a combination rule fulfills Property~$2$ with a certain value $\tau_1$.
Then, the same policy fulfills Property~$2$ with a higher value, e.g., $\tau_2>\tau_1$.

Third, consider a pair $(i,j)$ of interacting nodes, and let us examine~(\ref{eq:errnonneg}).
According to Property~$2$, the selection of $\tau$ relates only to the properties of the combination matrix, namely, to the behavior of $N p_N \bm{a}_{ij}$ for interacting nodes.
On the other hand, for finite sizes of the network, the error $\bm{e}_{ij}$ is small, but not zero.
As a result the quantity $N p_N [\widehat{{\bm A}}_{\mathcal{S}}]_{ij}$ will be greater than $N p_N \bm{a}_{ij}$, namely, the entries of the estimated combination matrix are, on average, shifted upward due to this additional (and positive) error.
As a result, one expects that, for finite values of $N$, increasing the values of $\tau$ may be beneficial for classification purposes.

The aforementioned issues show that there is freedom in selecting the threshold parameter to attain {\em exact} topology recovery {\em asymptotically}  (i.e., as~$N$ grows to infinity).
On the other hand, we remark that different threshold choices are expected to behave differently {\em for finite network sizes}.
In fact, the following trade-off arises: a higher threshold reduces the likelihood that a zero entry of the combination matrix gives rise to a ( false) threshold crossing, while concurrently increasing the likelihood that a nonzero entry gives rise to a (correct) threshold crossing.

\subsection{Nonparametric Strategies}
From the analysis conducted in the previous section, we have learned the following facts about tomography based on the thresholding operator.
First, a good threshold tuning requires some a-priori knowledge of the model (e.g., of the average degree, $N p_N$, or of the class of combination matrices to set the constant $\tau$).
Second, even with some good a-priori knowledge, it is not clear how the threshold should be optimized to maximize the performance, because a trade-off arises for finite network sizes, whose (nontrivial) solution would require an even more detailed knowledge of the underlying model.

For all these reasons, it is useful to verify the possibility of employing some {\em nonparametric} pattern recognition strategies, which work blindly (without any a-priori knowledge), and which are capable to automatically adapt the classification threshold to the empirical data.
In particular, in the forthcoming experiments we will consider a $k$-means clustering algorithm (with $k=2$) that will be fed with the entries of the truncated estimator matrix in~(\ref{eq:truncinvert}).
The clustering algorithm will attempt to find some separation threshold {\em empirically on the data}, and to split accordingly the matrix entries into two clusters (connected and non-connected). The cluster with higher arithmetic mean will be then elected as the cluster of connected nodes.



\subsection{Unknown Correlation Matrices}
Until now, we have implicitly assumed that the correlation matrices, $R_0$ and $R_1$, are perfectly known, and, hence, that the truncated estimator $\hat{\bm{A}}_{\mathcal{S}}$ in~(\ref{eq:truncinvert}) could be evaluated exactly.
However, in practice such correlation matrices are unknown, and must be estimated from the data.
For this reason, we will consider numerical simulations where we empirically estimate the truncated correlation matrices~$\left[R_0\right]_{\mathcal{S}}$ and~$\left[R_1\right]_{\mathcal{S}}$ from the observed data through the sample-average estimator (boldface notation is omitted to emphasize that the observed $y_n$ refers to a particular realization):
\beqa
\widehat{[R_0]}_{\mathcal{S}} & = & \frac{1}{n_{\max} + 1}
\sum_{n=0}^{n_{\max}} [y_{n}]_{\mathcal{S}} [y_n]^{\top}_{\mathcal{S}},
\label{eq:empir0}
\\
\widehat{[R_1]}_{\mathcal{S}} & = & \frac{1}{n_{\max}}
\sum_{n=0}^{n_{\max}-1} [y_{n+1}]_{\mathcal{S}} [y_n]^{\top}_{\mathcal{S}}.
\label{eq:empir1}
\eeqa
We remark that it is possible to optimize such estimates by exploiting prior information on the structural properties of the system, and such an optimization could boost the performance of the algorithm.
This estimation-tuning is outside the scope of this paper, but showing that our strategy works with the (perhaps) simplest correlation estimators is definitely encouraging.
In the next section, we will additionally display the performance of the algorithm under the ideal case of known correlations, where the exact computation of the truncated estimator in~(\ref{eq:truncinvert}) can be accomplished.
This ideal case provides a superior limit in performance also with respect to optimized correlation estimators.

\section{Numerical Experiments}
\label{sec:numexper}
\begin{figure} [t]
\begin{center}
\includegraphics[scale= 0.5]{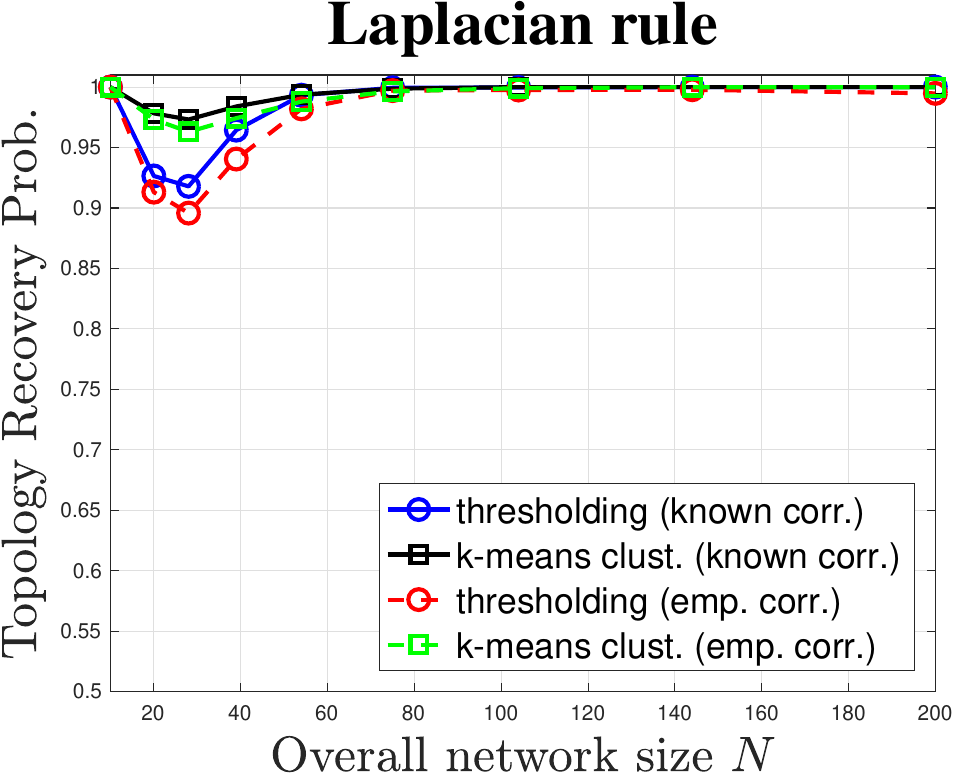}
\caption{Empirical recovery probability as a function of the size of the overall network, for the Laplacian combination rule -- see~\eqref{eq:LapMat}.
{\em Thresholding} stands for the tomography strategy where the entries of $\widehat{\bm{A}}_{\mathcal{S}}$ are thresholded with the threshold~$\eta_L$ determined in~\eqref{eq:etas}. {\em Empirical correlation} means that the truncated correlations were estimated as in~\eqref{eq:empir0} and~\eqref{eq:empir1}, whereas the curves with {\em known correlations} are also displayed as a superior limit in performance.}
\label{fig:Figsim1}
\end{center}
\end{figure}

We are now ready to present the results of the numerical experiments.
In Figure~\ref{fig:Figsim1}, we display the (empirically-estimated) topology-recovery probability, with reference to an overall network with number of nodes $N$ ranging from $10$ to $200$, and for the case of a Laplacian combination rule with $\rho=0.9$ and $\lambda=0.95$. The observable network is made up of $S=10$ nodes and the connections within the latent part (and between the latent and observable nodes) are drawn according to an Erd\H{o}s-R\'enyi model with connection probability $p_N=2\log(N)/N$.
We see that the probability of correct recovery gets close to $1$ as $N$ increases for all the considered scenarios: parametric {\em versus} $k$-means thresholding, and empirically estimated truncated correlation matrices (as in~\eqref{eq:empir0} and~\eqref{eq:empir1}) {\em versus} known truncated correlation matrices.
We notice that the recovery probability curve is not monotonic.
Such behavior matches perfectly our theoretical results, as we now explain.
First, when $N=S=10$ (first point in Figure~\ref{fig:Figsim1}), all the network is observed, and in view of~\eqref{eq:R1nR0n} (and the comments that follow this equation) the recovery probability must be equal to $1$.
Second, Theorem~$1$ ensures that a probability of correct recovery equal to $1$ must be also attained asymptotically (in $N$). Accordingly, since the error probability curve starts from the value $1$, and converges to $1$ as $N$ increases, the non-monotonic behavior exhibited in Figure~\ref{fig:Figsim1} makes sense.

Let us now compare the performance of the different strategies shown in Figure~\ref{fig:Figsim1}.
As one expects, the strategies that know the true correlation matrices outperform the strategies that do not know them.
A separate comment is called for while comparing the thresholding estimator and the clustering algorithm.
Perhaps unexpectedly, the latter strategy outperforms the former one.
However, this behavior matches well the theoretical considerations made in the previous section.
Indeed, in the simulations the threshold employed by the thresholding estimator is not optimized at all, whereas the nonparametric clustering algorithm might automatically adapt the threshold to the empirical evidence arising from the data, thus delivering a better performance.

\begin{figure} [t]
\begin{center}
\includegraphics[scale= 0.5]{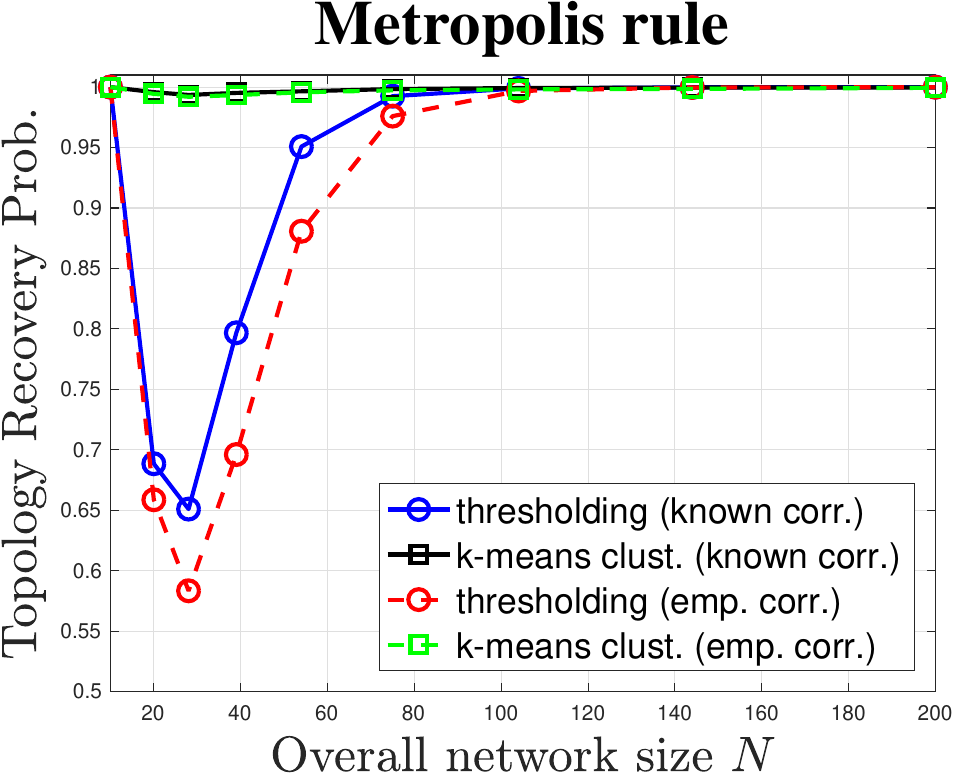}
\caption{Empirical recovery probability as a function of the size of the overall network, for the Metropolis combination rule -- see~\eqref{eq:MetroMat}.
{\em Thresholding} stands for the tomography strategy where the entries of $\widehat{\bm{A}}_{\mathcal{S}}$ are thresholded with the threshold~$\eta_M$ determined in~\eqref{eq:etas}. {\em Empirical correlation} means that the truncated correlations were estimated as in~\eqref{eq:empir0} and~\eqref{eq:empir1}, whereas the curves with {\em known correlations} are also displayed as a superior limit in performance.}
\label{fig:Figsim2}
\end{center}
\end{figure}

The above analysis is repeated for the case of a Metropolis combination rule with $\rho=0.9$, and the result is shown in Figure~\ref{fig:Figsim2}.
The same general conclusions that we draw for the Laplacian rule apply. However, we see here that the performance of the thresholding operator seems slightly worse.
One explanation for this behavior is the following.
The choice of the constant $\tau$ in~(\ref{eq:taus}) is perhaps over-conservative.
Indeed, such choice follows by estimating the asymptotic scaling law of the {\em maximal} degree (see Lemma~\ref{lem:policies}), whereas the nonzero entries of the Metropolis matrix in~(\ref{eq:MetroMat}) are determined only by the maximum over {\em pairs} of degrees. This means that, on average, the nonzero entries of the Metropolis matrix are higher than what is predicted by the chosen $\tau$. It is expected that for the Metropolis rule, a larger threshold can be used without affecting the identification of connected pairs, while reducing the errors corresponding to the disconnected pairs.






\subsection{Beyond Theorem~\ref{co:main}}
Theorem~\ref{co:main} establishes that, under certain technical conditions, it is possible to retrieve the topology of a subset $\mathcal{S}$, even when the majority of the network nodes are not observed. This appears to be a nontrivial result, since an observable measurement $\bm{y}_i(n)$, $i\in\mathcal{S}$, is subject to the influence of nodes from {\em all across} the network. This happens because the diffusion recursion in~(\ref{eq:1storder}) links the nodes through the overall $N\times N$ matrix $A$, which takes into account also the influences that unobserved nodes have on observed nodes.

The possibility of inferring the topology of a subnet by taking measurements from this subnet only, and by ignoring the unobserved components, is of paramount importance, in view of the accessability, probing and processing limitations that arise unavoidably in practical applications. In particular, it is tempting to think about a {\em sequential} reconstruction strategy, where a larger network is reconstructed through local tomography experiments over smaller network portions, and where each local experiment obeys some probing/processing constraints. We start by illustrating the perhaps simplest sequential reconstruction strategy.

\begin{figure} [t]
\begin{center}
\includegraphics[scale= 0.35]{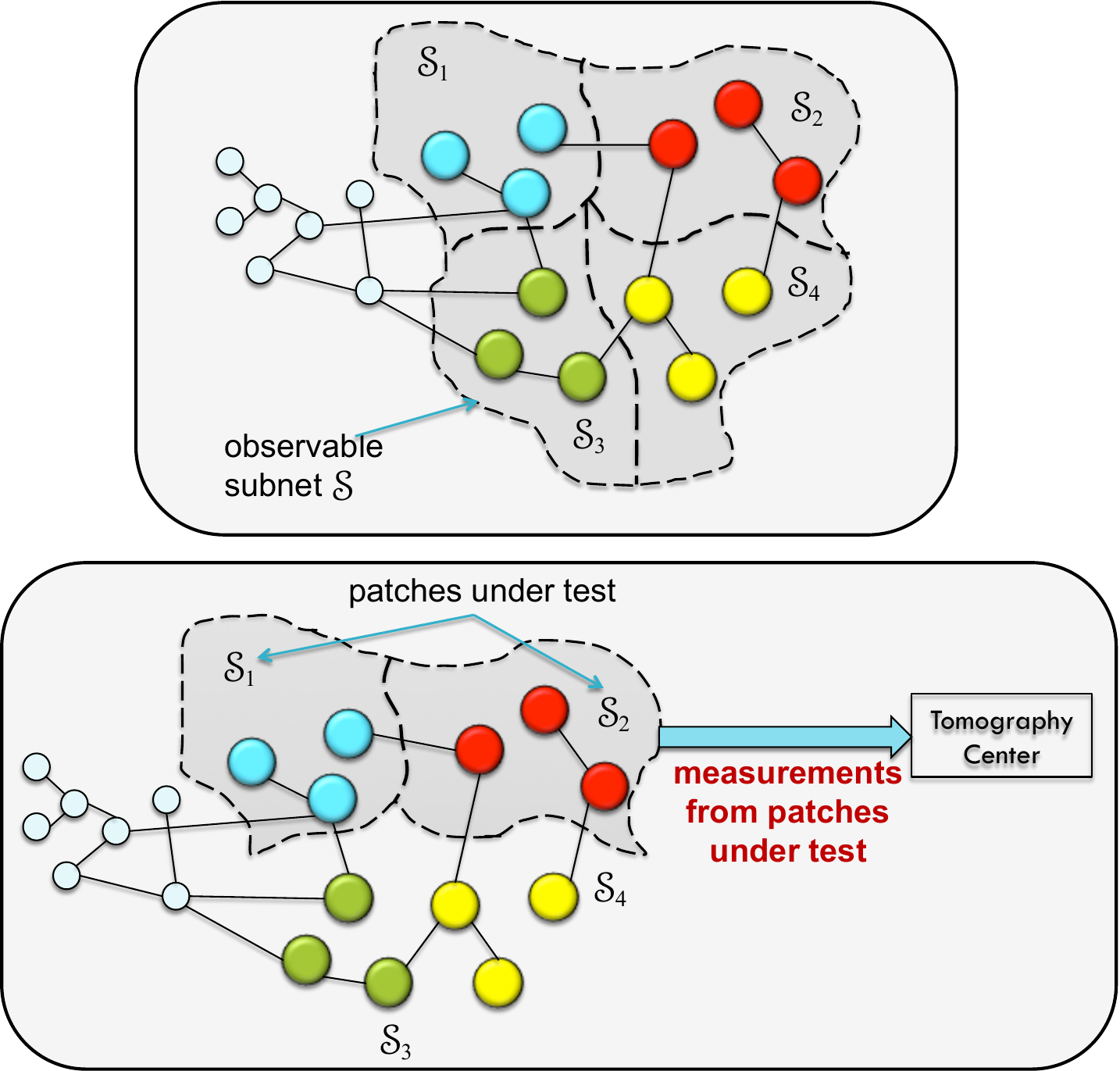}
\caption{Pictorial illustration of the sequential reconstruction through patching.}
\label{fig:patchespictorially}
\end{center}
\end{figure}

Assume that there is an observable subset of nodes $\mathcal{S}$, which is embedded in a larger network with many unobserved components. Due to probing and processing limitations, it is possible to probe and process at most $M$ nodes per single experiment.
Accordingly, the set $\mathcal{S}$ is divided into the ``patches'' $\mathcal{S}_1,\mathcal{S}_2,\ldots,\mathcal{S}_P$.
For simplicity, we assume that the patches do not overlap each other {\em and} that they cover completely the set $\mathcal{S}$ (i.e., the patches form a partition of $\mathcal{S}$).
Each local tomography experiment will correspond to probing the union of two patches, $\mathcal{S}_i\cup\mathcal{S}_j$.
For this reason, each patch has  cardinality $S_i\leq M/2$, for $i=1,2,\ldots, P$, which allows coping with the probing-and-processing constraint. The process is pictorially illustrated in Figure~\ref{fig:patchespictorially}.
Clearly, if a particular pair of nodes does not belong to the union of patches under test, we cannot make inference on that pair.
The maximum number of experiments that allows testing all pair of nodes is $P(P-1)/2$.
Per each experiment, we apply the local tomography strategy described in the previous section, namely: $i)$ we compute the empirical correlation matrices, $\widehat{[R_0]}_{\mathcal{S}_i\cup \mathcal{S}_j}$ and $\widehat{[R_1]}_{\mathcal{S}_i\cup \mathcal{S}_j}$, from which $ii)$ the truncated estimator $\widehat{A}_{\mathcal{S}_i\cup \mathcal{S}_j}$ is computed; and $iii)$ we apply the $k$-means algorithm to obtain an estimated subgraph $G_{\mathcal{S}_i\cup \mathcal{S}_j}$.
As more and more pairs of patches are tested, the connection profile of the network is reconstructed. A pseudo-code for the sequential reconstruction algorithm, nicknamed ``Patch\&Catch'', is shown in Algorithm~\ref{tb:algorithm}.

Before seeing the Patch\&Catch algorithm in operation, it is important to make a fundamental remark.
Proving that the sequential reconstruction strategy retrieves the topology of $G_{\mathcal{S}}$ consistently (as $N\rightarrow\infty$) seems a nontrivial task. In particular, we now explain why consistency of the Patch\&Catch algorithm does not come as a corollary of Theorem~\ref{co:main}.

Assume first that $G_{\mathcal{S}}$ is an arbitrary deterministic network.
In order to apply Theorem~\ref{co:main} to each local experiment, the unobserved component should be an Erd\H{o}s-R\'enyi graph. However, given a union-of-patches under test, $\mathcal{S}_i\cup\mathcal{S}_j$, the unobserved component is a mix of an Erd\H{o}s-R\'enyi graph and of a portion of $G_{\mathcal{S}}$ (refer to Figure~\ref{fig:patchespictorially}). Since the latter portion is not purely Erd\H{o}s-R\'enyi (because $G_{\mathcal{S}}$ is arbitrary), Theorem~\ref{co:main} does not directly apply.
On the other hand, if we assume that the whole graph (and, hence, also $G_{\mathcal{S}}$) is Erd\H{o}s-R\'enyi, then the network $G_{\mathcal{S}}$ would be not fixed as $N\rightarrow\infty$. In particular, since $\mathcal{S}$ has finite cardinality, it will become asymptotically disconnected, with high probability as $N\rightarrow\infty$.

In summary, we make no claim that the sequential reconstruction can grant consistent recovery. Therefore, the numerical results we are going to illustrate in this subsection must be intended as a preliminary test aimed at checking whether, in the finite network-size regime, a sequential reconstruction strategy might be successfully applicable.


\begin{algorithm}\caption{Patch\&Catch Sequential Tomography}
\label{tb:algorithm}
  \begin{algorithmic}[1]
    \Require Ensemble of patches~$\left\{\mathcal{S}_1,\mathcal{S}_2,\ldots,\mathcal{S}_P\right\}$ and observables~$\left\{[y_n]_{\mathcal{S}_i}\right\}$ over the patches $i=1,2,\ldots,P$, and for time epochs $n=0,1,\ldots,n_{max}$.
    \Ensure $\widehat{G}_{\mathcal{S}}$, estimate of the subnet $G_{\mathcal{S}}$ of observable nodes.
    \While{$i \leq P$}
     \While{$j < i$}
      \State $\widehat{[R_0]}_{\mathcal{S}_i\cup \mathcal{S}_j} =  \frac{1}{n_{\max} + 1}
      \sum_{n=0}^{n_{\max}} [y_{n}]_{\mathcal{S}_i\cup \mathcal{S}_j} [y_n]^{\top}_{\mathcal{S}_i\cup \mathcal{S}_j}$
      \State $\widehat{[R_1]}_{\mathcal{S}_i\cup \mathcal{S}_j}  =  \frac{1}{n_{\max}}
      \sum_{n=0}^{n_{\max}-1} [y_{n+1}]_{\mathcal{S}_i\cup \mathcal{S}_j} [y_n]^{\top}_{\mathcal{S}_i\cup \mathcal{S}_j}$
      \State $\widehat{A}_{\mathcal{S}_i\cup \mathcal{S}_j} =
      \widehat{[R_1]}_{\mathcal{S}_i\cup \mathcal{S}_j}
      \left(\widehat{[R_0]}_{\mathcal{S}_i\cup \mathcal{S}_j}\right)^{-1}$
      \State $\widehat{G}_{\mathcal{S}_i\cup \mathcal{S}_j} = \mbox{$k$-means}\left(\widehat{A}_{\mathcal{S}_i\cup \mathcal{S}_j}\right)‎$
      \State $j= j + 1$
     \EndWhile
      \State $j = 1$
      \State $i = i + 1$
    \EndWhile
  \end{algorithmic}
\end{algorithm}

We are now ready to see an application of the Patch\&Catch algorithm.
The overall network is made up of $N=300$ nodes, and is generated according to an Erd\H{o}s-R\'enyi graph with probability of connection $p_N=5 (\log N)/N$.
The combination matrix~$A$ is obtained via the Metropolis rule, and the system is observed over a time scale of  $n_{\max}=10^5$ samples.
We run the Patch\&Catch algorithm in a sub-region $\mathcal{S}$, assuming a strict probing constraint of $M=10$ nodes per experiment.

\begin{figure*}
{\centering{\bf ~~~~~~Sequential topology reconstruction}\par\medskip}
\begin{minipage}{.33\linewidth}
\centering
\includegraphics[scale=.5]{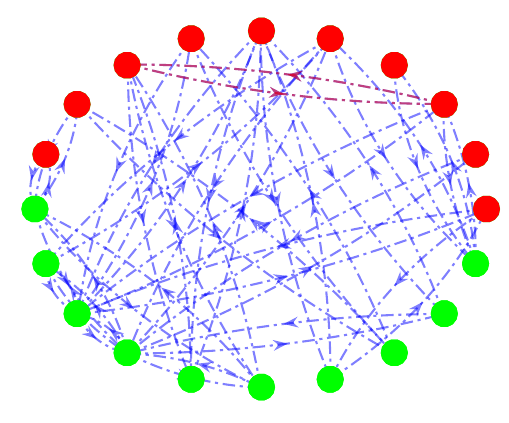}
\\
Probing experiment no. 1
\end{minipage}
\begin{minipage}{.33\linewidth}
\centering
\includegraphics[scale=.5]{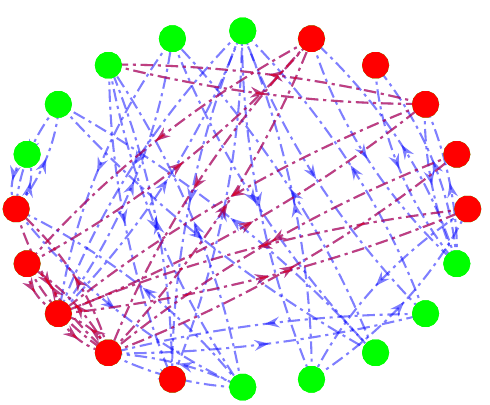}
\\
Probing experiment no. $2$
\end{minipage}
\begin{minipage}{.33\linewidth}
\centering
\includegraphics[scale=.5]{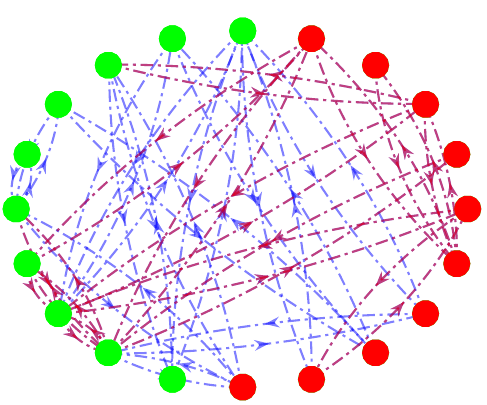}
\\
Probing experiment no. $3$
\end{minipage}
\\
\\
\\
\begin{minipage}{.33\linewidth}
\centering
\includegraphics[scale=.5]{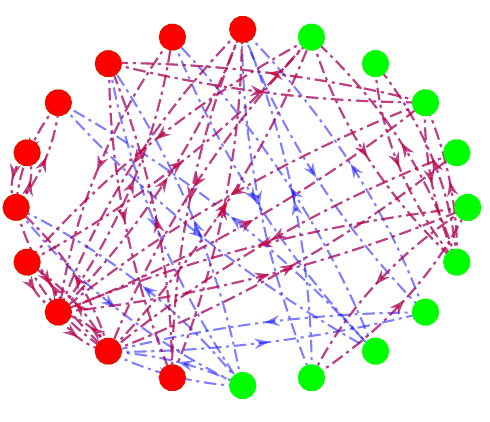}
\\
Probing experiment no. $4$
\end{minipage}
\begin{minipage}{.33\linewidth}
\centering
\includegraphics[scale=.5]{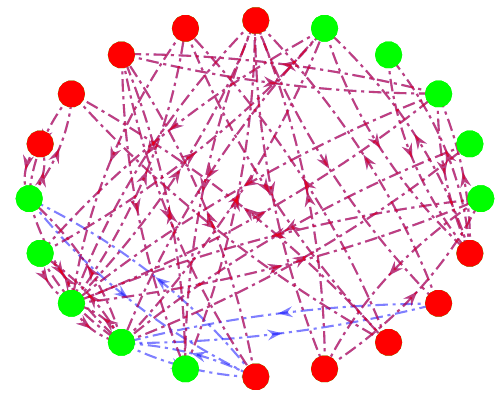}
\\
Probing experiment no. $5$
\end{minipage}
\begin{minipage}{.33\linewidth}
\centering
\includegraphics[scale=.5]{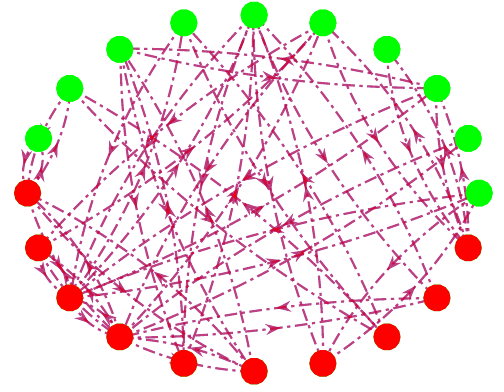}
\\
Probing experiment no. $6$
\end{minipage}
\caption{Illustration of the sequential graph reconstruction. We consider $S=20$ nodes with probing limit $M=10$. Each patch $\mathcal{S}_i$ has cardinality equal to $S_i=5$ nodes. At each experiment two patches are probed. The red nodes represent the nodes being probed at each experiment and the red edges represent the inferred edges up to the current experiment. All pairs were correctly classified.
}
\label{fig:tomography4}
\end{figure*}

In Figure~\ref{fig:tomography4} we consider a subset $\mathcal{S}$ of cardinality $S=20$, and we display the evolution of the algorithm for an increasing number of tested patches.
Since $M=10$, and choosing equal-sized patches, we get $P=S/(M/2)=4$ patches.
For each experiment, we depict the true graph of connections (blue edges), as well as the overall graph of connections estimated up to the current experiment (red edges). The network nodes that form the patches tested in the single experiment are highlighted in red.
We see from Figure~\ref{fig:tomography4} that the network is faithfully reconstructed, sequentially as the number of experiments grows, until the complete subnetwork topology is correctly retrieved after $P(P-1)/2=6$ experiments.

\begin{figure} [t]
\begin{center}
\includegraphics[scale= 0.5]{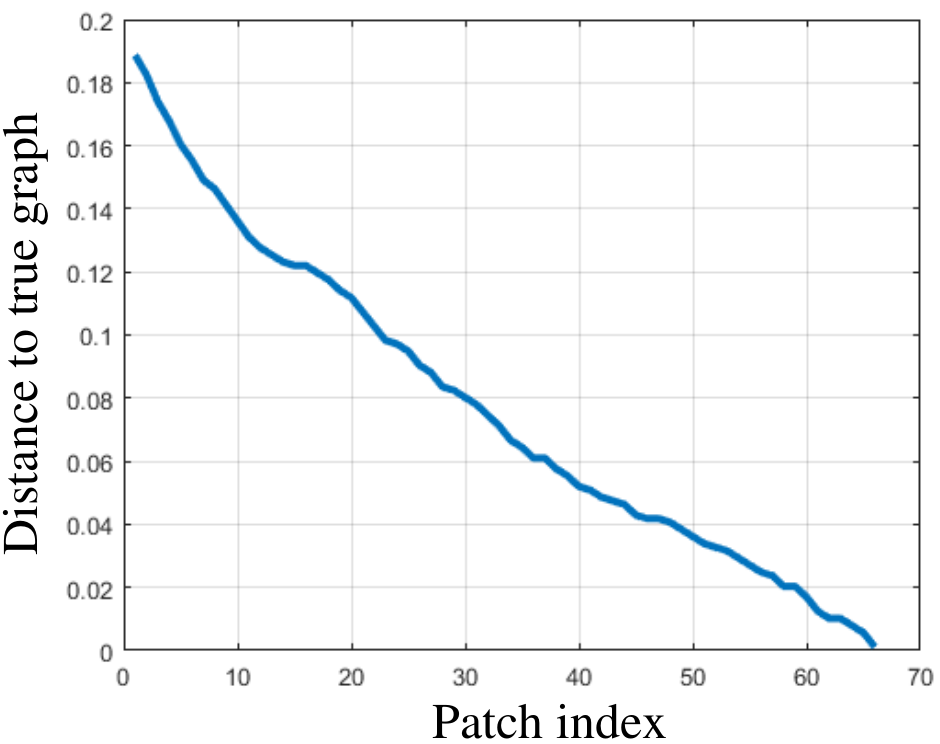}
\caption{Illustration of the monotonic decrease, as more patches are probed, of the distance between the true subnetwork $G_{\mathcal{S}}$ and the estimator $\widehat{G}_{\mathcal{S}}$. We consider $S=60$ nodes in $\mathcal{S}$ with probing limit $M=10$. Each patch has $S_i=5$ nodes, for a total number of $P=2 S/M=12$ patches. At each experiment two patches are probed, yielding a total of $P(P-1)/2=66$ experiments. The graph displays the distance between $G_{\mathcal{S}}$ and the estimated graph at the experiment $\ell=1,2,\ldots,66$. In the considered experiment, only two pairs were misclassified.}
\label{fig:tomography6}
\end{center}
\end{figure}

In Figure~\ref{fig:tomography6}, the same procedure is applied to a larger subset with $S=60$.
For this case, we illustrate the performance delivered by the Patch\&Catch algorithm in a more quantitative way. More precisely, we display the evolution, as more experiments are performed, of the normalized distance between the true graph $G_{\mathcal{S}}$ and the estimated graph $\widehat{G}_{\mathcal{S}}$, namely:
\beq
{\sf dist}(G_{\mathcal{S}},\widehat{G}_{\mathcal{S}})\dfz
\frac{2}{S(S-1)}
\sum_{i,j\in\mathcal{S}, i< j} |g_{ij}-\widehat{g}_{ij}|,
\eeq
and we assume that initially the estimated graph has no edges.
We see from Figure~\ref{fig:tomography6} that the aforementioned distance exhibits a desirable decreasing behavior: the discrepancies between the true graph and the estimated graph diminish progressively as more experiments are conducted.

Before concluding this section, it is useful to comment on two important aspects.
First, the algorithm can easily be generalized to account for overlapping patches. This would simply require to set a tie-break rule for managing the case where a particular pair, say $(h,k)$, is present in two distinct experiments.
Since usually the connection probability is small, one meaningful rule could be an AND rule, where the pair $\left(h,k\right)$ is labeled as connected only if so they are in both experiments. The simplest tie-break rule might be retaining just the first classification of one pair of nodes, ignoring the results possibly arising from subsequent experiments.
Second, in some applications, the reconstruction can be formed sequentially, by exploiting, at each experiment, the information coming from past experiments. For example, having ascertained the structure of a given subset of nodes might be informative of important network-level features of some nodes -- e.g., high degree nodes -- and hence, informative on their level of importance on the network. Likewise, some prior knowledge on a particular network structure (e.g., a tree structure), could help to optimize the formation of successive patches.

\section{Comparison with the results in~\cite{tomo}}
\label{sec:comparison}

It is useful to contrast the results of this work with the results in~\cite{tomo}.
As already explained in the introductory sections, the main differences can be summarized as follows.
In~\cite{tomo}, the network is homogeneous, since all the connections (also those in $\mathcal{S}$) obey a classic Erd\H{o}s-R\'enyi construction with connection probability $p_N$.
Also, the size of the network, $S$, scales linearly with $N$ as $\xi N$, meaning that the fraction of observable nodes, $\xi$, is constant (and greater than zero). Finally, the consistency result in [31] holds in a weak sense: it shows that the fraction of correctly identified edges converges to one in the limit and an independence approximation is used to control the error rate. In this work, consistency holds in a strong sense and the independence approximation is not required. The results obtained here generalize the above framework in several directions.

\subsection{The case of fixed $S$}
In this work, we prove that perfect recovery is achievable even in the extreme case that the number of observable nodes is fixed when $N$ diverges, namely, when the observable network portion is embedded into an infinitely large number of unobservable nodes.
We remark that the case of fixed $S$ cannot be addressed with the tools used in~\cite{tomo}. Let us now explain why.
The result proved in~\cite{tomo} relies essentially on the following result (Theorem~1 in~\cite{tomo}):
\beq
\sum_{j=1}^S e_{ij}\leq \rho,
\label{eq:sumerror}
\eeq
which reveals that the (column-wise) sum of the errors is limited, irrespectively of the network size. This result is obtained by exploiting matrix algebra tools.
It is shown in~\cite{tomo} how~(\ref{eq:sumerror}) leads to the useful conclusion that, on average, the off-diagonal entries of the error matrix scale as $1/S$, which further implies that\footnote{Actually, the result in~\cite{tomo} is formulated in terms of empirical fraction of errors. In the case that permutation invariance holds -- see Property~$2$ in~\cite{tomo} -- the result is easily rephrased as in~(\ref{eq:rough}).}:
\beq
\P[Np_N \bm{e}_{ij}>\epsilon]\lesssim \frac{N}{S} \,p_N, \qquad \textnormal{(Ref.~\cite{tomo})}
\label{eq:rough}
\eeq
where the symbol ``$\lesssim$'' here means that the quantity appearing on the left-hand side is upper bounded by a quantity that scales, asymptotically with $N$, as the quantity appearing on the right-hand side.
Equation~(\ref{eq:rough}) reveals two useful facts.
First, when $S/N$ stays constant as $N$ grows, and since $p_N$ goes to zero, we see that the magnified error vanishes. This is one fundamental conclusion ascertained in~\cite{tomo}.
At the same time, Eq.~(\ref{eq:rough}) highlights how, for fixed $S$, we are no longer in the position of establishing from~(\ref{eq:rough}) that the magnified error converges to zero, because the product $N p_N$ diverges with $N$.
In summary, the matrix-algebra tools taken in~\cite{tomo} are not powerful enough to address the challenging case when $S$ is fixed, namely, when the fraction of observable nodes goes to zero as $N$ grows.

On the other hand, in this work we show how this more challenging scenario can be addressed, by exploiting matrix-graph tools, i.e., by evaluating paths and distances over graphs.
One important benefit of the new approach is that the results now hold for {\em an arbitrary topology of the observable network portion}, while in~\cite{tomo} this latter component was constrained to be Erd\H{o}s-R\'enyi.

\subsection{The case of $S\sim\xi N$}
We notice that the results of this work can be applied to the case addressed in~\cite{tomo}. Indeed, when $G_{\mathcal{S}}$ is Erd\H{o}s-R\'enyi, we can repeat the proof of Theorem~1 by essentially skipping the homogenization-and-coupling step, because the overall graph is homogeneous {\em ab initio}. Then we would get, for any $i\neq j$ (also for the connected pairs, in this particular case):
\beq
\P[Np_N \bm{e}_{ij}>\epsilon]\lesssim p_N(Np_N)^{r_N + 2}.
\label{eq:magnifiederrprob}
\eeq
Therefore, both the {\em matrix-algebra} approach (used in~\cite{tomo}), and the {\em matrix-graph} approach (used here) lead to the result that the topology of the observable network portion can be reconstructed faithfully.
However, it must be remarked that the matrix-graph approach requires some additional conditions on the connection probability, $p_N$, which translate into a slightly more restrictive requirement in terms of sparsity.

On the other hand, and interestingly, the matrix-algebra approach and the matrix-graph approach lead to different estimates on {\em how} the error probability in~(\ref{eq:magnifiederrprob}) converges to zero.
Indeed, with the approach used in~\cite{tomo}, one is able to see that the rate of decay is at least in the order of $p_N$ (see~(\ref{eq:rough}), and observe that $N/S \sim 1/\xi$).
Moreover, in~\cite{tomo} it is shown that the decay rate is actually faster than $p_N$.
In contrast, with the approach adopted in the current work we get the upper bound in~(\ref{eq:magnifiederrprob}), which provides the (looser) asymptotic prediction that the decay rate is slower than $p_N$.
In summary, we conclude that, under the regime $S\sim\xi N$, and for a {\em full} Erd\H{o}s-R\'enyi construction, the results of~\cite{tomo} are more powerful in predicting the decay rate of the error probabilities.
It could be interesting at this point to ask whether it is possible to combine the matrix-algebra approach with the matrix-graph approach to obtain refined estimates.


\begin{appendices}

\section{Some useful lemmas}
\label{app:lemmas}

\begin{figure} [t]
\begin{center}
\includegraphics[scale= 0.75]{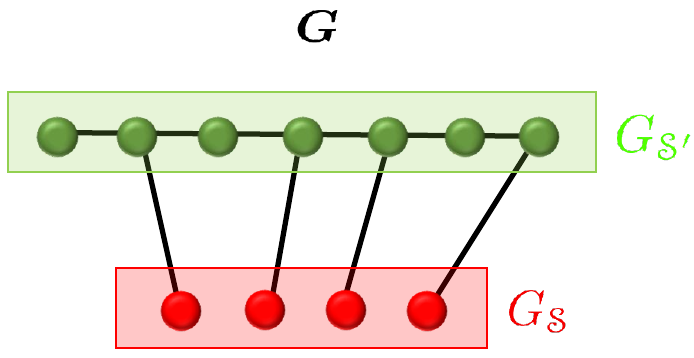}
\caption{If the subnetwork $G_{\mathcal{S}'}$ is connected and there is no node in $\mathcal{S}$ that is isolated in the network $G$, then $G$ is connected. Equation~\eqref{eq:connectedhigh} conforms to the corresponding contrapositive statement.}
\label{fig:connectedhigh2}
\end{center}
\end{figure}

\begin{IEEEproof}[Proof of Lemma~\ref{lem:connected}]
In order to prove that the partial Erd\H{o}s-R\'enyi graph $\bm{G}\sim\mathscr{G}^{\star}(N,p_N,G_{\mathcal{S}})$ is connected, it suffices to consider the worst case where the embedded graph, $G_{\mathcal{S}}$, is {\em internally} disconnected, i.e., where no edges exist between nodes in $\mathcal{S}$.
We note that the nodes in $\mathcal{S}$, even if disconnected, can still be connected to nodes belonging to the unobserved set, $\mathcal{S}'$. The latter property enables the possibility that the overall graph, $\bm{G}$, is connected, as we are going to show.

Since we are assuming that $G_{\mathcal{S}}$ is internally disconnected,
the overall graph is connected if both $\bm{G}_{\mathcal{S}'}$ is connected, and any node in $\mathcal{S}$ connects to some node in $\mathcal{S}'$. Refer to Figure~\ref{fig:connectedhigh2} for a graphical illustration. We prove Lemma~\ref{lem:connected} via the contrapositive statement:
the overall graph is not connected if either $\bm{G}_{\mathcal{S}'}$ is not connected, or if at least one node in $\mathcal{S}$ is not connected to $\mathcal{S}'$, namely, we have that:
\beqa\label{eq:connectedhigh}
& \{\bm{G}\textnormal{ not connected}\} & \\
& \subseteq &\nonumber\\
& \{\bm{G}_{\mathcal{S}'}\textnormal{ not connected}\} \bigcup \{\exists \textnormal{ an isolated node of $\bm{G}$ in }\mathcal{S}\}.&\nonumber
\eeqa
Therefore, applying the union bound we get:
\begin{eqnarray}
& \mathbb{P}[\bm{G}\textnormal{ is not connected}] & \\
& \leq & \nonumber\\
& \P[\bm{G}_{\mathcal{S}'}\textnormal{ is not connected}]+ \P[\exists  \textnormal{ an isolated node of $\bm{G}$ in $\mathcal{S}$}] & \nonumber\\
& = & \nonumber\\
& \P[\bm{G}_{\mathcal{S}'}\textnormal{ is not connected}]+ 1-(1-(1-p_N)^{N-S})^S.&\nonumber
\end{eqnarray}
Since $\bm{G}_{\mathcal{S}'}$ is a classic Erd\H{o}s-R\'enyi $\mathscr{G}^{\star}\left(N-S,p_N\right)$, we have that:
\beq
\lim_{N\rightarrow\infty}\P[\bm{G}_{\mathcal{S}'}\textnormal{ is not connected}]=0.
\eeq
Moreover, since $\mathcal{S}$ is fixed, we have that:
\beqa
(1-p_N)^{N-S} & = &
\left(1-\frac{\log N + c_N}{N}\right)^{N-S}\\
& \leq &
\left(1-\frac{\log N }{N}\right)^{N-S}\stackrel{N\rightarrow\infty}{\longrightarrow} 0.\nonumber
\eeqa
\end{IEEEproof}

\begin{IEEEproof}[Proof of Lemma~\ref{lem:policies}]
In order to prove the claim of the lemma, we must show that~(\ref{eq:dmaxa}) implies~(\ref{eq:bound}) with the choice $\tau=\gamma/e$.
Let us observe preliminarily that~(\ref{eq:dmaxa}) yields the following implication:
\beq
\{d_{\max}(\bm{G})< e Np_N , \bm{g}_{ij}=1\}
\subseteq
\{Np_N \bm{a}_{ij}>\tau , \bm{g}_{ij}=1\}.
\eeq
Therefore, we can write:
\beq
\P[Np_N \bm{a}_{ij}>\tau | \bm{g}_{ij}=1]\geq \P[d_{\max}(\bm{G})< e Np_N | \bm{g}_{ij}=1].
\eeq
Now, by trivial upper bounding techniques, we can obtain the following chain of inequalities:
\beqa
 \lefteqn{\P\left[\left.d_{\max}(\bm{G})\geq eNp_N \right| \bm{g}_{ij} = 1\right]}
\nonumber\\
& \leq &
\P\left[\left.\max_{n\notin \mathcal{S}} \sum_{k} \bm{g}_{nk} > eNp_N \right| \bm{g}_{ij} = 1\right]\\
&&+
\P\left[\left.\max_{n\in \mathcal{S}} \sum_{k} \bm{g}_{nk} > eNp_N \right| \bm{g}_{ij} = 1\right]\nonumber\\
&=&
\P\left[\left.\max_{n\notin \mathcal{S}} \sum_{k} \bm{g}_{nk} > eNp_N \right| \bm{g}_{ij} = 1\right]\\
&&+
\P\left[\left.\max_{n\in \mathcal{S}} \left(\sum_{k\in \mathcal{S}} \bm{g}_{nk}+\sum_{k\notin \mathcal{S}} \bm{g}_{nk}\right) > eNp_N \right| \bm{g}_{ij} = 1\right]
\nonumber\\
& \leq &
\P\left[\left.\max_{n\notin \mathcal{S}} \sum_{k} \bm{g}_{nk} > eNp_N \right| \bm{g}_{ij} = 1\right]\\
&&+
\P\left[\left.S+\max_{n\in \mathcal{S}}\sum_{k\notin \mathcal{S}} \bm{g}_{nk} > eNp_N \right| \bm{g}_{ij} = 1\right]\nonumber\\
&=&
\P\left[\left.\max_{n\notin \mathcal{S}} \sum_{k} \bm{g}_{nk} > eNp_N \right| \bm{g}_{ij} = 1\right]\\
&&+
\P\left[\left.\max_{n\in \mathcal{S}} \sum_{k\notin \mathcal{S}} \bm{g}_{nk} > eNp_N - S \right| \bm{g}_{ij} = 1\right]\nonumber\\
& \leq & \epsilon_N\stackrel{N\rightarrow\infty}{\longrightarrow} 0,
\eeqa
where the last inequality follows directly from Lemma~$1$ in~\cite{tomo}, since $S$ is fixed and since the subgraph formed by the edges $\bm{g}_{nk}$ with either $n\notin \mathcal{S}$ or $k\notin \mathcal{S}$ is Erd\H{o}s-R\'enyi with parameter $p_N$ defined in equation~\eqref{eq:pnscale}.
Actually, for $i,j\in \mathcal{S}$ (i.e., when $\bm{g}_{ij}=g_{ij}$ is deterministic) a simplified version of Lemma~$1$ in~\cite{tomo} does suffice, where the conditioning can be skipped.
\end{IEEEproof}

\section{Proof of Theorem~$1$}
\label{app:Theor1}
We first prove part $i)$.
It is shown in~\cite{tomo} that the entries of the error matrix defined in~(\ref{eq:errmatfirstdef0}) are nonnegative, i.e., $\bm{E}_{\mathcal{S}}\geq 0_{S\times S}$, and, hence, we can write, for $i,j\in \mathcal{S}$:
\beq
Np_N [\widehat{\bm{A}}_{\mathcal{S}}]_{ij}=Np_N \bm{a}_{ij} + N p_N \bm{e}_{ij}\geq Np_N \bm{a}_{ij}.
\eeq
Therefore, from Property~$2$ we get immediately the claim in part $i)$.
If we further show that the magnified error $Np_N \bm{e}_{ij}$ converges to zero in probability over the non-interacting pairs (i.e., if we prove part $ii)$ of the present theorem), then we can attain exact (with high probability) classification via inspection on the truncated estimator $\widehat{\bm{A}}_{\mathcal{S}}$: if $Np_N [\widehat{\bm{A}}_{\mathcal{S}}]_{ij}>\tau$, then classify $(i,j)$ as an interacting pair, otherwise classify it as non-interacting, where $\tau$ is the threshold characterizing the family~$\mathscr{C}_{\rho,\tau}$ of weight assignments from where $\bm{A}$ is obtained, in view of Property~$2$. As a result, and since the cardinality of the observable set is finite, part $iii)$ would follow if we are able to prove part $ii)$. The proof of part $ii)$ is demanding and will be developed through a sequence of five steps.

{\em Step~$1$: Relating the error to the distance between nodes belonging to $\mathcal{S}'$}.
It is shown in~\cite{tomo} that the error matrix in~(\ref{eq:errmatfirstdef0}) can be represented as:\footnote{During this first step the boldface notation will be skipped because we focus on properties that depend solely on the structure of the matrix, and not on the statistical model of the underlying graph.}
\begin{equation}
\boxed{
E_{\mathcal{S}}=A_{\mathcal{S}\mathcal{S}'}H B_{\mathcal{S}'\mathcal{S}}
}
\label{eq:Ematdefnew}
\end{equation}
where
\beq
B\dfz A^2,\qquad
H\dfz\left(I_{\mathcal{S}'}-B_{\mathcal{S}'}\right)^{-1}.
\label{eq:BHmats}
\eeq
From~(\ref{eq:Ematdefnew}) we can write, for $i,j\in \mathcal{S}$:
\beq
\boxed{
e_{ij} = \sum_{\ell,m\in \mathcal{S}'} a_{i \ell} h_{\ell m} b_{m j}
}
\label{eq:relevanterror}
\eeq
where~$e_{ij}$ is the error at the pair~$\left(i,j\right)$. Therefore, in order to control the size of the error $e_{ij}$, small values of the factors $h_{\ell m}$, for $\ell,m\in \mathcal{S}'$, would be desirable. In view of the definition for $H$ in~(\ref{eq:BHmats}), we have that:
\beq
h_{\ell m} = \left[\left(I_{\mathcal{S}'} - B_{\mathcal{S}'} \right)^{-1}\right]_{\ell m} = \left[\sum_{k=0}^{\infty} \left(B_{\mathcal{S}'}\right)^k\right]_{\ell m},
\label{eq:twistpower}
\eeq
as the matrix~$B_{\mathcal{S}'}=\left[A^2\right]_{\mathcal{S}'}$ is stable, since $\rho\left(B_{\mathcal{S}'}\right)<\left|\left|B_{\mathcal{S}'}\right|\right|_{\infty}<1$, from Property~$1$, where~$\rho\left(B_{\mathcal{S}'}\right)$ is the spectral radius of~$B_{\mathcal{S}'}$.
It is useful at this point to recall the following known fact from matrix algebra that relates the entries associated with the powers of a matrix with the distances between nodes on its underlying support graph.

Let $M\in\mathbb{S}_{+}^{N\times N}$ be a nonnegative symmetric matrix with positive diagonal entries, and let $G(M)$ be its underlying support graph.
Consider the powers of the matrix $M$, namely, $M^k$, for $k=1,2,\ldots$.
Then we have that~\cite{MatrixAnalysis}:
\beq
\delta_{\ell,m}(G(M))=r \Leftrightarrow \textnormal{the smallest $k$ with $[M^{k}]_{\ell m} > 0$ is $r$},
\label{eq:elementary}
\eeq
where~$\delta_{\ell,m}(G(M))$ represents the distance between the nodes $\ell$ and $m$ in the graph $G(M)$ as defined in Sec.~\ref{sec:graphnotation}. In fact, note that, if $\ell$ is not connected to $m$ in the support graph $G(M)$, then $M_{\ell m}=0$. If the smallest path connecting $\ell$ to $m$ has a length of two hops (in particular, $\ell$ is not connected to $m$, hence $M_{\ell m}=0)$, then there exists $k$ so that $M_{\ell k}>0$ and $M_{k m}>0$. Thus, $\left[M^2\right]_{\ell m}=\sum_{r} M_{\ell r} M_{r m}> M_{\ell k} M_{k m}>0$. Reasoning by induction one can establish~\eqref{eq:elementary}. The following observation follows: if $M$ is stable, i.e., $\rho(M)<1$, and if the distance $\delta_{\ell m}(G(M))=r$ is {\em large}, then $\left[M^k\right]_{\ell m}$ is {\em small} for all $k$ as for $k<r$ we have $\left[M^k\right]_{\ell m}=0$ and for $k\geq r$ the corresponding power $\left[M^k\right]_{\ell m}$ is {\em small} since $k$ is {\em large} and $M$ is stable.

Now, examining~(\ref{eq:twistpower}), and using~(\ref{eq:elementary}) with $M=B_{\mathcal{S}'}$, one might be tempted to conclude that a small $h_{\ell m}$ would result if nodes $\ell$ and $m$ are distant from each other.
The reasoning is correct, but note that, in general, the distance between $\ell$ and $m$ is dependent on the topology of the network $G_{\mathcal{S}}$, which is arbitrary.
In other words, by relying solely on the elementary observation in~\eqref{eq:elementary}, one would not be able to draw useful conclusions about the magnitude of the entries $h_{\ell m}$ (and, hence, of the entries in the error matrix~$E_{\mathcal{S}}$) in our context where $G_{\mathcal{S}}$ is arbitrary.


As a matter of fact, as stated in Theorem~\ref{lem:distancia} (proved in Appendix~\ref{sec:proof}) the distance affecting $h_{\ell m}$ is the one between $\ell$ and $m$ on a transformed graph, $G_{\mathcal{S}\nleftrightarrow \mathcal{S}}$, which is the graph obtained from $G$ by removing all the edges connecting nodes inside the observable subset $\mathcal{S}$, introduced in Sec.~\ref{sec:graphop} -- refer to Figure~\ref{fig:partial2} for a graphical illustration of the contrast between $G$ and $G_{\mathcal{S}\nleftrightarrow \mathcal{S}}$. Note that the edges possibly connecting nodes from $\mathcal{S}$ to nodes in $\mathcal{S}'$ are {\em not} removed in the graph $G_{\mathcal{S}\nleftrightarrow \mathcal{S}}$.




\begin{theorem}\label{lem:distancia}
Given two distinct nodes $\ell,m\in \mathcal{S}'$, we have that:
\begin{equation}\label{eq:hlbound}
\boxed{
\delta_{\ell,m}(G_{\mathcal{S}\nleftrightarrow \mathcal{S}})=r
\Rightarrow h_{\ell m} \leq \frac{\rho^{r}}{1-\rho^2}
}
\end{equation}
where $h_{\ell m}$ is the $(\ell,m)$-th entry of the matrix
\beq
H=\left(I_{\mathcal{S}'}-B_{\mathcal{S}'}\right)^{-1},
\eeq
$B_{\mathcal{S}'}=\left[A^2\right]_{\mathcal{S}'}$ and $0<\rho<1$ is an upper-bound for the maximum row-sum of the matrix $A$, in view of Property~\ref{pr:stability}, remarking that $A$ is a combination matrix satisfying Properties~\ref{pr:stability} and~\ref{pr:nondegeneracy}, i.e., obtained from any weight assignment in the class~$\mathscr{C}_{\rho,\tau}$, as, e.g., the Metropolis and the Laplacian weight assignment rules.\qed
\end{theorem}
In words, Theorem~\ref{lem:distancia} relates the magnitude of the entries of $H$ with the distance between nodes in a manner that does not depend on the subnetwork $G_{\mathcal{S}}$. We remark that we do {\em not} assume that the nodes in $G_{\mathcal{S}}$ are not connected among each other.
In fact, we impose no restrictions whatsoever on the topology of $G_{\mathcal{S}}$ to prove the main theorem.
Reference to the graph $G_{\mathcal{S}\nleftrightarrow \mathcal{S}}$ is used only when devising universal bounds on the terms $h_{\ell m}$, in view of Theorem~\ref{lem:distancia}.
In other words, we are able to rule out the role of the subnetwork topology $G_{\mathcal{S}}$ in as much as computing upper bounds for $H$.

In the next step, we will show in detail how~(\ref{eq:hlbound}) is helpful to control the size of the error in~(\ref{eq:relevanterror}).

\begin{figure} [t]
\begin{center}
\includegraphics[scale= 0.5]{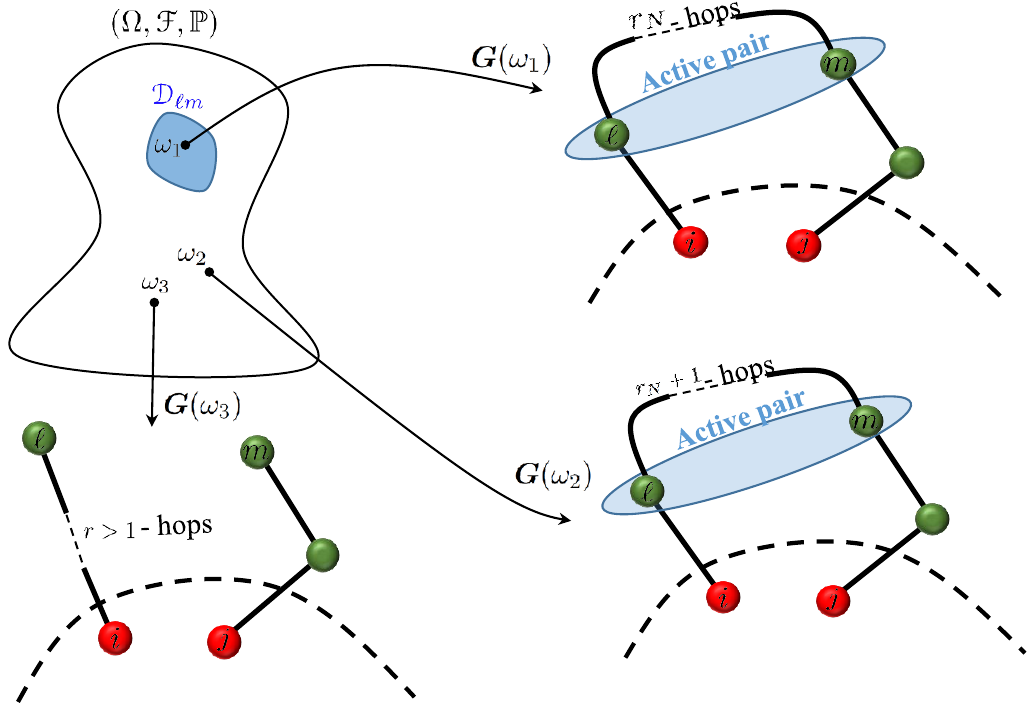}
\caption{Illustration of random realizations with $\omega_1\in\mathcal{D}_{\ell,m}$ and $\omega_2,\omega_3\in \mathcal{D}'_{\ell m}$. This emphasizes that other than the (edges of the) random graph $\bm{G}$, all other quantities, i.e., $r_N$, $i,j,\ell$ and $m$ are fixed (or deterministic).}
\label{fig:random}
\end{center}
\end{figure}

{\em Step~$2$: Large distances vs. small distances}.
The summation appearing in~(\ref{eq:relevanterror}) can be restricted to nodes that obey the conditions:
\beq
\ell\in \mathcal{N}_i(\bm{G}),\qquad
m\in \mathcal{N}_j^{(2)}(\bm{G}),
\label{eq:neighcond}
\eeq
namely, to nodes $\ell\in \mathcal{S}'$ that are neighbors of the node $i\in \mathcal{S}$ (so that $\bm{a}_{i \ell}>0$), and to nodes $m\in \mathcal{S}'$ that are second-order neighbors of the node $j\in \mathcal{S}$ (so that $\bm{b}_{m j}>0$). Henceforth, we refer to such pair $\left(\ell,m\right)$ as an {\em active pair}. Figure~\ref{fig:active} depicts the possible configurations of the active pairs. In words, the summation characterizing the error in equation~\eqref{eq:relevanterror} runs only over the active pairs.
In fact, the error in~(\ref{eq:relevanterror}) can be represented as:
\beqa
N p_N\bm{e}_{ij} & = &
N p_N \sum_{\ell,m\in \mathcal{S}'} \bm{a}_{i \ell} \bm{h}_{\ell m} \bm{b}_{m j} \\
& = &
N p_N \sum_{\ell,m\in \mathcal{S}'} \bm{a}_{i \ell} \bm{J}_{\ell m} \bm{h}_{\ell m} \bm{b}_{m j},
\label{eq:relevanterrornew}
\eeqa
where the randomness of the various quantities, arising from the randomness of the underlying random graph~$\bm{G}$, has been now emphasized through the boldface notation, and where we have introduced the variable:
\beqa
\bm{J}_{\ell m}&\dfz&\mathds{I}_{\{\bm{a}_{i\ell}>0,\,\bm{b}_{m j}>0\}}=\mathds{I}_{\left\{\ell\in \mathcal{N}_i(\bm{G}),\,m\in \mathcal{N}^{(2)}_j(\bm{G})\right\}},\nonumber\\
\label{eq:indicvar}
\eeqa
as the indicator of an active pair $\left(\ell,m\right)\in \mathcal{S}'\times \mathcal{S}'$, i.e.,~$\bm{J}_{\ell m}=1$ if~$\left(\ell,m\right)$ is an active pair and~$\bm{J}_{\ell m}=0$ otherwise. Now, in order to prove part $ii)$ of Theorem~$1$, we need to prove that, for two {\em non-interacting} nodes $i$ and $j$, and for any $\epsilon>0$:
\beq
\P[N p_N \bm{e}_{i j}>\epsilon]\stackrel{N\rightarrow\infty}{\longrightarrow} 0.
\label{eq:part2theo1}
\eeq
As stated in Theorem~\ref{lem:distancia}, in Step~$1$, the distance between nodes $\ell,m\in \mathcal{S}'$ on the aforementioned reference graph, $\bm{G}_{\mathcal{S}\nleftrightarrow \mathcal{S}}$, plays a role in the size of $h_{\ell m}$ and hence, in the magnitude of the error. In addition, we have seen that the relevant nodes are those obeying~(\ref{eq:neighcond}), i.e., the active pairs. It is therefore useful to introduce the following events.
For $\ell,m\in \mathcal{S}'$, with $\ell\neq m$, we define:
\beq
\mathcal{D}_{\ell,m}\dfz
\{\delta_{\ell,m}(\bm{G}_{\mathcal{S}\nleftrightarrow \mathcal{S}})\leq r_N, \ell\in \mathcal{N}_i(\bm{G}), m\in\mathcal{N}_j^{(2)}(\bm{G})\},
\label{eq:Dsmallm}
\eeq
where $r_N$ is a certain sequence of distances, with $r_N\rightarrow\infty$ as $N\rightarrow\infty$, in a way that will be specified later. The event in~(\ref{eq:Dsmallm}) certifies that the distance on the graph $\bm{G}_{\mathcal{S}\nleftrightarrow \mathcal{S}}$ between two distinct nodes, $\ell,m\in \mathcal{S}'$,  does not exceed a prescribed value $r_N$, {\em and}  also certifies the membership of the nodes $\ell$ and $m$ to the pertinent neighborhoods defined on the graph $\bm{G}$, i.e., it certifies that $\left(\ell,m\right)$ is an active pair. We remark that~$\mathcal{D}_{\ell,m}$ is, formally, a (measurable) set and that the only random object characterizing~$\mathcal{D}_{\ell,m}$ in equation~\eqref{eq:Dsmallm} is the random graph $\bm{G}$ -- refer to Remark~\ref{re:probspace} in Sec.~\ref{sec:graphnotation}.
The observable subset $\mathcal{S}$, the sequence~$r_N$ and the indexes~$i,j,\ell,m$ are fixed (or deterministic). Accordingly,~$\mathcal{D}_{\ell,m}$ represents the set of realizations of the partial Erd\H{o}s-R\'enyi random graph~$\bm{G}$ where the constraints of distance and neighborhood among the fixed nodes $i,j,\ell,m$ in equation~\eqref{eq:Dsmallm} are met. Refer to Figure~\ref{fig:random} for an illustration.


Likewise, for $\ell=m$ the set~$\mathcal{D}_{\ell,m}$ reduces to:
\beq
\mathcal{D}_{\ell,\ell}
\dfz
\{\ell\in \mathcal{N}_i(\bm{G}) \cap \mathcal{N}_j^{(2)}(\bm{G})\},
\label{eq:Dsmallll}
\eeq
where we see that the event $\mathcal{D}_{\ell,\ell}$ simply certifies the membership of the node $\ell$ to the pertinent neighborhoods defined on the graph $\bm{G}$.
We finally introduce the union event:
\beq
\mathcal{D}_{\textnormal{small}}\dfz\bigcup_{\ell,m\in \mathcal{S}'}\mathcal{D}_{\ell,m},
\label{eq:Dsmalltot}
\eeq
as the event where the distance is {\em small}, i.e., $\delta_{\ell m}\left(\bm{G}\right)\leq r_N$, for {\em at least} one active pair~$\left(\ell,m\right)\in S'\times S'$.
We can write:
\beqa
\P[N p_N \bm{e}_{i j}>\epsilon]
&=&
\P[N p_N \bm{e}_{i j}>\epsilon , \mathcal{D}_{\textnormal{small}}]\\
&&
+
\P[N p_N \bm{e}_{i j}>\epsilon , \mathcal{D}'_{\textnormal{small}}]\nonumber\\
&\leq&
\P[\mathcal{D}_{\textnormal{small}}]
+
\P[N p_N \bm{e}_{i j}>\epsilon , \mathcal{D}'_{\textnormal{small}}]\nonumber,
\label{eq:totproberr}
\eeqa
where~$\mathcal{D}'_{\textnormal{small}}$ is the complement of the event (or measurable set) $\mathcal{D}_{\textnormal{small}}$, i.e.,~$\mathcal{D}'_{\textnormal{small}}\cap \mathcal{D}_{\textnormal{small}}=\emptyset$, and can be read as the event (or set of realizations of $\bm{G}$) where the distances are {\em large}, i.e., $\delta_{\ell m}\left(\bm{G}\right)> r_N$, for all active pairs $\left(\ell,m\right)$. For the sake of a compact notation, we write
\begin{equation}
\P[N p_N \bm{e}_{i j}>\epsilon , \mathcal{D}'_{\textnormal{small}}]
\end{equation}
instead of
\begin{equation}
\P[\{N p_N \bm{e}_{i j}>\epsilon \} \cap \mathcal{D}'_{\textnormal{small}}].
\end{equation}
Then, the route that we follow to prove the claim in~(\ref{eq:part2theo1}) goes as follows: $i)$ (small-distance) we show that:
\beq
\P[\mathcal{D}_{\textnormal{small}}]\stackrel{N\rightarrow\infty}{\longrightarrow} 0,
\label{eq:smallrare}
\eeq
i.e., the occurrence of a small distance in at least one active pair $\left(\ell,m\right)$ is rare, with high probability, and $ii)$ (large-distance) we show that large distances\footnote{The terminology ``small distances'' and ``large distances'' will be often coined for simplicity to denote $\min_{\left(\ell,m\right) \textnormal{is active}}\delta_{\ell m}\left({\bm G}\right)\leq r_N$ and $\min_{\left(\ell,m\right) \textnormal{is active}}\delta_{\ell m}\left({\bm G}\right)> r_N$, respectively.} imply small errors, formally:
\beq
\P[N p_N \bm{e}_{i j}>\epsilon , \mathcal{D}'_{\textnormal{small}}]=0\quad \textnormal{for sufficiently large $N$}.
\label{eq:largeimplysmall}
\eeq
Equations~\eqref{eq:smallrare} and~\eqref{eq:largeimplysmall} will imply the desired result in equation~\eqref{eq:part2theo1} in view of equation~\eqref{eq:totproberr}. Let us start by proving~(\ref{eq:largeimplysmall}). From the definition in equation~(\ref{eq:Dsmalltot}), we have that
\beq
\mathcal{D}'_{\textnormal{small}}
=
\bigcap_{\ell,m\in \mathcal{S}'}\mathcal{D}'_{\ell,m}.
\label{eq:Dsmalltotcompl}
\eeq
Using~(\ref{eq:Dsmallm}) and~(\ref{eq:Dsmallll}), from~(\ref{eq:Dsmalltotcompl}) we conclude that the complementary event $\mathcal{D}'_{\textnormal{small}}$ can be compactly expressed through the indicator variables in~(\ref{eq:indicvar}) as follows:
\beq
\mathcal{D}'_{\textnormal{small}}
=
\left\{\bm{J}_{\ell m} \bm{J}^{(\delta)}_{\ell,m}=0~\textnormal{ for {\em all }}~\ell,m\in \mathcal{S}' \right\},
\label{eq:Dsmallnotthroughindic}
\eeq
where we have further introduced the indicator variable:
\beq
\bm{J}^{(\delta)}_{\ell,\ell}=1,\qquad
\bm{J}^{(\delta)}_{\ell,m}\dfz\mathds{I}_{\{\delta_{\ell,m}(\bm{G}_{\mathcal{S}\nleftrightarrow \mathcal{S}})\leq r_N\}}\quad \forall \ell\neq m.
\label{eq:indicvardelta}
\eeq
That is, the event $\mathcal{D}'_{\textnormal{small}}$ represents the set of all realizations of the random network $\bm{G}$, where each pair $\left(\ell,m\right)\in\mathcal{S}'\times \mathcal{S}'$ is either non-active or, if active, then $\ell$ is distant from $m$, i.e., $\delta_{\ell m}(\bm{G})> r_N$. We now show that, in view of~(\ref{eq:hlbound}), the occurrence of $\mathcal{D}'_{\textnormal{small}}$ implies an upper bound on the entries $\bm{h}_{\ell m}$, namely,
\beq
\boxed{
\mathcal{D}'_{\textnormal{small}}
\subseteq
\left\{ \bm{h}_{\ell m}\bm{J}_{\ell m} \leq
\frac{\rho^{r_N + 1}}{1-\rho^2}
\bm{J}_{\ell m}~\textnormal{ for {\em all }}~\ell,m\in \mathcal{S}' \right\}
}
\label{eq:hlbound2}
\eeq
Indeed, we know from~(\ref{eq:Dsmallnotthroughindic}) that the occurrence of $\mathcal{D}'_{\textnormal{small}}$ implies that the product $\bm{J}_{\ell m} \bm{J}^{(\delta)}_{\ell,m}$ is equal to zero for {\em all } $\ell,m\in \mathcal{S}'$.

Let us consider first the degenerate case $\ell=m$. Since $\bm{J}^{(\delta)}_{\ell,\ell}=1$, the variable $\bm{J}_{\ell m}$ must be equal to zero and~(\ref{eq:hlbound2}) holds trivially.

We switch to the case $\ell\neq m$. If $\bm{J}_{\ell m}=0$, i.e., $\left(\ell,m\right)$ is not an active pair, then~(\ref{eq:hlbound2}) holds trivially.
If, instead, $\bm{J}_{\ell m}=1$, then we must have $\bm{J}^{(\delta)}_{\ell,m}=0$, i.e.,
\beq
\delta_{\ell,m}(\bm{G}_{\mathcal{S}\nleftrightarrow \mathcal{S}})\geq r_N + 1.
\eeq
As a consequence, Eq.~(\ref{eq:hlbound2}) holds true in view of~(\ref{eq:hlbound}) (proved in Theorem~\ref{lem:distancia}).
Applying now~(\ref{eq:hlbound2}) to~(\ref{eq:relevanterrornew}), we conclude that, when $\mathcal{D}'_{\textnormal{small}}$ occurs, we must have that:
\beqa
N p_N \bm{e}_{ij} & \leq &
N p _N \frac{\rho^{r_N + 1}}{1-\rho^2}
\sum_{\ell\in \mathcal{S}'} \bm{a}_{i \ell}
\sum_{m\in \mathcal{S}'}\bm{b}_{m j}\bm{J}_{\ell m}\\
& \leq &
N p_N \frac{\rho^{r_N + 4}}{1-\rho^2}\nonumber,
\label{eq:errlargedistbound}
\eeqa
where the last inequality holds from the {\em row-sum} stability of $A\in\mathscr{C}_{\rho,\tau}$ (Property $1$):
\beq
\left|\left|A\right|\right|_{\infty} \leq \rho,\qquad \left|\left|B\right|\right|_{\infty} \leq \rho^2.
\eeq
Accordingly, from~(\ref{eq:errlargedistbound}) we have that:
\beq
\P[N p_N \bm{e}_{ij}>\epsilon , \mathcal{D}'_{\textnormal{small}}]
\leq
\P\left[N p_N \frac{\rho^{r_N + 4}}{1-\rho^2}>\epsilon\right],
\label{eq:probounderr}
\eeq
where we remark that the event appearing in the latter probability is in fact a {\em deterministic} event.
If we now find a sequence $r_N$ that drives to zero the quantity $N p_N \rho^{r_N+4}$, then, for sufficiently large $N$, the probabilities appearing in~(\ref{eq:probounderr}) are eventually zero.
We will illustrate how to make a proper selection of $r_N$ in the final step (i.e., Step~$5$) of this proof.
It suffices for now to assume that such a sequence exists, namely, that:
\beq
\boxed{
N p_N \rho^{r_N+4}\stackrel{N\rightarrow\infty}\longrightarrow 0
}
\label{eq:rNnotooslow}
\eeq
which, in view of~(\ref{eq:probounderr}), yields the desired claim in~(\ref{eq:largeimplysmall}).
Therefore, we conclude that, if nodes in $\mathcal{S}'$ forming active pairs, i.e., obeying~(\ref{eq:neighcond}), lie sufficiently far apart on the graph $\bm{G}_{\mathcal{S}\nleftrightarrow \mathcal{S}}$, then the magnified error can be driven to zero as $N\rightarrow\infty$. This observation corroborates the claim that large distances imply small values of the error.
In light of~(\ref{eq:totproberr}), the claim of Theorem~$1$ will be proved if we show that the occurrence of a small distance on {\em at least one} active pair $\left(\ell,m\right)$ is a rare event, namely, if we prove the claim in~(\ref{eq:smallrare}). We will address this challenge in the two forthcoming steps.

{\em Step~$3$: Relating partial Erd\H{o}s-R\'enyi to a standard Erd\H{o}s-R\'enyi via homogenization}.
Two sources of asymmetry make the proof of~(\ref{eq:smallrare}) challenging. First, we see that the events in~(\ref{eq:Dsmallm}) refer to different graphs, namely, $\bm{G}_{\mathcal{S}\nleftrightarrow \mathcal{S}}$ and $\bm{G}$ -- the graph $\bm{G}$ for the neighborhood constraint and $\bm{G}_{\mathcal{S}\nleftrightarrow \mathcal{S}}$ for the distance constraint.
Second, both the local disconnection implied by $\bm{G}_{\mathcal{S}\nleftrightarrow \mathcal{S}}$, and the partial Erd\H{o}s-Renyi construction implied by $\bm{G}$ (recall Figure~\ref{fig:partial2}), introduce additional non-homogeneity across nodes that makes the estimation of the probability of the event $\mathcal{D}_{\ell m}$ in~\eqref{eq:Dsmallm}, and hence the estimation of the probability in~(\ref{eq:smallrare}), rather intricate.

In order to overcome this issue, Theorem~$3$ further ahead states that, without loss of generality, we can replace the events~$\mathcal{D}_{\ell m}$ in~(\ref{eq:Dsmallm}), by the events
\beq
\widetilde{\mathcal{D}}_{\ell,m}\dfz
\{\delta_{\ell,m}(\widetilde{\bm{G}})\leq r_N, \ell\in \mathcal{N}_i(\widetilde{\bm{G}}), m\in\mathcal{N}_j^{(2)}(\widetilde{\bm{G}})\},
\label{eq:DsmallmHC}
\eeq
where~$\widetilde{\bm{G}}\sim\mathscr{G}^{\star}(N,\widetilde{p}_N)$ is a standard Erd\H{o}s-R\'enyi graph with $\widetilde{p}_N=S p_N$ (for sufficiently large $N$), in that if we prove the convergence
\begin{equation}\label{eq:convhomogeneous}
\P[\widetilde{\mathcal{D}}_{\textnormal{small}}]\longrightarrow 0,
\end{equation}
then, the convergence in~\eqref{eq:smallrare} holds, where we have defined
\beq
\widetilde{\mathcal{D}}_{\textnormal{small}}\dfz\bigcup_{\ell,m\in \mathcal{S}'}\widetilde{\mathcal{D}}_{\ell,m}.
\label{eq:DsmalltotHC}
\eeq
This is accomplished, in the proof of Theorem~\ref{th:couplingpart}, via constructing a graph $\widetilde{G}$ that is Erd\H{o}s-R\'enyi and that is coupled with $G$ in the sense that
\begin{equation}
\mathcal{D}_{\ell,m} \subseteq \widetilde{\mathcal{D}}_{\ell,m}
\end{equation}
for all $\ell,m\in \mathcal{S}'$ and hence,
\beq
\mathcal{D}_{\textnormal{small}}
\subseteq
\widetilde{\mathcal{D}}_{\textnormal{small}}.
\label{eq:couplimplic}
\eeq
Therefore, the induced coupling yields:
\beq
\P[\mathcal{D}_{\textnormal{small}}]
\leq
\P[\widetilde{\mathcal{D}}_{\textnormal{small}}],
\label{eq:theorem4preview}
\eeq
implying that if one is able to prove that the probability for the homogeneous case vanishes, equation~\eqref{eq:convhomogeneous}, so does the probability for the original (non-homogeneous) case. We refer to this coupling procedure as homogenization as the conditions characterizing the event~$\widetilde{\mathcal{D}}_{\textnormal{small}}$ refer to the same graph~$\widetilde{\bm{G}}$ and, further the graph~$\widetilde{\bm{G}}$ is a standard Erd\H{o}s-R\'enyi. That is, the inhomogeneity that characterizes the events~$\mathcal{D}_{\textnormal{small}}$ is not present in the new event~$\widetilde{\mathcal{D}}_{\textnormal{small}}$.

We now state Theorem~\ref{th:couplingpart} and prove it in Appendix~\ref{sec:HC}.

\begin{theorem}[Coupling and homogenization]\label{th:couplingpart}
Let $\bm{G}\sim\mathscr{G}^{\star}(N,p_N;G_{\mathcal{S}})$ be a partial Erd\H{o}s-R\'enyi random graph, and let $\widetilde{\bm{G}}$ be a pure Erd\H{o}s-R\'enyi random graph $\widetilde{\bm{G}}\sim\mathscr{G}^{\star}(N,\widetilde{p}_N)$ where, for $N$ sufficiently large:
\beq
\widetilde{p}_N=S p_N=\frac{\log N + \widetilde{c}_N}{N}.
\eeq

If $i,j\in \mathcal{S}$ are non-interacting ($g_{ij}=0$), then we have that:
\beq
\boxed{
\P[\mathcal{D}_{\textnormal{small}}]
\leq
\P[\widetilde{\mathcal{D}}_{\textnormal{small}}]
}
\label{eq:TheoCouplingMain}
\eeq\qed
\end{theorem}

\begin{figure} [t]
\begin{center}
\includegraphics[scale= 0.7]{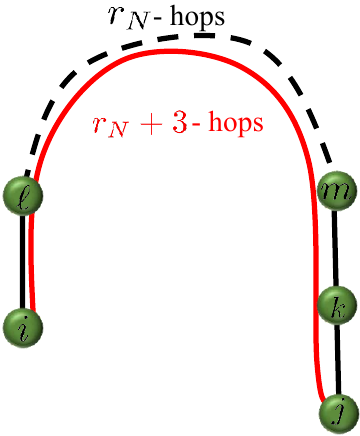}
\caption{If there is a path $\mathcal{P}_{\ell m}$ between $\ell$ and $m$ of $r_N$ hops (dashed path), then there is a path of length $r_N+3$ hops (red path) connecting $i$ to $j$: one can reach $\ell$ in one hop from $i$ (since $\ell \in \mathcal{N}_i(\widetilde{\bm{G}})$), follow the path $\mathcal{P}_{\ell m}$ to reach $m$ and reach $j$ in two more hops from $m$ (since $m \in \mathcal{N}^{(2)}_j(\widetilde{\bm{G}})$). Therefore, the distance between $i$ and $j$ cannot exceed $r_N+3$.}
\label{fig:boundistance}
\end{center}
\end{figure}

{\em Step~$4$: Managing the small-distance pairs}.
The final step to prove Theorem~$1$, in view of inequality~\eqref{eq:TheoCouplingMain}, consists in proving that~(\ref{eq:smallrare}) holds true on the homogenized graph $\widetilde{\bm{G}}\sim\mathscr{G}^{\star}(N,\widetilde{p}_N)$, namely, that:
\beq
\P[\widetilde{\mathcal{D}}_{\textnormal{small}}]\stackrel{N\rightarrow\infty}{\longrightarrow} 0.
\eeq
Using~(\ref{eq:DsmallmHC}) and~(\ref{eq:DsmalltotHC}), we observe that:
\begin{equation}\label{eq:UBprob2}
\widetilde{\mathcal{D}}_{\textnormal{small}}\subseteq \left\{\delta_{i,j}(\widetilde{\bm{G}})\leq r_N+3\right\},
\end{equation}
i.e., if the distance between $\ell,m$ of an active pair $\left(\ell,m\right)$ is bounded by $r_N$, then, since $\ell$ and $m$ are neighbor and second order neighbor of $i$ and $j$, respectively, the distance between the nodes $i$ and $j$ cannot exceed $r_N+3$. Refer to Figure~\ref{fig:boundistance} for an illustration.
Therefore,
\begin{equation}\label{eq:UBprob}
\mathbb{P}\left[\widetilde{\mathcal{D}}_{\textnormal{small}}\right] \leq \mathbb{P}\left[\delta_{i,j}(\widetilde{\bm{G}})\leq r_N+3\right],
\end{equation}
and the estimation of~$\mathbb{P}\left[\delta_{i,j}(\widetilde{\bm{G}})\leq r_N+3\right]$ amounts to a standard analysis of distance scaling on Erd\H{o}s-R\'enyi random graphs as the graph~$\widetilde{G}$ is a pure Erd\H{o}s-R\'enyi. In fact, Lemma~\ref{lemma:dist2new} (included in Appendix~\ref{app:smalldist} for completeness) asserts that
\begin{equation}\label{eq:upper}
\mathbb{P}\left[\delta_{i,j}(\widetilde{\bm{G}})\leq r_N+3\right] \leq \widetilde{p}_N \, (N \widetilde{p}_N)^{r_N+2}\left(\frac{1}{1-1/(N \widetilde{p}_N)}\right).
\end{equation}
Now, since by assumption $r_N\rightarrow\infty$ and in the Erd\H{o}s-R\'enyi regime that we are assuming we have $Np_N\rightarrow \infty$, as $N$ grows large, we conclude from~\eqref{eq:upper} that~$\mathbb{P}\left[\widetilde{\mathcal{D}}_{\textnormal{small}}\right]$ vanishes if we are able to choose a sequence $r_N$ yielding:
\beq
\boxed{
\widetilde{p}_N (N \widetilde{p}_N)^{r_N + 2} \stackrel{N\rightarrow\infty}{\longrightarrow} 0
}
\label{eq:rNnotoofast}
\eeq
Note that the requirement~\eqref{eq:rNnotooslow} implies that $r_N$ cannot diverge {\em too slow}, whereas the requirement in~\eqref{eq:rNnotoofast} implies that $r_N$ cannot diverge {\em too fast}. The next step illustrates how to choose a sequence~$r_N$, with $r_N\rightarrow \infty$, yielding both~\eqref{eq:rNnotooslow} and~\eqref{eq:rNnotoofast}.

\begin{remark}[More on homogenization]
Ultimately, the homogenization in Theorem~\ref{th:couplingpart} reduces the estimation of $\mathbb{P}\left[\mathcal{D}_{\textnormal{small}}\right]$ to a simple analysis of distance scaling between only one pair of nodes $\left(i,j\right)$ in a pure Erd\H{o}s-R\'enyi random graph, in view of the subset inequalities~\eqref{eq:couplimplic} and~\eqref{eq:UBprob2}. Note that, to prove the convergence~\eqref{eq:smallrare}, one may be tempted to directly apply the following inequality (instead of invoking the extra homogenization, inequality~\eqref{eq:couplimplic}, granted by Theorem~\ref{th:couplingpart})
\begin{equation}\label{eq:useless}
\mathcal{D}_{\textnormal{small}}\subseteq \left\{\delta_{i,j}(\bm{G})\leq r_N+3\right\},
\end{equation}
for the original heterogeneous event~$\mathcal{D}_{\textnormal{small}}$ and with $\bm{G}$ in the RHS instead of the pure Erd\H{o}s-R\'enyi $\widetilde{\bm{G}}$. But the probability~$\mathbb{P}\left[\left\{\delta_{i,j}(\bm{G})\leq r_N+3\right\}\right]$ does not converge to zero as $N$ grows large in this case as the distance~$\delta_{i,j}(\bm{G})$ depends on $G_{\mathcal{S}}$ which is arbitrary. Therefore, the inequality~\eqref{eq:useless} is not useful to establish the convergence~\eqref{eq:smallrare}. In this line of thought, one can attempt to find a reference graph $\bm{G}_{\textnormal{ref}}$, if any, so that
\begin{equation}\label{eq:useless2}
\mathcal{D}_{\textnormal{small}}\subseteq \left\{\delta_{i,j}(\bm{G}_{\textnormal{ref}})\leq r_N+3\right\},
\end{equation}
and for which $\mathbb{P}\left[\left\{\delta_{i,j}(\bm{G}_{\textnormal{ref}})\leq r_N+3\right\}\right]$ converges to zero. Another natural candidate is $\bm{G}_{\textnormal{ref}}:=\bm{G}_{\mathcal{S}\nleftrightarrow \mathcal{S}}$, but the referred inequality does not hold in this case, i.e.,
\begin{equation}
\mathcal{D}_{\textnormal{small}}\nsubseteq \left\{\delta_{i,j}(\bm{G}_{\mathcal{S}\nleftrightarrow \mathcal{S}})\leq r_N+3\right\}.
\end{equation}
Refer to Figure~\ref{fig:counterex} for a graphical counter-example on this.
\begin{figure} [t]
\begin{center}
\includegraphics[scale= 0.45]{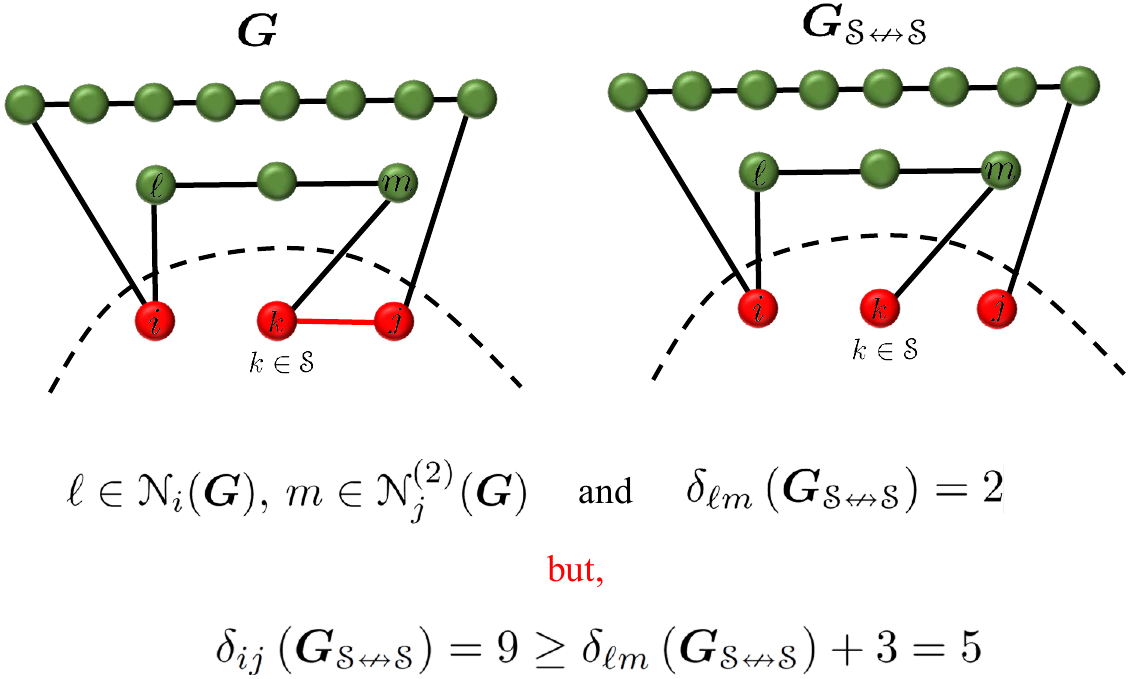}
\caption{Counter-example: the inequality~\eqref{eq:useless2} does not hold for $\bm{G}_{\textnormal{ref}}:=\bm{G}_{\mathcal{S}\nleftrightarrow \mathcal{S}}$.}
\label{fig:counterex}
\end{center}
\end{figure}
One can attempt to simply fill in the gap of $\bm{G}_{\mathcal{S}\nleftrightarrow \mathcal{S}}$ by considering the reference graph
\begin{equation}
\bm{G}_{\textnormal{ref}} := \widetilde{\bm{G}}_{\mathcal{S}} \oplus \bm{G}_{\mathcal{S}\nleftrightarrow \mathcal{S}}
\end{equation}
where~$\widetilde{\bm{G}}_{\mathcal{S}}\sim\mathscr{G}^{\star}(S,\widetilde{p}_N)$, but this also does not work as the inequality~\eqref{eq:useless2} is not satisfied (for the same reason as $\bm{G}_{\mathcal{S}\nleftrightarrow \mathcal{S}}$ fails to meet it).

The homogenization in Theorem~\ref{th:couplingpart} provides a careful construction of a reference graph that allows us to lip this difficulty and rigorously reduce the computation of $\mathbb{P}\left[\mathcal{D}_{\textnormal{small}}\right]$ to a simple distance-scaling of a particular pair of nodes in an Erd\H{o}s-R\'enyi random graph. Such construction may be of independent interest.
\end{remark}

{\em Step~$5$: Choosing the sequence $r_N$}.
In the above steps (specifically, in Step~$1$ and Step~$4$), we have maintained that a certain sequence of distances, $r_N$, exists that fulfills the two conditions in~(\ref{eq:rNnotooslow}) and~(\ref{eq:rNnotoofast}).
We start by examining these conditions in more detail.
Taking the logarithm of the functions appearing in~(\ref{eq:rNnotooslow}), we get:
\beqa
\log\left(N p_N \rho^{r_N+4}\right) & = \!\!\! & \!\!\! \log(\log N + c_N) + (r_N+4)\log(\rho)\nonumber\\
& \stackrel{N\rightarrow \infty}{\longrightarrow} & -\infty,
\label{eq:logrNnotooslow}
\eeqa
where we have used the expression for $p_N$ in~(\ref{eq:pconncond}).
Observing that $\log(\rho)<0$ since $\rho<1$, and letting:
\beq
\alpha\dfz |\log(\rho)|, \qquad \omega_N\dfz\log(\log N + c_N),
\label{eq:constdefin}
\eeq
from~(\ref{eq:logrNnotooslow}) we can write:
\beq
\boxed{
\omega_N - \alpha \, r_N - 4\alpha\stackrel{N\rightarrow \infty}{\longrightarrow} -\infty
}
\label{eq:logrNnotooslow2}
\eeq
Let us switch to the analysis of~(\ref{eq:rNnotoofast}). Taking the logarithm of the functions appearing in~(\ref{eq:rNnotoofast}) we can write:
\beqa
\log\left(\widetilde{p}_N (N \widetilde{p}_N)^{r_N + 2}\right) & = & \log\left(\frac{(N \widetilde{p}_N)^{r_N + 3}}{N}\right)
\\
& = &
(r_N + 3)\log(N \widetilde{p}_N) - \log N \nonumber\\
& \stackrel{N\rightarrow \infty}{\longrightarrow} & -\infty,\nonumber
\label{eq:logrNnotoofast}
\eeqa
which, since $\widetilde{p}_N=S p_N$, in view of~(\ref{eq:pconncond}) and~(\ref{eq:constdefin}), yields:
\beq
\boxed{
(r_N + 3)[\log S + \omega_N] - \log N,
\stackrel{N\rightarrow \infty}{\longrightarrow} -\infty
}
\label{eq:logrNnotoofast2}
\eeq
We first show why the assumption in~(\ref{eq:cnconststatement}) is related to~(\ref{eq:logrNnotooslow2}) and~(\ref{eq:logrNnotoofast2}).
From~(\ref{eq:logrNnotooslow2}) and~(\ref{eq:logrNnotoofast2}) we conclude that, for sufficiently large $N$, we must necessarily have:
\beq
\frac{\omega_N}{\alpha}<r_N<\frac{\log N}{\omega_N},
\label{eq:rNfirsteq}
\eeq
which in turn implies:
\beq
\frac{\omega_N^2}{\log N}<\alpha
\Leftrightarrow
\frac{[\log(\log N + c_N)]^2}{\log N}<|\log(\rho)|,
\label{eq:necessityond}
\eeq
where in the last step we used the definitions in~(\ref{eq:constdefin}).
Now, if we want to guarantee the verification of~(\ref{eq:necessityond}) irrespectively of the particular value of $0 < \rho < 1$, we need to enforce condition~(\ref{eq:cnconststatement}).

Next we illustrate how to choose a sequence $r_N$ that, under assumption~(\ref{eq:cnconststatement}), fulfills simultaneously~(\ref{eq:logrNnotooslow2}) and~(\ref{eq:logrNnotoofast2}). We set:
\beq
r_N=\left\lfloor
\frac 1 2 \frac{\log N}{\omega_N}
\right\rfloor.
\label{eq:candidaterN}
\eeq
Substituting~(\ref{eq:candidaterN}) into~(\ref{eq:logrNnotooslow2}), and observing that $\lfloor x \rfloor > x-1$, where $\lfloor x \rfloor$ stands for the greatest integer smaller than or equal to $x$, we can write:
\beqa
\omega_N - \alpha \,
\left\lfloor
\frac 1 2 \frac{\log N}{\omega_N}
\right\rfloor
- 4\alpha
&< \!\!\!\!\! & \!\!\!\!\!\!
\omega_N -  \frac \alpha 2 \frac{\log N}{\omega_N}
+\alpha- 4\alpha
\nonumber\\
&= \!\!\!\!\! & \!\!\!\!\!\!
\omega_N
\left(
1 -
\underbrace{\frac \alpha 2 \frac{\log N}{\omega_N^2}}_{\rightarrow\infty \textnormal{ from~(\ref{eq:cnconststatement})}}
-\frac{3\alpha}{\omega_N}
\right)\nonumber\\
& \stackrel{N\rightarrow \infty}{\longrightarrow} & -\infty,
\eeqa
which shows that the condition in~(\ref{eq:logrNnotooslow2}) is met with the choice in~(\ref{eq:candidaterN}).

Likewise, substituting~(\ref{eq:candidaterN}) into~(\ref{eq:logrNnotoofast2}), and observing that $\lfloor x \rfloor \leq x$, we have:
\beqa
\lefteqn{
\left(
\left\lfloor
\frac 1 2 \frac{\log N}{\omega_N}\right\rfloor
+ 3
\right)
[\log S + \omega_N] - \log N
}\nonumber\\
&\leq&
\left(
\frac 1 2 \frac{\log N}{\omega_N} + 3
\right)[\log S + \omega_N] - \log N
\nonumber\\
&=&
\frac{\log S} {2} \frac{\log N}{\omega_N} + 3\log S
+\frac{1} {2} \log N + 3\,\omega_N
 - \log N
\nonumber\\
&=&
\frac{\log N}{\omega_N}
\left[
\frac{\log S} {2}  + 3\log S\frac{\omega_N}{\log N}-
\frac{\omega_N} {2} + 3\frac{\omega_N^2}{\log N}
\right]\nonumber\\
& \stackrel{N\rightarrow \infty}{\longrightarrow} & -\infty,
\eeqa
with the convergence holding true because $\omega_N\rightarrow\infty$ as $N\rightarrow\infty$, while $\omega_N^2/\log N$ and $\omega_N/\log N$ vanish in view of~(\ref{eq:cnconststatement}). We have in fact shown that the condition in~(\ref{eq:logrNnotoofast2}) is met with the choice in~(\ref{eq:candidaterN}).

\begin{figure*} [t]
\begin{center}
\includegraphics[scale= 0.8]{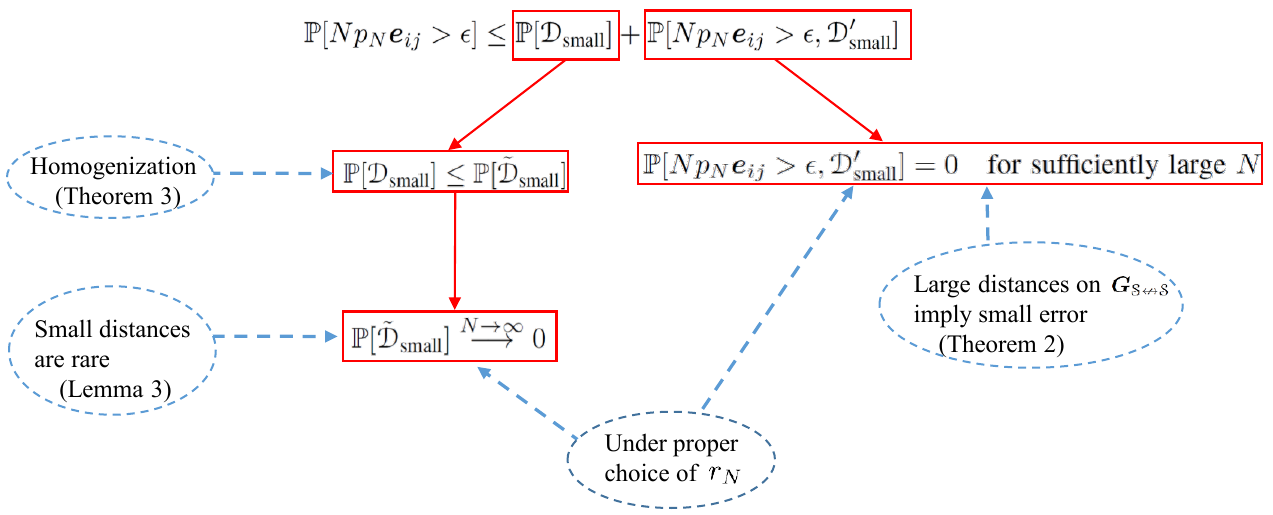}
\caption{Diagram-summary of the proof.}
\label{fig:diagramsummary}
\end{center}
\end{figure*}

Refer to Figure~\ref{fig:diagramsummary} for a summary of the proof of Theorem~$1$.

\section{Proof of Theorem~\ref{lem:distancia}}\label{sec:proof}
First, we start by observing that the matrix $H$ in~(\ref{eq:BHmats}) is not sensitive to the submatrix $A_{\mathcal{S}}$.
Note that this is not immediately clear via simple inspection, since computation of $H$ involves the matrix $B=A^2$.
However, note that
\begin{equation}
A=\left[\begin{array}{cc} A_{\mathcal{S}} & A_{\mathcal{S}\mathcal{S}'}\\ A_{\mathcal{S}'\mathcal{S}} & A_{\mathcal{S}'}\end{array}\right]
\end{equation}
and from the rules for block-matrix multiplication, we can write:
\beq
\boxed{
B_{\mathcal{S}'} = [A^2]_{\mathcal{S}'} = A_{\mathcal{S}'\mathcal{S}}A_{\mathcal{S}\mathcal{S}'} + (A_{\mathcal{S}'})^2
}
\label{eq:magicprop}
\eeq
which highlights that the matrix $B_{\mathcal{S}'}$ does {\em not} depend on the submatrix $A_{\mathcal{S}}$.
As a corollary to this observation, it follows that the matrix $H$ in~(\ref{eq:BHmats})
is not a function of the particular submatrix $A_{\mathcal{S}}$, and, hence, it is insensitive to the particular topology of the subnetwork connecting the observed agents. Since we are to devise bounds for the terms $h_{\ell m}$,
we can assume without loss of generality that all the entries of $A_{\mathcal{S}}$ are equal to zero\footnote{Assuming that the diagonal entries are equal to zero contradicts our rules for constructing a combination policy. However, this is immaterial, because setting the $A_{\mathcal{S}}$ block to zero is only a mathematical expedient to compute suitable bounds, without any physical meaning.}, namely, that $A_{\mathcal{S}} =0_{S \times S}$.
In other words, we can remove the edges among the observable agents as far as computing bounds on $H$ goes.
This will imply
that an appropriate distance between nodes in $\mathcal{S}'$ (which will play a role in bounding $h_{\ell m}$) is given by $\delta_{\ell,m}(G_{\mathcal{S}\nleftrightarrow \mathcal{S}})$, namely, by the distance between $\ell$ and $m$ on the graph $G_{\mathcal{S}\nleftrightarrow \mathcal{S}}$ where the edges among the observed agents in $\mathcal{S}$ have been removed.


Theorem~\ref{lem:distancia}, proved next, establishes an exponential bound on $h_{\ell m}$, which holds if $\ell$ and $m$ are distant on $G_{\mathcal{S}\nleftrightarrow \mathcal{S}}$ (not necessarily on $G$, and hence not dependent on $G_{\mathcal{S}}$).




\begin{IEEEproof}[Proof of Theorem~\ref{lem:distancia}]
We remind that:
\beq
H=(I_{\mathcal{S}'}-B_{\mathcal{S}'})^{-1}=\sum_{k=0}^{\infty} (B_{\mathcal{S}'})^{k},
\label{eq:hb}
\eeq
since~$A$ is row-sum stable and~$B=A^2$ (refer, e.g., to~\cite{tomo,Sayed}).
Let now $\widetilde{A}$ be the matrix obtained from $A$ by replacing the submatrix $A_{\mathcal{S}}$ with the void matrix, $0_{S\times S}$, and let accordingly $\widetilde{B}=\widetilde{A}^2$.
Since, in view of~(\ref{eq:magicprop}), modifying the submatrix $A_{\mathcal{S}}$ does not alter the submatrix $B_{\mathcal{S}'}$, we can safely write:
\beq
\boxed{
B_{\mathcal{S}'}=\widetilde{B}_{\mathcal{S}'}
}
\label{eq:Btilde}
\eeq
Moreover, it is known that, for any two nonnegative matrices $Q$ and $R$
\beq
[QR]_{\mathcal{S}'}=
\underbrace{Q_{\mathcal{S}'\mathcal{S}}R_{\mathcal{S}\mathcal{S}'}}_{\geq 0} + Q_{\mathcal{S}'} R_{\mathcal{S}'}\geq Q_{\mathcal{S}'} R_{\mathcal{S}'},
\label{eq:ide}
\eeq
with entry-wise inequality. Taking $Q=R=\widetilde{B}$, and reasoning by induction, we have then:
\beq
(B_{\mathcal{S}'})^n=(\widetilde{B}_{\mathcal{S}'})^n \leq [\widetilde{B}^n]_{\mathcal{S}'}=[\widetilde{A}^{2 n}]_{\mathcal{S}'},
\label{eq:ineq}
\eeq
where the first equality follows from~(\ref{eq:Btilde}).
Rephrasing~(\ref{eq:ineq}) on an entry-wise basis, we get, for all $\ell, m\in \mathcal{S}'$:
\beq
[(B_{\mathcal{S}'})^n]_{\ell m} \leq [\widetilde{A}^{2n}]_{\ell m}.
\label{eq:entrywiseineq}
\eeq
We recall that the support graph of $\widetilde{A}$ is given by $G_{\mathcal{S}\nleftrightarrow \mathcal{S}}$.
Since by assumption $\delta_{\ell,m}(G_{\mathcal{S}\nleftrightarrow \mathcal{S}})=r$, then in view of~\eqref{eq:elementary}, the smallest $k$ yielding $[\widetilde{A}^{k}]_{\ell m}>0$ is $k=r$. In view of~(\ref{eq:entrywiseineq}), this property implies that one can consider only the terms $[(B_{\mathcal{S}'})^n]_{\ell m}$ for $2 n\geq r$.
Since $r$ is not necessarily an even number, we could in general consider all the terms for which $n\geq \lceil r/2\rceil$, where $\lceil x\rceil$ stands for the smallest integer that is greater than or equal to $x$.
With this choice, the series in~(\ref{eq:hb}) can be truncated as (the term $n=0$ is zero because $\ell\neq m$):
\beqa
h_{\ell m}
&=&
\sum_{n=1}^{\infty}
\left[
\left(B_{\mathcal{S}'}\right)^n
\right]_{\ell m}
=
\sum_{n \geq \lceil r/2 \rceil}^{\infty}
\left[\left(B_{\mathcal{S}'}\right)^n\right]_{\ell m}\nonumber\\
& \stackrel{(a)}{\leq} &
\sum_{n \geq \lceil r/2 \rceil}^{\infty}
[A^{2n}]_{\ell m}
\nonumber\\
&\stackrel{(b)}{\leq}&
\sum_{n \geq \lceil r/2 \rceil}^{\infty}
\rho^{2n}
\stackrel{(c)}{=}
\frac{\rho^{2 \lceil r/2\rceil}}{1-\rho^2}
\stackrel{(d)}{\leq}
\frac{\rho^{r}}{1-\rho^2},
\label{eq:specdecMAT}
\eeqa
where inequality $(a)$ follows by using~(\ref{eq:entrywiseineq}) with $A$ in place of $\widetilde{A}$; inequality $(b)$ follows due to~(\ref{eq:prop1stab}) as we know that $\sum_{\ell=1}^N[A^{2n}]_{\ell m}\leq\rho^{2n}$; equality $(c)$ is the partial sum of the geometric series; and inequality $(d)$ follows from the known bound $\lceil r/2\rceil \geq r/2$.
\end{IEEEproof}



\section{Homogenization}
\label{sec:HC}

In this section, we prove that if small distances are rare over a pure (i.e., standard) Erd\H{o}s-R\'enyi graph $\widetilde{G}$, i.e.,
\beq\label{eq:conv1}
\P[\widetilde{\mathcal{D}}_{\textnormal{small}}]\stackrel{N\rightarrow\infty}{\longrightarrow} 0,
\eeq
then small distances are also rare over the {\em partial} Erd\H{o}s-R\'enyi~$\bm{G}$, i.e.,
\beq\label{eq:conv2}
\P[\mathcal{D}_{\textnormal{small}}]\stackrel{N\rightarrow\infty}{\longrightarrow} 0,
\eeq
where we recall the definitions
\beq
\widetilde{\mathcal{D}}_{\textnormal{small}}\dfz\bigcup_{\ell,m\in \mathcal{S}'}\widetilde{\mathcal{D}}_{\ell,m},\qquad \mathcal{D}_{\textnormal{small}}\dfz\bigcup_{\ell,m\in \mathcal{S}'}\mathcal{D}_{\ell,m},
\label{eq:DsmalltotHC2}
\eeq
with
\beqa
\widetilde{\mathcal{D}}_{\ell,m}&\dfz&
\{\delta_{\ell,m}(\widetilde{\bm{G}})\leq r_N, \ell\in \mathcal{N}_i(\widetilde{\bm{G}}), m\in\mathcal{N}_j^{(2)}(\widetilde{\bm{G}})\},\nonumber\\
{\mathcal{D}}_{\ell,m}&\dfz&
\{\delta_{\ell,m}(\bm{G}_{\mathcal{S}\nleftrightarrow \mathcal{S}})\leq r_N, \ell\in \mathcal{N}_i(\bm{G}), m\in\mathcal{N}_j^{(2)}(\bm{G})\}.\nonumber\\
\label{eq:DsmallmHC2}
\eeqa

This is a relevant assertion as estimating the probability~$\P[\mathcal{D}_{\textnormal{small}}]$ is rather intricate. In fact, by examining the events~$\mathcal{D}_{\ell m}$ in~\eqref{eq:DsmallmHC2}, two sources of asymmetry stick out (as opposed to the characterization of~$\widetilde{\mathcal{D}}_{\ell m}$). First, the pertinent distance,~$\delta_{\ell,m}(\bm{G}_{\mathcal{S}\nleftrightarrow \mathcal{S}})$, is computed with respect to the graph $\bm{G}_{\mathcal{S}\nleftrightarrow \mathcal{S}}$, while the conditions on the neighborhood memberships $\ell\in\mathcal{N}_i(\bm{G})$ and $m\in\mathcal{N}_j^{(2)}(\bm{G})$ characterizing the active pairs are defined in terms of the original graph, $\bm{G}$. Second, both graphs $\bm{G}$ and $\bm{G}_{\mathcal{S}\nleftrightarrow \mathcal{S}}$ are non-homogeneous. That is, the connections among nodes in $\mathcal{S}$ for the graph $\bm{G}$ are given by $G_{\mathcal{S}}$ -- whose topology is arbitrary and hence, it has a nature that differs from the rest of the network $\bm{G}$ -- while in $\bm{G}_{\mathcal{S}\nleftrightarrow \mathcal{S}}$ the connections among nodes in $\mathcal{S}$ are absent.


Therefore, in order to estimate~$\P[\mathcal{D}_{\ell m}]$, and hence~$\P[\mathcal{D}_{\textnormal{small}}]$, in the proof of Theorem~\ref{th:couplingpart}, we appropriately modify the structure of the partial Erd\H{o}s-R\'enyi $\bm{G}$ in such a way that the resulting transformed graph $\widetilde{\bm{G}}$ fulfills the following properties: $i)$ $\widetilde{\bm{G}}$ is a homogeneous (i.e., classic) Erd\H{o}s-R\'enyi graph; $ii)$ the original event $\mathcal{D}_{\ell,m}$ on $\bm{G}$ implies its counterpart $\widetilde{\mathcal{D}}_{\ell,m}$ defined on the new graph~$\widetilde{\bm{G}}$, i.e.,
\begin{equation}\label{eq:subsetcouple}
\mathcal{D}_{\ell,m}\subseteq \widetilde{\mathcal{D}}_{\ell,m}
\end{equation}
for all $\ell,m\in \mathcal{S}'$, and hence,
\begin{equation}\label{eq:subinclu}
\mathcal{D}_{\textnormal{small}}\subseteq \widetilde{\mathcal{D}}_{\textnormal{small}}.
\end{equation}
This further yields the desired inequality
\beq
\P[\mathcal{D}_{\textnormal{small}}]
\leq
\P[\widetilde{\mathcal{D}}_{\textnormal{small}}].
\label{eq:theorem4preview3}
\eeq
As a result, showing the convergence in~\eqref{eq:conv1} for the homogeneous system implies the convergence in~\eqref{eq:conv2} for the original heterogeneous partial Erd\H{o}s-R\'enyi. We refer to this {\em coupling} procedure simply as homogenization.

\begin{figure*} [t]
\begin{center}
\includegraphics[scale= 0.65]{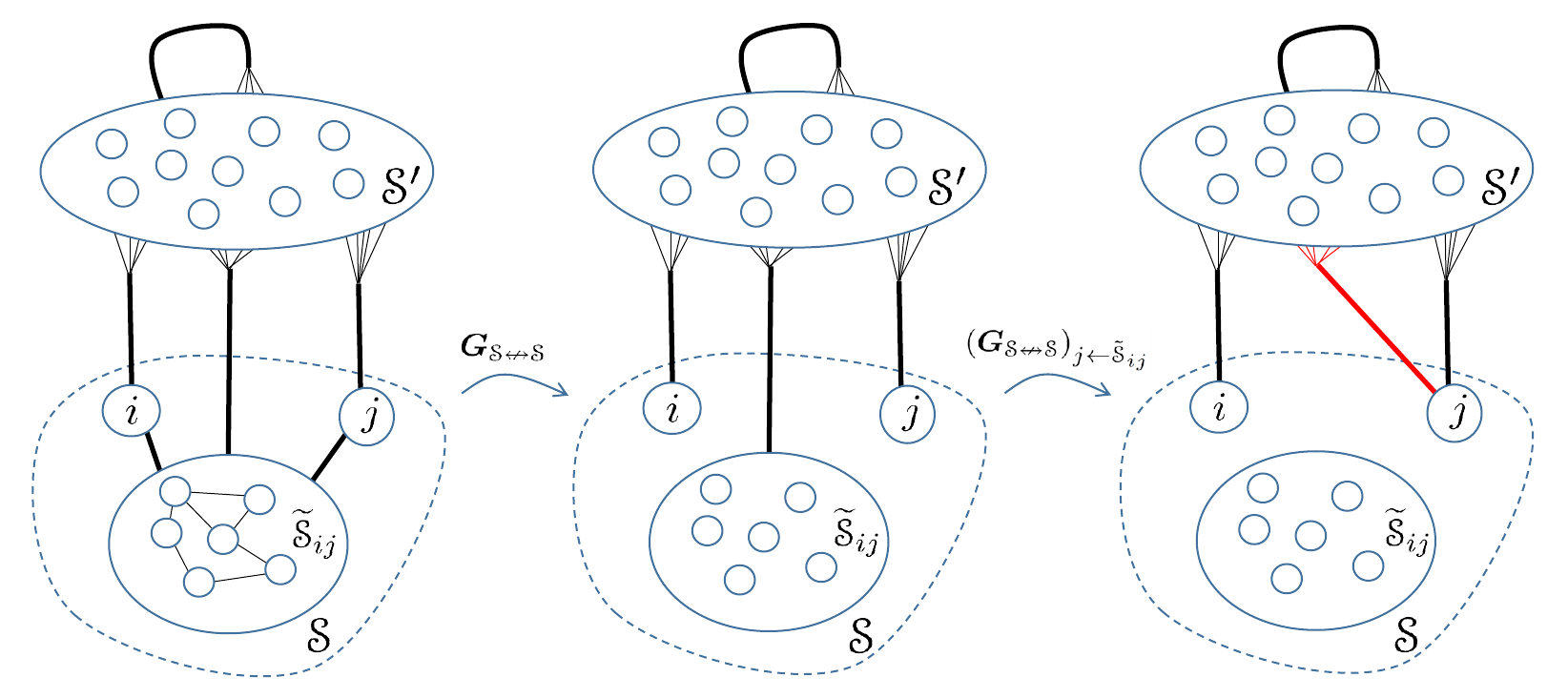}
\caption{The original graph $\bm{G}$ (leftmost panel); the locally-disconnected graph, $\bm{G}_{\mathcal{S}\nleftrightarrow \mathcal{S}}$ (middle panel); the transformed graph, $\overline{\bm{G}}\dfz(\bm{G}_{\mathcal{S}\nleftrightarrow \mathcal{S}})_{j\leftarrow\widetilde{\mathcal{S}}_{ij}}$, obtained from $\bm{G}_{\mathcal{S}\nleftrightarrow \mathcal{S}}$ when all the connections from $\widetilde{\mathcal{S}}_{ij}$ to $\mathcal{S}'$ are inherited by the node $j$.}
\label{fig:sequence_operations}
\end{center}
\end{figure*}

\begin{remark}
It is tempting at first glance to simply replace the graph~$\bm{G}_{\mathcal{S}\nleftrightarrow \mathcal{S}}$ in~\eqref{eq:DsmallmHC2} with the standard Erd\H{o}s-R\'enyi~$\widetilde{\bm{G}}$ by reinforcing the coupling $\widetilde{\bm{G}}_{\mathcal{S}'}=\bm{G}_{\mathcal{S}'}$. This, in fact, \emph{contracts} the distance, i.e., $\delta_{\ell m}(\widetilde{\bm{G}})\leq \delta_{\ell m}(\bm{G}_{\mathcal{S}\nleftrightarrow \mathcal{S}})$ which yields the implication
\begin{equation}
\delta_{\ell m}(\bm{G}_{\mathcal{S}\nleftrightarrow \mathcal{S}})\leq r_N \Rightarrow \delta_{\ell m}(\widetilde{\bm{G}})\leq r_N
\end{equation}
at the same time as the new graph $\widetilde{\bm{G}}$ is homogeneous, but the neighborhood constraint~$m\in\mathcal{N}_j^{(2)}(\widetilde{\bm{G}})$ is jeopardized as it is not implied by its counterpart on $\bm{G}$, since the condition ~$m\in\mathcal{N}_j^{(2)}(\bm{G})$ also depends on $G_{\mathcal{S}}$, which is not Erd\H{o}s-R\'enyi, but arbitrary (hence, the subset inclusion~\eqref{eq:subinclu} does not follow from this simple homogenization). The structure modification on the original graph $\bm{G}$ is carefully performed in the proof of Theorem~\ref{th:couplingpart} to both grant the contraction of the distances at the same time as preserving the neighborhood constraints.
\end{remark}

\begin{figure} [t]
\begin{center}
\includegraphics[scale= 0.4]{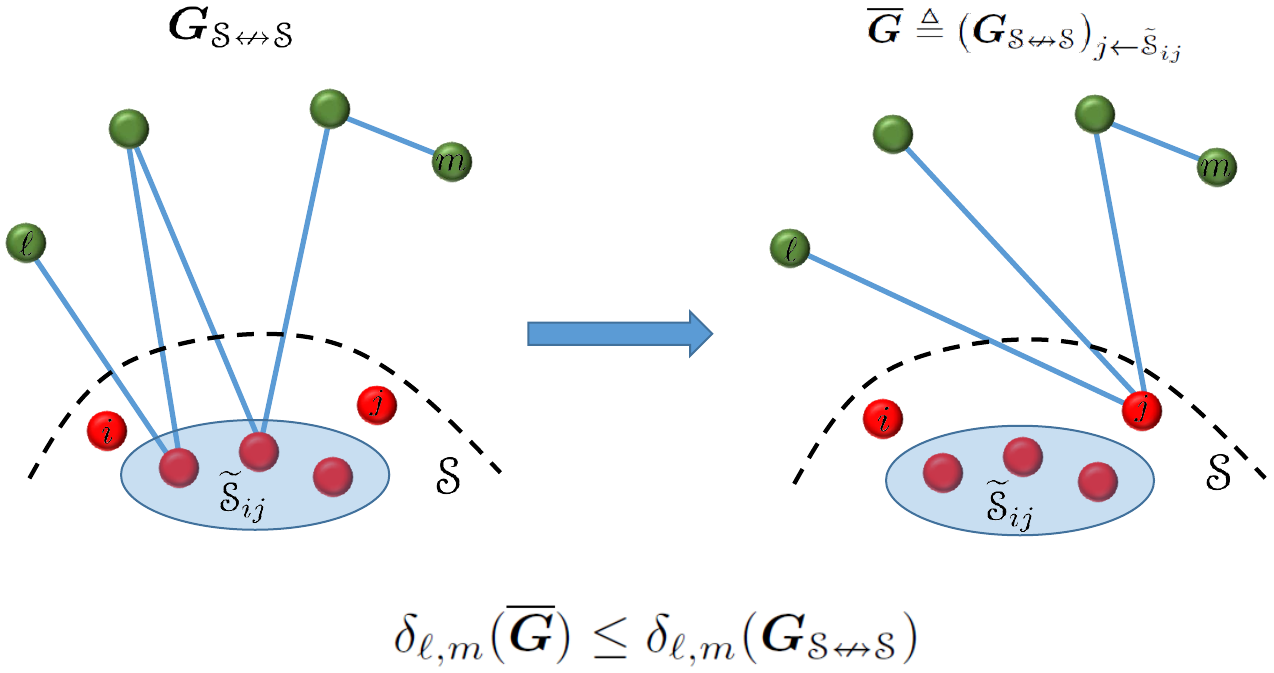}
\caption{Illustration of a particular path (blue color) connecting $\ell$ to $m$ in the graph $\bm{G}_{\mathcal{S}\nleftrightarrow \mathcal{S}}$ on the LHS of the figure. The edges linking to nodes in $\widetilde{\mathcal{S}}_{ij}$ on the graph $\bm{G}_{\mathcal{S}\nleftrightarrow \mathcal{S}}$, link only to $j$ on the graph $\overline{G}$. A path that crosses $M\geq 1$ nodes in $\widetilde{\mathcal{S}}_{ij}$ in the graph $\bm{G}_{\mathcal{S}\nleftrightarrow \mathcal{S}}$, only crosses $j$ in the new graph $\overline{G}$.}
\label{fig:quotientout}
\end{center}
\end{figure}

\begin{IEEEproof}[Proof of Theorem~\ref{th:couplingpart}]
In Figure~\ref{fig:sequence_operations}, middle panel, we display the graph $\bm{G}_{\mathcal{S}\nleftrightarrow \mathcal{S}}$.
Moreover, we denote by $\widetilde{\mathcal{S}}_{ij}$ the set $\mathcal{S}$ deprived of the nodes $i$ and $j$, namely,  $\widetilde{\mathcal{S}}_{ij}\dfz \mathcal{S}\setminus \left\{i,j\right\}$.
The basic trick that allows homogenization is defining a new graph where {\em all the connections from $\widetilde{\mathcal{S}}_{ij}$ to $\mathcal{S}'$} are inherited by the node $j$. The transformed graph is denoted by (and is displayed in the rightmost panel of Figure~\ref{fig:sequence_operations}):
\beq
\boxed{
\overline{\bm{G}}\dfz(\bm{G}_{\mathcal{S}\nleftrightarrow \mathcal{S}})_{j\leftarrow\widetilde{\mathcal{S}}_{ij}}
}
\label{eq:doubletransform}
\eeq
This operation achieves the twofold goal of ensuring that $i)$ the distance $\delta_{\ell,m}(G_{\mathcal{S}\nleftrightarrow \mathcal{S}})$ between any two nodes $\ell$ and $m$ in $\mathcal{S}'$ is reduced, namely,
\beq
\boxed{
\delta_{\ell,m}(\overline{\bm{G}})
\leq
\delta_{\ell,m}(\bm{G}_{\mathcal{S}\nleftrightarrow \mathcal{S}})
}
\label{eq:distimplic2}
\eeq
and $ii)$ if node $m$ is second-order neighbor of $j$ on the {\em original} graph $\bm{G}$, so is on the transformed graph, namely,
\beq
\boxed{
m\in\mathcal{N}^{(2)}_j(\bm{G})\Rightarrow
m\in\mathcal{N}^{(2)}_j(\overline{\bm{G}})
}
\label{eq:N2implic2}
\eeq
Note that equations~\eqref{eq:distimplic2} and~\eqref{eq:N2implic2} induce the desired coupling between the transformed graph, $\overline{\bm{G}}$, and the graphs $\bm{G}_{\mathcal{S}\nleftrightarrow \mathcal{S}}$, $\bm{G}$, in that for all~$\ell, m\in \mathcal{S}'$:
\beqa
& \left\{
\delta_{\ell,m}(\bm{G}_{\mathcal{S}\nleftrightarrow \mathcal{S}})\leq r_N, \ell\in \mathcal{N}_i(\bm{G}), m\in\mathcal{N}_j^{(2)}(\bm{G})
\right\} &\nonumber\\
& \subseteq &\nonumber\\
& \left\{
\delta_{\ell,m}(\overline{\bm{G}})\leq r_N, \ell\in \mathcal{N}_i(\overline{\bm{G}}), m\in\mathcal{N}_j^{(2)}(\overline{\bm{G}})
\right\}. &
\label{eq:overlineGimplic}
\eeqa
At this point, we observe that~$\overline{\bm{G}}$ is still not homogeneous (in particular, the nodes in $\widetilde{\mathcal{S}}_{ij}$ on the graph $\overline{G}$ are isolated) and hence, the proof is not finished. Before proceeding on this point, we first justify equations~(\ref{eq:distimplic2}) and~(\ref{eq:N2implic2}).

The inequality~(\ref{eq:distimplic2}) stems from the following observation.
The only modification in $\bm{G}_{\mathcal{S}\nleftrightarrow \mathcal{S}}$ to get $\overline{\bm{G}}$ is related to $\widetilde{\mathcal{S}}_{ij}$.
Therefore, if there exists a path from $\ell\in \mathcal{S}'$ to $m\in \mathcal{S}'$ on $\bm{G}_{\mathcal{S}\nleftrightarrow \mathcal{S}}$, which flows through $\widetilde{\mathcal{S}}_{ij}$, such path (or a shortened version thereof) is also present in $\overline{\bm{G}}$, but now via $j$. Refer to Figure~\ref{fig:quotientout} for an illustration.
\end{IEEEproof}
Indeed, each path on $\bm{G}_{\mathcal{S}\nleftrightarrow \mathcal{S}}$ hopping across $\widetilde{\mathcal{S}}_{ij}$ is mapped into a path traversing node $j$ (instead of traversing the corresponding nodes in $\widetilde{\mathcal{S}}_{ij}$), since the node $j$ has inherited all connections between $\widetilde{\mathcal{S}}_{ij}$ and $\mathcal{S}'$.

The neighborhood implication~(\ref{eq:N2implic2}) results from the following observation.
If on the graph $\bm{G}$, the node $m$ is connected to $j$ through an intermediate node belonging to $\widetilde{\mathcal{S}}_{ij}$, then it is connected to $j$ in one step on the graph $\overline{\bm{G}}$.
One difficulty might arise if, on graph $\bm{G}$, node $m$ is connected to $j$ through node $i$, because on $\overline{\bm{G}}$ nodes $i$ and $j$ are disconnected. This is not a problem, however, because to prove our result we need to examine only the case that $i$ and $j$ are disconnected on the original graph $\bm{G}$ (as stated in the theorem).
We remark that, for the case that $\ell=m$, condition~(\ref{eq:distimplic2}) is redundant and
\beq
\left\{
\ell\in \mathcal{N}_i(\bm{G})\cap \mathcal{N}_j^{(2)}(\bm{G})
\right\}
\subseteq
\left\{
\ell\in \mathcal{N}_i(\overline{\bm{G}}) \cap \mathcal{N}_j^{(2)}(\overline{\bm{G}})
\right\}.
\label{eq:overlineGimplic2}
\eeq

\begin{figure} [t]
\begin{center}
\includegraphics[scale= 0.75]{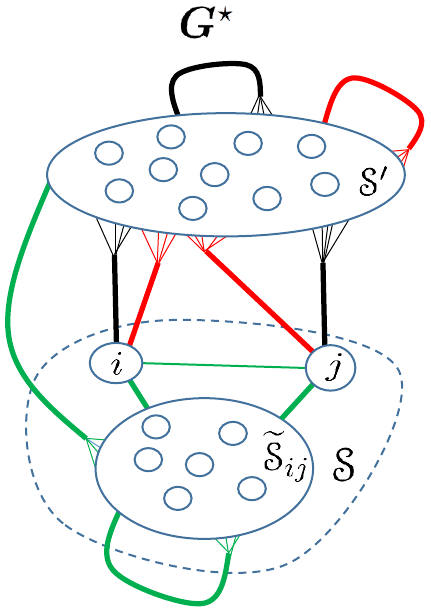}
\caption{Homogenizing the graph~$\overline{G}$. In the graph~$\overline{G}$, the node $j$ is connected to a particular node in $\mathcal{S}'$ with probability~$\overline{p}_{N}=1-(1-p_N)^{S-1}$ whereas, e.g., the nodes in $\widetilde{\mathcal{S}}_{ij}$ are not connected among each other. New Bernoulli realizations are performed so that the probability of any pair of nodes $(u,v)\in \mathcal{S}\times \mathcal{S}$ being connected is $p^{\star}_{N}=1-(1-p_N)^{S-1}$ in $\bm{G}^{\star}$. This is formalized in equations~\eqref{eq:bernoullihomog2} and~\eqref{eq:bernoullihomog}. The final graph $\bm{G}^{\star}$ is Erd\H{o}s-R\'enyi.}
\label{fig:homogenization}
\end{center}
\end{figure}

Now, we return to the observation that the transformed graph, $\overline{\bm{G}}$, is still asymmetrical, because, apart from the fact that the set $\mathcal{S}$ contains disconnected nodes, the probability that the node $j$ is connected to a node in $\mathcal{S}'$ is now augmented as~$j$ inherited all the connections from $\widetilde{\mathcal{S}}_{ij}$ to $\mathcal{S}'$.
Since under the partial Erd\H{o}s-R\'enyi construction, these connections follow a standard Bernoulli law, we conclude that the probability of $j$ being connected to a particular node in $\mathcal{S}'$, in the new random graph $\overline{G}$, is simply given by (recall that $\widetilde{\mathcal{S}}_{ij}$ does not contain node $i$):
\beq
1-(1-p_N)^{S-1}.
\label{eq:superjconn}
\eeq
To see why, $j$ is {\em not connected} to a particular node in $\mathcal{S}'$, say $k\in \mathcal{S}'$, in the graph $\overline{G}$, if and only if, $j$ is not connected to $k$ in $G$, i.e., $g_{jk}=0$, and $g_{\alpha k}=0$ for all $\alpha\in \widetilde{\mathcal{S}}_{ij}$. In other words,
\begin{eqnarray}
\mathbb{P}\left[\overline{g}_{jk}=0\right] & = & \mathbb{P}\left[g_{j k}=0,\, g_{\alpha k}=0\,\forall{\alpha \in \widetilde{\mathcal{S}}_{ij}}\right]\nonumber\\
& = & (1-p_N)^{S-1}.
\end{eqnarray}
Hence, $\overline{g}_{j k}=1$ with probability $\overline{p}_N=1-(1-p_N)^{S-1}$.
But now {\em homogenizing} the transformed graph $\overline{\bm{G}}$ is an easy task. It suffices to augment the connection probabilities of all the remaining pairs (including those in $\mathcal{S}$), in order to match the connection probability in~(\ref{eq:superjconn}). More formally, resorting to a simple coupling between Bernoulli random variables, we can define a new (random) graph~$G^{\star}$ from $\overline{G}$ as
\begin{equation}
g^{\star}_{uv}=\max\left\{\overline{g}_{uv},q_{uv}\right\}\label{eq:bernoullihomog2}
\end{equation}
where $\left\{q_{uv}\right\}_{u<v}$ are i.i.d. Bernoulli random variables with
\begin{equation}
\mathbb{P}\left[q_{uv}=1\right]\!\!=\!\!\left\{\begin{array}{cl}  \!\! 1-(1-p_N)^{S-1}, \! \! & \!\!  \mbox{If }u\in \widetilde{\mathcal{S}}_{ij}\\
                                                          \!\! 1-(1-p_N)^{S-1},  \! \! & \!\!   \mbox{If }(u,v)=(i,j)\\
                                                          \!\! 1-(1-p_N)^{S-2},  \! \! & \!\!   \mbox{If }u\in\left\{i\right\}\cup\mathcal{S}' \mbox{ and }v\in \mathcal{S}' \\
                                                          \!\! 0,                \! \! & \!\!   \mbox{If }u=j \mbox{ and }v\in \mathcal{S}'.  \end{array}\right.\label{eq:bernoullihomog}
\end{equation}
The resulting graph~$G^{\star}$ is Erd\H{o}s-R\'enyi with~$p^{\star}_{N}=1-(1-p_N)^{S-1}$. Figure~\ref{fig:homogenization} graphically summarizes the idea.
Moreover, since $1-(1-p_N)^{S-1}\leq S p_N$, and in order to obtain a random graph whose connection probability is explicitly given by~\eqref{eq:pconncond}, we can further define a graph $\widetilde{G}$ with connection probability given by
\beq
\widetilde{p}_N=S p_N=S \frac{\log N + c_N}{N}=
\frac{\log N + \widetilde{c}_N}{N},
\eeq
where $\widetilde{c}_N=(S-1)\log N + S c_N$, and with the coupling $g^{\star}_{uv}\leq\widetilde{g}_{uv}$ (realization-wise) for all $u,v$ -- this can be easily obtained via a standard coupling between Bernoulli random variables. Therefore $\overline{G}\subseteq \widetilde{G}$, i.e., $\overline{G}$ is a subgraph of $\widetilde{G}$, realization-wise. Since fleshing out a graph with new connections can only decrease distances and favor membership to any neighborhood, the implications shown in~(\ref{eq:overlineGimplic}) and~(\ref{eq:overlineGimplic2}) hold true with $\overline{\bm{G}}$ replaced by $\widetilde{\bm{G}}$. This implies, in view of~(\ref{eq:DsmallmHC}) and~(\ref{eq:DsmalltotHC}), that:
\beq
\mathcal{D}_{\textnormal{small}}\subseteq \widetilde{\mathcal{D}}_{\textnormal{small}},
\eeq
which in turn implies the claim of the theorem.

\section{Managing small distances}
\label{app:smalldist}

\begin{lemma}\label{lemma:dist2new}
Let $\bm{G}$ be a {\em pure} Erd\H{o}s-R\'enyi random graph $\mathscr{G}(N,p_N)$. We have that
\beq
\boxed{
\P[\delta_{i,j}(\bm{G})\leq r]
\leq  p_N \, (N p_N)^{r-1}\left(\frac{1}{1-1/(N p_N)}\right)
}
\label{eq:convergence2}
\eeq
\end{lemma}
\begin{IEEEproof}

Since the event $\{\delta_{i,j}(\bm{G})=r\}$ signifies that the {\em shortest} path connecting $i$ to $j$ has length equal to $r$, there must exist a path connecting $i$ to $j$ obeying the following conditions: $i)$ all intermediate nodes are visited only once through the path (otherwise the path itself could be squeezed to one of a shorter length); $ii)$ along the path, one cannot spend one or more steps lingering on the same node.
Accordingly, we can write:
\beq
\{\delta_{i,j}(\bm{G})=r\}\subseteq
\mathcal{E}\dfz\bigcup_{\mathcal{M}} \left\{\bm{g}_{i n_1}\bm{g}_{n_1 n_2}\cdots \bm{g}_{n_{r-1} j}=1\right\},
\label{eq:uniondist}
\eeq
where the set~$\mathcal{M}$ is defined as $\mathcal{M}\dfz \mathcal{M}_1 \cap \mathcal{M}_2$ with
\beqa
\mathcal{M}_1 \!\!\! & \!\!\! \dfz \!\!\! & \!\!\! \{n:=\left(n_{1},\ldots,n_{r-1}\right)\in\mathbb{N}^{r-1}: n_u\neq n_v\, \forall{u,v}\},
\label{eq:Msetdef1}\\
\mathcal{M}_2 \!\!\! & \!\!\! \dfz \!\!\! & \!\!\! \{n:=\left(n_{1},\ldots,n_{r-1}\right)\in\mathbb{N}^{r-1}: n_k\neq i,j\,\forall{k}\}.
\label{eq:Msetdef1}
\eeqa
It is useful to remark that the event $\mathcal{E}$ in~(\ref{eq:uniondist}) does not coincide with the event that the shortest path has length equal to $r$, because the possibility of having paths longer than $r$ is not ruled out. The event $\mathcal{E}$ in~(\ref{eq:uniondist}) simply underlies the existence of at least one path of length $r$ with the necessary characteristics, which explains the one-sided implication in~(\ref{eq:uniondist}), and yields $\P[\delta_{\ell,m}(\bm{G})= r ]\leq\P[\mathcal{E}]$.
We have
\beqa
\P\left[
\mathcal{E}
\right]\leq
\sum_{\mathcal{M}}
\P[
\bm{g}_{i n_1}\bm{g}_{n_1 n_2}\cdots \bm{g}_{n_{r-1} j}
=1]
=
M \; p_N^{r}.
\label{eq:proboundMh}
\eeqa
where we recall that~$M$ stands for the cardinality of the set~$\mathcal{M}$ in view of the notation in Sec.~\ref{sec:symbols}, where sets are represented by calligraphic letters and the corresponding cardinalities are represented by normal font letters. Observe that
\begin{equation}
M=(N-2)(N-3)\ldots(N-r)\leq (N-2)^{r-1}.
\end{equation}
Therefore,
\begin{equation}
\mathbb{P}\left[\delta_{i,j}(\bm{G})=r\right] \leq (N-2)^{r-1}p_N^{r}\leq p_N\left(Np_N\right)^{r-1}
\end{equation}
and as a result,
\beqa
\P[\delta_{i,j}(\bm{G})\leq r] & = &
\sum_{\alpha=1}^r
\P[\delta_{i,j}(\bm{G}) = \alpha]\leq
p_N \sum_{\alpha=1}^{r} (N p_N)^{\alpha-1}\nonumber\\ & = & p_N \sum_{\alpha=0}^{r-1} (N p_N)^{\alpha}
\leq
\frac{p_N\,(N p_N)^{r-1}}{1-1/(N p_N)}.
\label{eq:finalboundsmalldist}
\eeqa
\end{IEEEproof}

\end{appendices}

\section*{Acknowledgment}
Short and limited versions of this work appear in the conference publications [51] and [52].

\small
\bibliographystyle{myIEEEtran}
\bibliography{IEEEabrv,biblio,biblio_new}

\section*{}

\textbf{Augusto Santos} received the B.Sc. and M.Sc. from Instituto Superior T\'ecnico (IST), Lisbon, Portugal, in 2007 and 2008, respectively.
He received his Ph.D. from IST and Carnegie Mellon University (CMU), Pittsburgh, USA, in 2014.
All degrees were obtained in electrical and computer engineering.
He held a post-doctoral scholar position at CMU during 2015-2017 and another at the Adaptive Systems Laboratory at \'Ecole Polytechnique F\'ed\'erale de Lausanne (EPFL), Lausanne, Switzerland, during 2017-2019.

\vspace*{5mm}

\textbf{Vincenzo Matta} received the Laurea degree (cum laude) in
electronic engineering and the Ph.D. degree in information
engineering from the University of Salerno, Fisciano, Italy, in
2001 and 2005, respectively. He is currently an Associate
Professor with the Department of Information and Electrical
Engineering and Applied Mathematics, University of Salerno.
His research interests cover the wide area of statistical signal
processing and information theory, with current emphasis on:
Adaptation and learning over networks; the interplay between
inference, communications and security in distributed systems;
multiobject/multisensor tracking and data fusion; detection of
gravitational waves. He has published more than 100 articles on
international journals and proceedings of international
conferences. He serves as a Senior Area Editor for the IEEE
Signal Processing Letters, and formerly served as an Associate
Editor for the IEEE Transactions on Aerospace and Electronic
Systems, for the IEEE Signal Processing Letters, and for the
IEEE Transactions on Signal and Information Processing over networks

\vspace*{5mm}

\textbf{Ali H. Sayed} is Dean of Engineering at EPFL, Switzerland. He has also served as distinguished professor and former chairman of electrical engineering at UCLA. He is a member of the US National Academy of Engineering and recognized as a highly-cited researcher. An author of over 530 scholarly publications and six books, his research involves several areas including adaptation and learning, data and network sciences, and multi-agent systems. Dr. Sayed has received several awards. He is a Fellow of IEEE, EURASIP, and the American Association for the Advancement of Science (AAAS). He is serving as President of the IEEE Signal Processing Society.

\end{document}